\documentclass[12pt]{article}
\usepackage[top=2.0cm, left=2.0cm, bottom=3.0cm, right=2.0cm]{geometry}
\usepackage[english]{babel}%
\usepackage{graphicx}
\usepackage{amsfonts}
\usepackage{amsmath}
\usepackage{amssymb}

\def\qed{{\bf \hfill $\Box$}\endtrivlist}

\usepackage{tikz}
\usepackage{caption}
\usepackage{rotating}
\usetikzlibrary{shapes.geometric, arrows}
\tikzstyle{process} = [rectangle, minimum width=3cm, minimum height=1cm, text centered, draw=black, fill=white]
\tikzstyle{decision} = [diamond, minimum width=3cm, minimum height=1cm, text centered, draw=black, fill=white]
\tikzstyle{arrow} = [thick,->,>=stealth]

\begin{document}
\sloppy

\newtheorem{theorem}{Theorem}
\newtheorem{cor}{Corollary}
\newtheorem{example}{Example}

\renewcommand{\thesection}{\arabic{section}}
\renewcommand{\thetheorem}{\arabic{section}.\arabic{theorem}}
\renewcommand{\theequation}{\arabic{section}.\arabic{equation}}
\renewcommand{\theexample}{\arabic{section}.\arabic{example}}
\renewcommand{\thecor}{\arabic{section}.\arabic{cor}}

\title{Solution of Interpolation Problems \\ via the Hankel Polynomial Construction}

\author{
  Alexei Yu. Uteshev\footnote{The corresponding author}\\
  \textit{alexeiuteshev@gmail.com} \and
  Ivan Baravy\\
  \textit{ivan\_baravy@apmath.spbu.ru}
}

\date{St.\,Petersburg State University\\ St.\,Petersburg, Russia}

\maketitle
\thispagestyle{empty}

\begin{abstract}
We treat the interpolation problem $ \{f(x_j)=y_j\}_{j=1}^N $ for polynomial and rational functions. Developing the approach by C.Jacobi, we represent the interpolants by virtue of the Hankel polynomials generated by the
sequences $ \{\sum_{j=1}^N x_j^ky_j/W^{\prime}(x_j) \}_{k\in \mathbb N} $ and $ \{\sum_{j=1}^N x_j^k/(y_jW^{\prime}(x_j)) \}_{k\in \mathbb N} $; here $ W(x)=\prod_{j=1}^N(x-x_j) $. The obtained results are applied for the error correction problem, i.e. the problem of reconstructing the polynomial  from a redundant set of its values some of which are probably erroneous. The problem of evaluation of the resultant of polynomials $ p(x) $ and $ q(x) $ from the set of values $ \{p(x_j)/q(x_j) \}_{j=1}^N $ is also tackled within the framework of this approach.

\textbf{Keywords:} Polynomial interpolation, rational interpolation, Hankel matrices and polynomials, error correction, Berlekamp---Massey algorithm, resultant interpolation

\textbf{MSC 2010:} 68W30, 30E05, 16D05, 12Y05, 26C15.

\end{abstract}


%

\section{Introduction}

\setcounter{equation}{0}
\setcounter{theorem}{0}
\setcounter{example}{0}

The title of the paper describes ends and means of the present research: solution of the univariate interpolation problems with the aid of the \textbf{Hankel polynomials}, i.e. polynomials in the variable $ x $ with the following representation
in the determinantal form
\begin{equation}
\mathcal H_{k} (x) =
\left|
\begin{array}{lllll}
c_0     & c_1     & c_2 & \ldots & c_{k}  \\
c_1     & c_2     & c_3 &\ldots & c_{k+1}    \\
\vdots        &    &     & \ddots& \vdots    \\
c_{k-1} & c_{k} & c_{k+1} & \ldots   & c_{2k-1} \\
1       & x       & x^2 &  \ldots & x^{k}
\end{array} \right|
  \label{HankelPOL}
\end{equation}
Here the entries $ c_0,c_1,\dots, c_{2k-1} $ (generators of the polynomial) are the elements of some (finite or infinite) field.

Given the table of values for the variables $ x $ and $ y $
\begin{equation}
  \begin{array}{c|c|c|c|c}
    x & x_1 & x_2 & \ldots & x_N \\
\hline
    y & y_1 & y_2 & \ldots & y_N
  \end{array},
  \label{table}
\end{equation}
with distinct nodes $ x_1,\dots,x_N $, we treat the interpolation problem in a sense of finding a function $ f(x) $ such that $ f(x_j)=y_j $ for $ j\in \{1,\dots, N\} $. Systematic exploration of the problem was started since XVII century; for the historical overview and the review of numerous applications of the relevant theory we refer to \cite{Mejiering} and the references cited therein.

Depending on the concrete objectives, the solution to the problem can be looked for in various classes of interpolants like  algebraic polynomial, trigonometric polynomial, sum of exponentials, rational functions  etc.
The two interpolation problems dealt with  in the present paper are polynomial and rational ones.

In comparison with the polynomial interpolation problem, the rational interpolation one has its beginning a century and a half later, with the first explicit formula due to Cauchy \cite{Cauchy}. Further its development was made by Kronecker \cite{Kronecker81_2} and Netto \cite{Netto} (we briefly discuss it in Section \ref{rat-interp}).  The interest to the problem revives in the second part of the XX century and is connected with its application in Control Theory (recovering of the transfer function from the frequency responses) and Error Correcting Codes; in the latter case the problem is treated in finite fields.  We refer to  \cite{Becker&Lab,DAndrea15,VZG} for recent developments and further references on the rational interpolation problem as well as to its generalization, known as rational \textbf{Hermite's} or osculatory \textbf{rational interpolation} problem, where the values of some derivatives for $ f(x) $ are assumed to be specified at some nodes.

 In Section \ref{rat-interp} we develop  an approach to the rational interpolation problem originated in 1846 by Carl Jacobi \cite{Jacobi46} and nearly forgotten since then. Within the framework of this approach, the numerator and the denominator of interpolation fraction $ f(x)=p(x)/q(x), \deg p(x)+ \deg q(x) = N-1  $ are constructed in the form of the Hankel polynomials generated by the sequences of values
\begin{equation}
\widetilde \tau_k=\sum_{j=1}^N \frac{1}{y_j} \frac{x_j^k}{W^{\prime}(x_j)} \quad \mbox{ and } \quad   \tau_k=\sum_{j=1}^N y_j \frac{x_j^k}{W^{\prime}(x_j)} \quad \mbox{ for } k\in \{0,1,\dots \}
\label{posti}
\end{equation}
correspondingly; here $ W(x)=\prod_{j=1}^N(x-x_j) $. Aside from theoretical results on existence and uniqueness of solution, we focus ourselves on the computational aspects of the suggested approach.
Indeed, computation of a parameter dependent determinant  in case of its large order is not a trivial task. Fortunately a specific structure of the involved determinants
helps much: there exists a recursive procedure for the Hankel polynomial computation. It is based on the identity linking the Hankel determinants of three successive orders:
\begin{equation}
A_k \mathcal H_{k-2} (x)+ (B_k-x) \mathcal H_{k-1}(x) + 1/A_k \mathcal H_{k}(x) \equiv 0 \, ,
\label{INT_JJ}
\end{equation}
here $ A_k $ and $ B_k $ are some constants. This identity was first deduced  in 1836 by aforementioned Jacobi \cite{Jacobi36} and completely proved in 1854 by his disciple Ferdinand Joachimsthal \cite{Joach}. We
present this result in Section \ref{section-hankel}. Formula (\ref{INT_JJ}), backed also up with  some extra formulas for evaluation of $ A_k $ and $ B_k $,  allows  one to recursively compute any Hankel polynomial. We discuss also an opportunity to extend this recursive procedure to degenerate case when, for instance, $ A_k= 0 $. As a matter of fact,  formula (\ref{INT_JJ}) should be treated as  an origination of the algorithm which is now known as the \textbf{Berlekamp---Massey algorithm} \cite{blahut}; it was suggested for the  decoding procedure in BCH or Reed-Solomon codes  and for finding  the minimal polynomial of a linear recurrent sequence. Application of this algorithm to Jacobi's approach for rational interpolation problem provides one with an opportunity to efficiently compute not only a single interpolant but the whole family of fractions with all the possible combinations of degrees for the numerator and the denominator satisfying the restriction $ \deg p(x)+ \deg q(x) = N-1  $.

In between of Sections \ref{section-hankel} and \ref{rat-interp} with Jacobi related results, the present paper contains two sections which are focused to polynomial interpolation. Section \ref{SPoly_interp}
is devoted to the classical polynomial interpolation problem and contains the result which, at first glance, can be judged as a trivial theoretical corollary for the results of Section \ref{rat-interp} on the rational interpolation problem. Indeed, representation for the polynomial interpolant in the form of the Hankel   determinant generated by the first sequence from  (\ref{posti}), even if accompanied with an efficient procedure for its  computation, is hardly competitive with the traditional  (Lagrange or Newton) schemes for the interpolant construction.

Justification to the proposed approach is given in Section \ref{SPoly_Error} where we consider the problem of interpolation of redundant but corrupted table (\ref{table}). Assuming that the $ x $-values are ``true'' and corresponding (observations, measurements) $ \{y_{j}\}_{j=1}^N $ are ``noisy'', the problem is to find a polynomial $ p(x) $ of a degree $ n <N-1 $ (generically, $ n \ll N $), such that  $ \{p(x_j)\}_{j=1}^N $ approximate the given data set $ \{ y_j \}_{j=1}^N $ as close as possible. The most commonly used method for solving the problem, if treated over $ \mathbb R $, consists in minimization of the sum
$$ \sum_{j=1}^N (p(x_j)-y_j)^2  $$
by appropriate selection of (real) coefficients of $ p(x) $; this is known as the \textbf{polynomial least squares method}. Under an assumption of the  normality of distribution for errors, the method yields the maximum likelihood estimator. However, the method is sensitive to the occurrence of \emph{outliers}, i.e. incidental systematic error (flaws). We are treating the problem of finding these errors in the following statement: The table (\ref{table}) has been originally generated by a polynomial $ p(x) $ of a degree $ n <N-1 $, after that up to $ E $ values of $ y $ has been probably corrupted. Find the corrupted nodes and the original polynomial. In finite fields and for the above mentioned BCH or Reed-Solomon codes, solution of this problem is a cornerstone of the error correction algorithm by Berlekamp and Welch \cite{Ber&Wel}.
It turns out that the sequences of the Hankel polynomials generated by (\ref{posti}) solve the problem if $ n<N-2\,E $. The sequence $ \{ \mathcal H_{\ell}(x) \}_{\ell=1}^{N-1} $ generated by the sequence $ \{ \tau_k \} $ contains at least one polynomial coinciding with the\footnote{We utilize the notion from the Coding Theory.} \textbf{error locator polynomial}, i.e. the polynomial with the set of zeros coinciding with the set of all the nodes $ x_j $ corresponding to corrupted values $ y_j $. Similar sequence of the Hankel polynomials generated by $ \{ \widetilde \tau_k \} $ contains a polynomial which equals the product of the original polynomial $ p(x) $ by the error locator polynomial.

In Section \ref{ResInterp} we deal with the question which relates to the uniqueness problem for the rational interpolation. Having the table (\ref{table}) generated by some rational fraction $ p(x)/q(x) $ with $ \deg p(x)+\deg q(x) = N-1 $ is it possible to establish that polynomials $ p(x) $ and $ q(x) $ do not possess a common zero --- avoiding, as an intermediate step, their explicit representation? The question is equivalent to possibility of expressing the \textbf{resultant} of polynomials $ p(x) $ and $ q(x) $ in terms of the entries of the table (\ref{table}). We prove that the resultant can be expressed in the form of an appropriate  Hankel determinant generated by any of the sequences (\ref{posti}).

~\\
\indent\textbf{Remark 1.1.} The results of the paper are formulated and exemplified for the case of an infinite field. Nevertheless, nearly all of these results will work in finite fields (with some additional assumptions on their orders).

\section{The Hankel Determinants and Polynomials}\label{section-hankel}

\setcounter{equation}{0}
\setcounter{theorem}{0}
\setcounter{example}{0}

For a (finite or infinite) sequence of complex numbers
\begin{equation}
\{c_j\}_{j=0}^{\infty}=\left\{ c_0,c_1,\dots \right\}
\label{seq0}
\end{equation}
matrix of the form
\begin{equation}
\left[c_{i+j-2} \right]_{i,j=1}^{k} =
\left[
\begin{array}{lllll}
c_0     & c_1     & c_2 & \ldots & c_{k-1}  \\
c_1     & c_2     & c_3 & \ldots & c_{k}    \\
\vdots        &     &    & \ddots&   \vdots       \\
c_{k-2} & c_{k-1} & c_k & \ldots & c_{2k-3} \\
c_{k-1} & c_{k}   & c_{k+1} & \ldots & c_{2k-2}
\end{array} \right]_{k \times k}
\label{Han_matr}
\end{equation}
is called the \textbf{Hankel matrix} of the order $ k $ generated by the sequence (\ref{seq0}); its determinant will be denoted by $ H_k (\{c\}) $ or simply $ H_k $ if it will not cause misunderstandings.

The following result  is a particular case of the Sylvester determinantal identity \cite{Gantmacher}. 

\begin{theorem}  The Hankel determinants of three successive orders are linked by the equality:
\begin{equation}
H_{k-2} H_k = H_{k-1}
\left|
\begin{array}{llcll}
c_0 & c_1 & \dots & c_{k-3} & c_{k-1} \\
c_1 & c_2 & \dots & c_{k-2} & c_{k} \\
\vdots & & & & \vdots \\
c_{k-3} & c_{k-2} & \dots & c_{2k-6} & c_{2k-4} \\
c_{k-1} & c_{k} & \dots & c_{2k-4} & c_{2k-2}
\end{array} \right|-
\left|
\begin{array}{llcll}
c_0 & c_1 & \dots & c_{k-3} & c_{k-1} \\
c_1 & c_2 & \dots & c_{k-2} & c_{k} \\
\vdots & & & & \vdots \\
c_{k-3} & c_{k-2} & \dots & c_{2k-6} & c_{2k-5} \\
c_{k-1} & c_{k} & \dots & c_{2k-4} & c_{2k-3}
\end{array} \right|^2 \enspace .
\label{sylv1}
\end{equation}
\end{theorem}

\begin{theorem} \label{Han_minors} If the Hankel matrix is generated by the sequence (\ref{seq0}) where
$$ \left\{c_j=\sum_{\ell=1}^n \alpha_{\ell}^j \beta_{\ell} \right\}_{j=0}^{\infty} \quad \mbox{ for some} \ \{\alpha_{\ell},\beta_{\ell} \}_{\ell=1}^n \subset \mathbb C $$
then any its minor of the order $ k > n $ equals zero.
\end{theorem}

\textbf{Proof} will be illustrated for the specialization $ n=3, k=4 $. An arbitrary $ 4 $th order minor of the Hankel matrix (\ref{Han_matr}) is the determinant of the matrix
$$
\left(\begin{array}{cccc}
c_{i_1+j_1} & c_{i_1+j_2} & c_{i_1+j_3} & c_{i_1+j_4} \\
c_{i_2+j_1} & c_{i_2+j_2} & c_{i_2+j_3} & c_{i_2+j_4} \\
c_{i_3+j_1} & c_{i_3+j_2} & c_{i_3+j_3} & c_{i_3+j_4} \\
c_{i_4+j_1} & c_{i_4+j_2} & c_{i_4+j_3} & c_{i_4+j_4}
\end{array}
\right)
$$
for some sequences of subscripts $ i_1<i_2<i_3< i_4 $ and $ j_1<j_2<j_3< j_4 $. This matrix can be represented as the product
$$
=\left(
\begin{array}{ccc}
\alpha_1^{i_1} & \alpha_2^{i_1} & \alpha_3^{i_1} \\
\alpha_1^{i_2} & \alpha_2^{i_2} & \alpha_3^{i_2} \\
\alpha_1^{i_3} & \alpha_2^{i_3} & \alpha_3^{i_3} \\
\alpha_1^{i_4} & \alpha_2^{i_4} & \alpha_3^{i_4}
\end{array}
\right)
\left(
\begin{array}{ccc}
\beta_1& 0 & 0 \\
0 & \beta_2 & 0 \\
0 & 0 & \beta_3
\end{array}
\right)
\left(
\begin{array}{cccc}
\alpha_1^{j_1} & \alpha_1^{j_2} & \alpha_1^{j_3} & \alpha_1^{j_4} \\
\alpha_2^{j_1} & \alpha_2^{j_2} & \alpha_2^{j_3} & \alpha_2^{j_4} \\
\alpha_3^{j_1} & \alpha_3^{j_2} & \alpha_3^{j_3} & \alpha_3^{j_4} \\
\end{array}
\right)
$$
Due to the Cauchy-Binet formula, the determinant of this product equals zero.  \qed

If we replace the last row of the Hankel matrix of the order $ k+1 $   with the row of powers of $x$, then the corresponding determinant
\begin{equation}
\mathcal H_k(x; \{c\}) =
\left|
\begin{array}{lllll}
c_0     & c_1     & c_2 & \ldots & c_{k}  \\
c_1     & c_2     & c_3 &\ldots & c_{k+1}    \\
\vdots        &    &     & \ddots& \vdots    \\
c_{k-1} & c_{k} & c_{k+1} & \ldots   & c_{2k-1} \\
1       & x       & x^2 &  \ldots & x^{k}
\end{array} \right|_{(k+1) \times (k+1)}
\label{Hkx}
\end{equation}
or simply $ \mathcal H_k(x) $, is called the $ k $th \textbf{Hankel polynomial} \cite{Henrici} generated by the sequence (\ref{seq0}). Expansion of (\ref{Hkx}) by its last row yields
\begin{equation}
\mathcal H_k(x)\equiv h_{k0} x^{k}+h_{k1}x^{k-1}+h_{k2}x^{k-2}+\dots \quad {\rm with } \ h_{k0}=H_k \, .
\label{Hk}
\end{equation}
Thus $ \deg \mathcal H_k(x) =k $ if and only if $ H_k \ne 0 $. We will also utilize an alternative representation for the Hankel polynomial in the form of the Hankel determinant resulting from linear transformation of the columns of the determinant (\ref{Hkx}):
\begin{equation}
\mathcal H_k(x; \{c\})=(-1)^k
\left|
\begin{array}{lllll}
c_1 -c_0x    & c_2-c_1x & \ldots & c_{k}-c_{k-1}x  \\
c_2-c_1x      & c_3-c_2x &\ldots & c_{k+1}-c_kx    \\
\vdots            &     & \ddots& \vdots    \\
c_{k}-c_{k-1}x & c_{k+1}-c_kx & \ldots   & c_{2k-1}-c_{2k-2}x
\end{array} \right|_{k \times k}=(-1)^k H_k (\{c_{j+1}-c_jx\}_{j=0}^{\infty})
\label{Hkx1}
\end{equation}

It turns out that there exists an explicit linear relation between any three consecutive Hankel polynomials generated by arbitrary sequence (\ref{seq0}).

\begin{example} \label{Ex2} For the sequence
$$ \{1,1,2,-1,-9,-142,-2051,-29709,-430018,-6224467,\dots \} $$
find the Hankel polynomials $ \{\mathcal H_k(x)\}_{k=1}^5 $
\end{example}

\textbf{Solution.} One has:
$$
\mathcal H_1(x)=x-1,\ \mathcal H_2(x)=x^2+3\,x-5,\  \mathcal H_3(x)=-22\,x^3+164\,x^2+316\,x-666,\
$$
$$
\mathcal H_4(x)=19656\,x^4-278356\,x^3-97864\,x^2+93808\,x+468,
$$
$$
\mathcal H_5(x)=4638712(x^5-14\,x^4-7\,x^3+2\,x^2-3\,x+8) \ .
$$
It can be verified that
$$
-22\, \mathcal H_1(x)+\left(\frac{115}{11}-x\right)\mathcal H_2(x) -\frac{1}{22} \mathcal H_3(x) \equiv 0,
$$
$$
-\frac{9828}{11}\, \mathcal H_2(x) +\left( \frac{27887}{4158}-x \right) \mathcal H_3(x)-\frac{11}{9828} \mathcal H_4(x) \equiv 0,
$$
and
$$
\frac{44603}{189} \mathcal H_3(x)+ \left( -\frac{61}{378}-x \right) \mathcal H_4(x)+\frac{189}{44603} \mathcal H_5(x) \equiv 0 \ .
$$
\qed

\begin{theorem}[Jacobi, Joachimsthal] \label{TJoach} Any three consecutive Hankel polynomials
$$ \mathcal H_{k-2}(x), \mathcal H_{k-1}(x), \mathcal H_{k}(x) $$
are linked by the following identity
\begin{equation}
H_k^2\mathcal H_{k-2}(x) + \left(H_kh_{k-1,1}-H_{k-1}h_{k1}-H_kH_{k-1}x\right)\mathcal H_{k-1}(x) + H_{k-1}^2 \mathcal H_{k}(x)  \equiv 0
\label{Joach_iden}
\end{equation}
which will be referred to as the \textbf{JJ-identity}.
\end{theorem}

\textbf{Proof}.  We modernize slightly the style of original proof given in \cite{Joach}. First consider the case where the
generating sequence (\ref{seq0}) is given as
\begin{equation}
 c_j=\sum_{\ell=1}^m \lambda_{\ell}^j, \quad \mbox{\rm for } \ j \in \{0,\dots,2k-1\}
 \label{seq1}
 \end{equation}
and for arbitrary distinct $ \lambda_1,\dots,\lambda_m $ with $ m > k $. Hence, $ c_0=m $.

\textbf{Lemma 2.1} \label{Tvspom1} \emph{The following equalities are valid:}
 \begin{equation}
 \sum_{\ell=1}^m \lambda_{\ell}^j \mathcal H_k(\lambda_{\ell}) =
 \left\{ \begin{array}{ll}
 0 & if \ j \in \{0,\dots, k-1\}, \\
 H_{k+1} & if \ j = k \, .
 \end{array} \right.
 \label{eqty100}
 \end{equation}

\textbf{Proof of Lemma 2.1.}
$$
\lambda_1^j \mathcal H_k(\lambda_1) + \lambda_2^j \mathcal H_k(\lambda_2) + \dots + \lambda_m^j \mathcal H_k(\lambda_m)
$$
$$
=\lambda_1^j\left|
\begin{array}{llll}
c_0     & c_1     & \ldots & c_{k}  \\
c_1     & c_2     & \ldots & c_{k+1}    \\
\vdots        &         & \ddots& \vdots    \\
c_{k-1} & c_{k} & \ldots   & c_{2k-1} \\
1       & \lambda_1       & \ldots & \lambda_1^{k}
\end{array}
\right|+
\lambda_2^j\left|
\begin{array}{llll}
c_0     & c_1     & \ldots & c_{k}  \\
c_1     & c_2     & \ldots & c_{k+1}    \\
\vdots        &         & \ddots& \vdots    \\
c_{k-1} & c_{k} & \ldots   & c_{2k-1} \\
1       & \lambda_2       & \ldots & \lambda_2^{k}
\end{array}
\right|+\dots+
\lambda_m^j\left|
\begin{array}{llll}
c_0     & c_1     & \ldots & c_{k}  \\
c_1     & c_2     & \ldots & c_{k+1}    \\
\vdots        &         & \ddots& \vdots    \\
c_{k-1} & c_{k} & \ldots   & c_{2k-1} \\
1       & \lambda_m       & \ldots & \lambda_m^{k}
\end{array}
\right|
$$
Using the linear property of the determinant convert this linear combination of the determinants into a single one:
$$
=
\left|
\begin{array}{llll}
c_0     & c_1     & \ldots & c_{k}  \\
c_1     & c_2     & \ldots & c_{k+1}    \\
\vdots        &         & \ddots& \vdots    \\
c_{k-1} & c_{k} & \ldots   & c_{2k-1} \\
c_j       &   c_{j+1}     & \ldots & c_{j+k}
\end{array}
\right| \, .
$$
If $ j < k $ then the last determinant possesses two identical rows. Therefore, in this case, it is just zero. For $ j=k $ the obtained determinant coincides with $ H_{k+1} $. \qed

\textbf{Proof of Theorem \ref{TJoach} (continued)}.

Assuming that $ H_{k-1}\ne 0 $ (i.e. $ \deg \mathcal  H_{k-1}(x)= k-1 $) divide $ \mathcal H_k(x) $ by $ \mathcal H_{k-1}(x) $:
\begin{equation}
\mathcal H_k(x) \equiv Q(x) \mathcal H_{k-1}(x) +R(x) \, .
\label{ident2}
\end{equation}
Here the coefficients of the quotient
$$ Q(x)=Q_0+Q_1x $$
can be determined from those of $ \mathcal H_k(x) $ and  $ \mathcal H_{k-1}(x) $ via the undetermined coefficients method:
\begin{equation}
Q_1=\frac{H_k}{H_{k-1}}\, ,\ Q_0=\frac{H_{k-1}h_{k1}-H_kh_{k-1,1}}{H_{k-1}^2} \, ,
\label{L,M}
\end{equation}
or, alternatively, $ Q(x) $ can be represented in the determinantal form as
\begin{equation}
Q(x) \equiv  - \left|\begin{array}{ccc} H_{k-1} & 0 & H_k \\ h_{k-1,1} & H_{k-1} & h_{k1} \\ x & 1 & 0 \end{array} \right|
\Bigg/ H_{k-1}^2\enspace .
\label{alterL,M}
\end{equation}
To find the coefficients of the remainder $ R(x) $
\begin{equation}
R(x) = R_0+R_1x+\dots+ R_{k-2}x^{k-2}
\label{eqr}
\end{equation}
substitute $ x = \lambda_1,\dots,x=\lambda_m $ into (\ref{ident2}):
\begin{equation}
\left\{\begin{array}{ll}
\mathcal H_k(\lambda_1) &= \left(Q_1\lambda_1+Q_0 \right) \mathcal H_{k-1}(\lambda_1) + \left( R_0+R_1\lambda_1+\dots+ R_{k-2}\lambda_1^{k-2}\right) , \\
\mathcal H_k(\lambda_2) &= \left(Q_1\lambda_2+Q_0 \right) \mathcal H_{k-1}(\lambda_2) + \left( R_0+R_1\lambda_2+\dots+ R_{k-2}\lambda_2^{k-2}\right) , \\
\dots &  \dots \, , \\
\mathcal H_k(\lambda_m) &= \left(Q_1\lambda_m+Q_0 \right) \mathcal H_{k-1}(\lambda_m) + \left( R_0+R_1\lambda_m+\dots+ R_{k-2}\lambda_m^{k-2}\right) \, .
\end{array}
\right.
\label{syst101}
\end{equation}
Summation of  these equalities yields
$$
\mathcal H_k(\lambda_1)+\mathcal H_k(\lambda_2) + \dots +  \mathcal H_k(\lambda_m)
$$
$$
=Q_1\left( \lambda_1\mathcal H_{k-1}(\lambda_1)+\lambda_2\mathcal H_{k-1}(\lambda_2)+\dots+\lambda_m \mathcal  H_{k-1}(\lambda_m) \right)+
Q_0\left( \mathcal H_{k-1}(\lambda_1)+ \mathcal H_{k-1}(\lambda_2)+\dots+\mathcal H_{k-1}(\lambda_m) \right)
$$
$$
+ (c_0R_0+c_1R_1+\dots+c_{k-2}R_{k-2}) \, .
$$
Due to (\ref{eqty100}), one gets
$$
0=c_0R_0+c_1R_1+\dots+c_{k-2}R_{k-2} \, .
$$
Next multiply every equality (\ref{syst101}) by corresponding  $ \lambda_{\ell} $ for $ \ell\in \{1,\dots, m\} $ and sum up the obtained equalities. The resulting equality looks similar to the previously deduced:
$$
0=c_1R_0+c_2R_1+\dots+c_{k-1}R_{k-2} \, .
$$
In a similar way, with the aid of multiplication of (\ref{syst101}) by equal powers of $ \lambda_1,\dots, \lambda_{\ell} $, one obtains
\begin{eqnarray*}
0 & = & c_2R_0+c_3R_1+\dots+c_{k}R_{k-2} \, , \\
&  &  \dots \, , \\
0&= & c_{k-3}R_0+c_{k-2}R_1+\dots+c_{2k-5}R_{k-2} \ .
\end{eqnarray*}
Multiplication of equalities (\ref{syst101}) by $ \lambda_1^{k-2},\dots, \lambda_m^{k-2} $ yields something different:
$$
0=H_k Q_1+ c_{k-2}R_0+c_{k-1}R_1+\dots+c_{2k-4}R_{k-2}  \, .
$$
Unifying the obtained relations for $ R_0,\dots,R_{k-2} $ with (\ref{eqr}), one gets the linear system:
$$
\left\{\begin{array}{cccccc}
c_0R_0 &+c_1R_1 &+\dots & +c_{k-2}R_{k-2} & & =0, \\
c_1R_0 &+c_2R_1 &+\dots & +c_{k-1}R_{k-2} & & =0, \\
\dots & & & & & \dots \\
c_{k-3}R_0 & +c_{k-2}R_1 & +\dots &+c_{2k-5}R_{k-2} & & =0, \\
c_{k-2}R_0 & +c_{k-1}R_1 & +\dots &+c_{2k-4}R_{k-2} &+H_kQ_1 & =0,\\
R_0& +R_1x &+\dots &+ R_{k-2}x^{k-2} & -R(x) & =0.
\end{array}
\right.
$$
Consider it as a system of homogeneous equations with respect to the variables $ R_0,R_1,\dots,R_{k-2},1  $.
Since it possesses a nontrivial solution, its determinant necessarily vanishes:
$$
\left|
\begin{array}{ccccc}
c_0 & c_1 & \dots & c_{k-2} & 0 \\
c_1 & c_2 & \dots & c_{k-1} & 0 \\
\vdots & & & & \vdots \\
c_{k-3} & c_{k-2} & \dots & c_{2k-5} & 0 \\
c_{k-2} & c_{k-1} & \dots & c_{2k-4} & H_kQ_1 \\
1 & x & \dots & x^{k-2} & -R(x)
\end{array}
\right|=0 \, .
$$
Expansion of the determinant by its last column yields:
$$
R(x) H_{k-1} + H_kQ_1 \mathcal H_{k-2}(x) \equiv 0 \, .
$$
Together with the already obtained expression (\ref{L,M}) for $ Q_1 $, this confirms the validity of (\ref{Joach_iden}) for the particular case of generating sequence given by (\ref{seq1}).

Consider now the case of arbitrary generating sequence (\ref{seq0}). For any given sequence of complex numbers $ c_1,\dots,c_{2k-1} $ it is possible to find complex numbers $ \lambda_1,\dots, \lambda_{\ell} $ with $ \ell > 2k-1 $ such that the equations (\ref{seq1}) are consistent. These numbers can be chosen to be the zeros of a polynomial of the degree $ \ell $ whose
first $ 2k-1 $ Newton sums \cite{Uspensky} coincide with $ \{ c_j \}_{j=1}^{2k-1} $.

To complete the proof of (\ref{Joach_iden}), one should fill one gap in the arguments of the previous paragraph. Whereas the numbers $ c_1,\dots,c_{2k-1} $ can be chosen arbitrarily, the number $ c_0 $ takes the positive integer value, namely $ \ell $.
Thus the validity of (\ref{Joach_iden}) is proved only for any positive integer $ c_0 $. However, this equality is an algebraic one in $ c_0 $. Being valid for an infinite set of integers, it should be valid for any $ c_0 \in \mathbb C $.  \qed

The relationship (\ref{Joach_iden}) gives rise to a more symmetric form which was demonstrated in solution to Example \ref{Ex2}:

\begin{cor} If $H_{k} \ne 0, H_{k-1} \ne 0 $  then the JJ-identity can be written down as
\begin{equation}
\frac{H_k}{H_{k-1}}\mathcal H_{k-2}(x)- \left(x-\frac{h_{k-1,1}}{H_{k-1}}+\frac{h_{k1}}{H_{k}} \right)\mathcal H_{k-1}(x)+\frac{H_{k-1}}{H_{k}}\mathcal H_{k}(x) \equiv 0 \enspace .
\label{Joach_iden1}
\end{equation}
\end{cor}

The JJ-identity permits one to generate the recursive procedure for computation of the Hankel polynomials. Indeed, assume that
the expressions for $ \mathcal H_{k-2}(x) $ and $ \mathcal H_{k-1}(x) $ are already computed and
\begin{equation}
\mathcal H_{k-1}(x) \equiv  h_{k-1,0} x^{k-1}+h_{k-1,1}x^{k-2}+\dots+ h_{k-1,k-1} \quad {\rm with } \ h_{k-1,0}=H_{k-1} \, .
\label{Hk-1}
\end{equation}
Then in (\ref{Joach_iden}) all the constants are also evaluated except for $ H_k $ and  $ h_{k1} $ for which one has just only their
determinantal representations:
$$ H_k =
\left|
\begin{array}{lllll}
c_0 & c_1 & \dots & c_{k-2} & c_{k-1} \\
c_1 & c_2 & \dots & c_{k-1} & c_{k} \\
\vdots & & & & \vdots \\
c_{k-2} & c_{k-1} & \dots & c_{2k-4} & c_{2k-3} \\
c_{k-1} & c_{k} & \dots & c_{2k-3} & c_{2k-2}
\end{array}
\right| \quad \mbox{\rm and } \quad
 h_{k1} = -
 \left|
\begin{array}{lllll}
c_0 & c_1 & \dots & c_{k-2} & c_{k} \\
c_1 & c_2 & \dots & c_{k-1} & c_{k+1} \\
\vdots & & & & \vdots \\
c_{k-2} & c_{k-1} & \dots & c_{2k-4} & c_{2k-2} \\
c_{k-1} & c_{k} & \dots & c_{2k-3} & c_{2k-1}
\end{array}
\right| \, .
 $$
 These determinants differs from the transposed determinantal representation for $ \mathcal H_{k-1}(x) $ only in their last columns. Expansions by the entries of the last columns have the same values for corresponding cofactors, and, therefore, the following formulas
\begin{equation}
\left\{\begin{array}{rcl}
h_{k0}=H_k&=&c_{k-1}h_{k-1,k-1}+c_{k}h_{k-1,k-2}+\dots+c_{2k-2}h_{k-1,0}, \\
h_{k1}&=&-(c_{k}h_{k-1,k-1}+c_{k+1}h_{k-1,k-2}+\dots+c_{2k-1}h_{k-1,0})
\end{array}
\right.
\label{hk0hk1}
\end{equation}
allow one to evaluate  $ h_{k0} $ and  $ h_{k1} $ via the already computed coefficients of $ \mathcal H_{k-1}(x) $.

However the just outlined algorithm for recursive computation of $ \mathcal H_{k}(x) $ fails
for the case where $ H_{k-1}=0 $. We now wish to modify the procedure in order to cover this case.

\begin{theorem} \label{TJoach1} Let $ H_{k-2} \ne 0, H_{k-1}=0 $. If $ h_{k-1,1}=0 $ then
\begin{equation}
\mathcal H_{k-1}(x) \equiv 0
\label{HanHk01}
\end{equation}
and
\begin{equation}
\mathcal H_k(x) \equiv \frac{h_{k2}}{H_{k-2}} \mathcal H_{k-2}(x) \, .
\label{HanHk0}
\end{equation}
Otherwise
\begin{equation}
\mathcal H_{k-1}(x) \equiv \frac{h_{k-1,1}}{H_{k-2}}\mathcal H_{k-2}(x)
\label{HanHk11}
\end{equation}
and
\begin{equation}
\mathcal H_k(x) \equiv \frac{H_kH_{k-2}h_{k-1,1}\mathcal H_{k-3}(x)- \left|\begin{array}{cccc} H_{k-2} & 0 & 0 & H_k \\ h_{k-2,1} & H_{k-2} & 0 & h_{k1} \\
h_{k-2,2} & h_{k-2,1} & H_{k-2} & h_{k2} \\
x^2 & x &  1 & 0 \end{array} \right| \mathcal H_{k-2}(x)}{H_{k-2}^3} \, .
\label{HanHk1}
\end{equation}
\end{theorem}

\textbf{Proof.} If $ H_{k-2}\ne 0 $ and $ H_{k-1}=0, h_{k-1,1}=0 $ then all the $ (k-1) $th order minors of the matrix
\begin{equation}
\left(\begin{array}{llllll}
c_0 & c_1 & \dots & c_{k-3} & c_{k-2} & c_{k-1} \\
c_1 & c_2 & \dots & c_{k-2} & c_{k-1} & c_{k} \\
\vdots & & & & & \vdots \\
c_{k-3} & c_{k-2} & \dots & c_{2k-6} & c_{2k-5} & c_{2k-4} \\
c_{k-2} & c_{k-1} & \dots & c_{2k-5} & c_{2k-4} & c_{2k-3}
\end{array}
\right)_{(k-1)\times k}
\label{Hk-1k}
\end{equation}
are zero. This statement is based on the following

\textbf{Lemma 2.2}  \label{ThE-R} \cite{Bocher}.
\emph{If in a given matrix a certain} $ \mathfrak r $th \emph{order minor is not zero, and all the} $ (\mathfrak r+1) $th \emph{order minors containing that}
$ \mathfrak r $th \emph{order minor are zero, then all the} $ (\mathfrak r+1) $th \emph{order minors are zero (and, therefore, the rank of the matrix equals} $ \mathfrak r $).

The rank of the matrix (\ref{Hk-1k}) equals $ k-2 $, therefore all the coefficients of the polynomial $ \mathcal H_{k-1}(x) $ are also zero since they are the $ (k-1) $th minors of this matrix.

To establish the validity of (\ref{HanHk0}), it is sufficient to prove that the remainder of the division of $ \mathcal H_{k}(x) $ by $ \mathcal H_{k-2}(x) $ is identically zero. This will be done later, while now we intend to prove (\ref{HanHk11}). Since $ H_{k-1}=0 $, the identity (\ref{Joach_iden}) can be rewritten as
\begin{equation}
H_k(H_k \mathcal H_{k-2}(x)+h_{k-1,1}\mathcal H_{k-1}(x)) \equiv 0 \, .
\label{ttt1}
\end{equation}
If $h_{k-1,1}\ne 0 $ then the Sylvester identity (\ref{sylv1}) leads one to
$$
H_kH_{k-2} = -  h_{k-1,1}^2\, ,
$$
wherefrom it follows that
$$
H_k \ne 0 \quad \mbox{ and } \quad H_k=-  h_{k-1,1}^2/H_{k-2} \, .
$$
Consequently the identity (\ref{ttt1}) leads one to
$$
\mathcal H_{k-1}(x) \equiv - \frac{H_k}{h_{k-1,1}} \mathcal H_{k-2}(x) \equiv \frac{h_{k-1,1}}{H_{k-2}}\mathcal H_{k-2}(x) \,
$$
which proves (\ref{HanHk11}).

We now intend to prove (\ref{HanHk0}) and (\ref{HanHk1}). Divide $ \mathcal H_k(x) $ by $ \mathcal H_{k-2}(x) $:
\begin{equation}
\mathcal H_k(x) \equiv Q(x) \mathcal H_{k-2}(x) +R(x) \enspace .
\label{rem_quo1}
\end{equation}
Here the quotient $ Q(x) $ equals $ h_{k2}/H_{k-2} $ if $ h_{k-1,1}=0 $, while for the case  $ h_{k-1,1}\ne 0 $ one has
$$
Q(x)\equiv Q_0+Q_1x+Q_2x^2
$$
with coefficients determined by the equalities:
\begin{equation}
Q_2=\frac{H_k}{H_{k-2}},\ Q_1=\frac{H_{k-2}h_{k1}-H_kh_{k-2,1}}{H_{k-2}^2},\
Q_0= \frac{H_{k-2}^2h_{k2}-H_kH_{k-2}^2h_{k-2,2}-H_kh_{k-2,1}^2}{H_{k-2}^3} \enspace .
\label{q0q1q2}
\end{equation}

To find the coefficients of the remainder
\begin{equation}
R(x) = R_0+R_1x+\dots+ R_{k-3}x^{k-3}
\label{eqr1}
\end{equation}
use the arguments similar to those used in the proof of Theorem \ref{TJoach}. First consider the case where the
generating sequence (\ref{seq0}) is given as (\ref{seq1}). Substitute $ x=\lambda_{\ell} $ into (\ref{rem_quo1})
\begin{equation}
\mathcal H_k(\lambda_{\ell}) = (Q_0+Q_1\lambda_{\ell}+Q_2\lambda_{\ell}^2) \mathcal H_{k-2}(\lambda_{\ell}) +
R_0+R_1\lambda_{\ell}+\dots+ R_{k-3}\lambda_{\ell}^{k-3} \quad \mbox{ for } \ \ell\in \{1,\dots, m\}
\label{syst1010}
\end{equation}
and sum up the obtained equalities. Due to (\ref{eqty100}) one gets
$$
0=c_0R_0+c_1R_1+\dots+c_{k-3}R_{k-3} \enspace .
$$
Similar equalities result from multiplication  of (\ref{syst1010}) by $ \lambda_{\ell}^{j} $ for $ j\in \{1,\dots,k-5\} $ and further summation by $ \ell $:
\begin{eqnarray*}
0 & = & c_1R_0+c_2R_1+\dots+c_{k-2}R_{k-3}, \\
\dots & & \dots \\
0 & = & c_{k-5}R_0+c_{k-4}R_1+\dots+c_{2k-8}R_{k-3}.
\end{eqnarray*}
Multiplication of  (\ref{syst1010}) by $ \lambda_{\ell}^{k-4} $ and summation yields
$$
0=Q_2H_{k-1}+c_{k-4}R_0+c_{k-3}R_1+\dots+c_{2k-7}R_{k-3} \enspace ,
$$
and, since $ H_{k-1}=0 $, the obtained equality looks similar to the previous ones. Multiplication of  (\ref{syst1010}) by $ \lambda_{\ell}^{k-3} $ and summation
leads to
$$
0=Q_2 \sum_{j=1}^m \lambda_j^{k-1} \mathcal H_{k-2}(\lambda_j)+c_{k-3}R_0+c_{k-2}R_1+\dots+c_{2k-6}R_{k-3}
$$
with
$$
\sum_{\ell=1}^m \lambda_{\ell}^{k-1} \mathcal H_{k-2}(\lambda_{\ell})=
\left|
\begin{array}{llll}
c_0 & c_1 & \dots & c_{k-2} \\
c_1 & c_2 & \dots & c_{k-1} \\
\vdots & & & \vdots \\
c_{k-3} & c_{k-2} & \dots & c_{2k-5} \\
\displaystyle \sum_{\ell=1}^m \lambda_{\ell}^{k-1} & \displaystyle \sum_{\ell=1}^m \lambda_{\ell}^{k} & \dots & \displaystyle \sum_{\ell=1}^m \lambda_{\ell}^{2k-3}
\end{array}
\right|=
\left|
\begin{array}{llll}
c_0 & c_1 & \dots & c_{k-2} \\
c_1 & c_2 & \dots & c_{k-1} \\
\vdots & & & \vdots \\
c_{k-3} & c_{k-2} & \dots & c_{2k-5} \\
c_{k-1} & c_{k} & \dots & c_{2k-3}
\end{array}
\right|=-h_{k-1,1} \, .
$$
If $ h_{k-1,1} = 0 $ then the obtained linear system of equalities with respect to $ R_0,\dots,R_{k-3} $, namely
$$
c_i R_0+c_{i+1}R_1+\dots+c_{k+i-3}R_{k-3}=0 \quad \mbox{ for } \quad i \in \{0,\dots,k-3\}
$$
implies that $ \{ R_j=0 \}_{j=0}^{k-3} $ (since the determinant of the system equals $ H_{k-2} \ne 0 $). This proves (\ref{HanHk0}).
For the case $ h_{k-1,1} \ne 0 $, unify all the obtained relationships with (\ref{eqr1}) and compose the linear system with respect to $ R_0,\dots,R_{k-3}, 1 $. Since it is consistent, its determinant should vanish:
$$
\left|
\begin{array}{llllc}
c_0 & c_1 & \dots & c_{k-2} & 0 \\
c_1 & c_2 & \dots & c_{k-1} & 0\\
\vdots & & & \vdots & \vdots \\
c_{k-3} & c_{k-2} & \dots & c_{2k-6} & - h_{k-1,1}q_2 \\
1 & x & \dots & x^{k-3} & - R(x)
\end{array}
\right|=0 \enspace .
$$
Expansion of the determinant by its last column and usage the first formula from (\ref{q0q1q2}) leads one to the formula
$$ R(x) H_{k-2} \equiv \frac{h_{k-1,1}H_k}{H_{k-2}} \mathcal H_{k-3}(x)
$$
which completes the proof of (\ref{HanHk1}). \qed

~\\
\indent\textbf{Remark~2.1.} Formulas of Theorem \ref{TJoach1}  allow one to organize the recursive computation of $ \mathcal H_k(x) $ if  the polynomials  $ \mathcal H_{k-2}(x) $ and
$ \mathcal H_{k-3}(x) $ are already computed. The involved factors, such as $ h_{k-2,1} $, $ h_{k-2,2} $, $ h_{k-1,1} $ and $ h_{k,1} $, can also be evaluated either automatically as the coefficients of the Hankel polynomials, or by formulas (\ref{hk0hk1}). The only exception is the value for $ h_{k2} $. For its evaluation, we suggest the following representation
\begin{equation}
h_{k2}=-\frac{1}{H_{k-2}} \left|
\begin{array}{llll}
c_0 & c_1 & \dots & c_{k-3} \\
c_1 & c_2 & \dots & c_{k-2} \\
\vdots & & & \vdots \\
c_{k-3} & c_{k-2} & \dots & c_{2k-5} \\
c_{k} & c_{k+1} & \dots & c_{2k-2}
\end{array}
\right|^2=-\frac{\left(c_{2k-2}h_{k-2,0}+c_{2k-3}h_{k-2,1}+\dots+ c_{k}h_{k-2,k-2} \right)^2}{H_{k-2}}
\label{hk2}
\end{equation}
which is valid under assumption $ H_{k-1} = 0 $. We do not give here the proof of this formula.

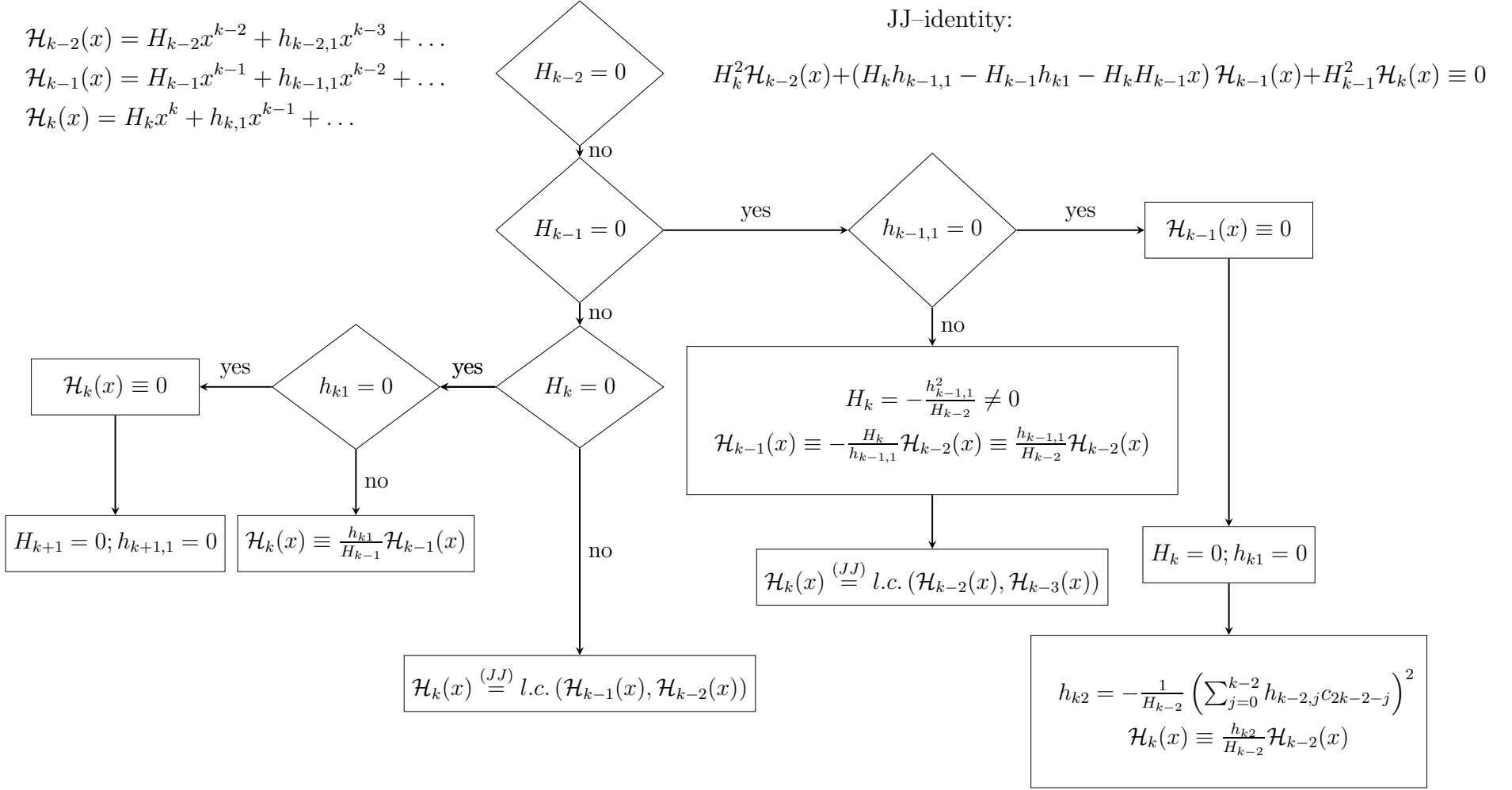
\begin{figure}[h!]
\begin{sideways}
\begin{minipage}{\textwidth}

\vspace*{4cm}%
\vspace*{-3cm}%
\hspace*{-6.5cm}%
\noindent%
\parbox{8cm}{
\begin{equation*}
\begin{split}
&\mathcal{H}_{k-2}(x) = H_{k-2}x^{k-2} + h_{k-2,1}x^{k-3} + \ldots\\
&\mathcal{H}_{k-1}(x) = H_{k-1}x^{k-1} + h_{k-1,1}x^{k-2} + \ldots\\
&\mathcal{H}_{k  }(x) = H_{k  }x^{k  } + h_{k  ,1}x^{k-1} + \ldots
\end{split}
\end{equation*}
}

\vspace*{-3.0cm}%
\hspace*{5.5cm}%
\noindent%
\parbox{8cm}{
\begin{center}
JJ--identity:
\begin{equation*}
H_k^2\mathcal{H}_{k-2}(x) + \left( H_kh_{k-1,1} - H_{k-1}h_{k1} - H_kH_{k-1}x     \right)\mathcal{H}_{k-1}(x) + H_{k-1}^2\mathcal{H}_k(x) \equiv 0
\end{equation*}
\end{center}
}

\vspace*{-2cm}%
\hspace*{-6.5cm}%
\scalebox{0.94}{
\noindent%
\begin{tikzpicture}[node distance=2.8cm]

\node (dec0c) [decision, xshift=-3cm, yshift=-3cm] {$H_{k-2}=0$};

\node (dec1c) [decision, below of=dec0c] {$H_{k-1}=0$};
\node (dec1d) [decision, right of=dec1c, xshift=3.5cm] {$h_{k-1,1}=0$};
\node (pro1e) [process,  right of=dec1d, xshift=2.5cm] {$\mathcal{H}_{k-1}(x) \equiv 0$};

\node (dec2c) [decision, below of=dec1c] {$H_{k}=0$};
\node (dec2b) [decision, left of=dec2c, xshift=-1.2cm] {$h_{k1}=0$};
\node (pro2a) [process,  left of=dec2b, xshift=-1.5cm] {$\mathcal{H}_k(x) \equiv 0$};
\node (pro2e) [process,  below of=pro1e, yshift=-3cm] {$H_k=0; h_{k1}=0$};

\node (pro3a) [process,  below of=pro2a] {$H_{k+1}=0; h_{k+1,1}=0$};
\node (pro3e) [process,  below of=pro2e] {\parbox{6.8cm}{\begin{eqnarray*}&h_{k2}=-\frac{1}{H_{k-2}}\left( \sum_{j=0}^{k-2}h_{k-2,j}c_{2k-2-j}\right)^2\\ &\mathcal{H}_k(x) \equiv \frac{h_{k2}}{H_{k-2}}\mathcal{H}_{k-2}(x)\end{eqnarray*}}};

\node (pro4b) [process,  below of=dec2b] {$\mathcal{H}_k(x) \equiv \frac{h_{k1}}{H_{k-1}}\mathcal{H}_{k-1}(x)$};
\node (pro4c) [process, below of=dec2c, yshift=-2.5cm] {$\mathcal{H}_k(x)\stackrel{(JJ)}{=}l.c.\left(\mathcal{H}_{k-1}(x),\mathcal{H}_{k-2}(x)\right)$};
\node (pro4d) [process, below of=dec1d, yshift=-0.6cm] {\parbox{8.5cm}{\begin{eqnarray*}&H_k=-\frac{h^2_{k-1,1}}{H_{k-2}} \neq 0\\ &\mathcal{H}_{k-1}(x) \equiv -\frac{H_k}{h_{k-1,1}}\mathcal{H}_{k-2}(x) \equiv \frac{h_{k-1,1}}{H_{k-2}}\mathcal{H}_{k-2}(x)\end{eqnarray*}}};

\node (pro5d) [process, below of=pro4d] {$\mathcal{H}_k(x)\stackrel{(JJ)}{=}l.c.\left(\mathcal{H}_{k-2}(x),\mathcal{H}_{k-3}(x)\right)$};

\draw [arrow] (dec0c) -- node[anchor=west] {no} (dec1c);
\draw [arrow] (dec1c) -- node[anchor=west] {no} (dec2c);
\draw [arrow] (dec1c) -- node[anchor=south] {yes} (dec1d);
\draw [arrow] (dec1d) -- node[anchor=west] {no} (pro4d);
\draw [arrow] (dec1d) -- node[anchor=south] {yes} (pro1e);
\draw [arrow] (pro1e) -- (pro2e);
\draw [arrow] (pro2e) -- (pro3e);
\draw [arrow] (dec2c) -- node[anchor=west] {no} (pro4c);
\draw [arrow] (dec2c) -- node[anchor=south] {yes} (dec2b);
\draw [arrow] (dec2b) -- node[anchor=south] {yes} (pro2a);
\draw [arrow] (dec2b) -- node[anchor=west] {no} (pro4b);
\draw [arrow] (pro2a) -- (pro3a);
\draw [arrow] (dec2c) -- node[anchor=south] {yes} (dec2b);
\draw [arrow] (pro4d) -- (pro5d);


\end{tikzpicture}
}
~\\
\caption{Flowchart for the Hankel polynomial computation.\\$(l.c. = $ linear combination)}
\label{flowchart}
\end{minipage}
\end{sideways}
\end{figure}

The flowchart for the procedure of the Hankel polynomial computation based on the results
of Theorems \ref{TJoach} and \ref{TJoach1} is displayed in Fig.~\ref{flowchart}.

\newpage
\thispagestyle{empty}
\newpage

\section{Polynomial Interpolation}\label{SPoly_interp}

\setcounter{equation}{0}
\setcounter{theorem}{0}
\setcounter{example}{0}
\setcounter{cor}{0}

The classical polynomial interpolation problem is formulated as follows:

\textbf{Problem 1}. Find a polynomial  of the form
\begin{equation}
p(x)= p_{0}x^{N-1} + p_{1}x^{N-2} + \ldots + p_{N-1}
\label{polyF}
\end{equation}
satisfying the table (\ref{table}), i.e.
\begin{equation}
p(x_j) = y_j \quad \mbox{ for } \ j \in \{1,\dots N \} \, .
\label{condit0}
\end{equation}

Problem 1 always possesses a unique solution which can be found via resolving the system of linear equations with the Vandermonde matrix or, in other terms, via the determinant evaluation
\begin{equation}
p(x)\equiv
\left|\begin{array}{cccccc}
1 & x_1 & x_1^2 & \dots & x_1^{N-1} & y_1 \\
1 & x_2 & x_2^2 & \dots & x_2^{N-1} & y_2 \\
\vdots & & & & \vdots & \vdots \\
1 & x_N & x_N^2 & \dots & x_N^{N-1} & y_N \\
1 & x & x^2 & \dots & x^{N-1} & 0 \\
\end{array}
\right| \Bigg/
\prod_{1\le j < k \le N} (x_k-x_j) \, .
\label{Vander_interp}
\end{equation}
This computation is usually performed by virtue of some auxiliary constructions like representation of the interpolant in Lagrange or Newton form. Lagrange's method consists in the following steps. First construct the polynomials
\begin{equation}
W(x)=\prod_{j=1}^N (x-x_j) \,
\label{W}
\end{equation}
and
\begin{equation}
W_k(x)= \frac{W(x)}{x-x_k} \equiv \prod_{j=1 \atop j \ne k}^N (x-x_j) \quad \mbox{ for } \ k\in \{1,\dots, N\} \, .
\label{Wk}
\end{equation}
Then the interpolation polynomial can be represented as
\begin{equation}
p(x) \equiv \sum_{j=1}^N y_j \frac{W_j(x)}{W_j(x_j)}=\sum_{j=1}^N y_j \frac{W_j(x)}{W^{\prime}(x_j)} \, .
\label{lagrange-poly}
\end{equation}
However this representation does not immediately provide one with the canonical form for the interpolation polynomial, i.e. an explicit expression for the coefficients in (\ref{polyF}). In order to extract them, let us prove first a preliminary result

\begin{theorem} \label{ThPrelim} Let $ G(x)\equiv G_0x^{N-1}+G_1x^{N-2}+\dots+G_{N-1} $ be an arbitrary polynomial of the degree at most $ N-1 $. Then the following equalities are valid
\begin{equation}
\sum_{j=1}^N \frac{G(x_j)}{W^{\prime}(x_j)} =
\left\{
\begin{array}{ll} 0 & if \  \deg G < N-1; \\
G_0 & if \ \deg G = N-1.
\end{array} \right.
\label{Eu-Lag1}
\end{equation}
\end{theorem}

\textbf{Proof.} Use the following Lagrange formula:
$$
\frac{G(x)}{W(x)}\equiv \frac{G(x_1)}{W^{\prime}(x_1)(x-x_1)}  + \frac{G(x_2)}{W^{\prime}(x_2)(x-x_2)}+\dots+ \frac{G(x_N)}{W^{\prime}(x_N)(x-x_N)}
$$
provided that $ \{ G(x_j)\ne 0 \}_{j=1}^N $.
From this the expansion of the fraction $ G(x)/W(x) $ in the Laurent series in negative powers of $ x $ can be derived:
$$
\frac{G(x)}{W(x)}\equiv \left(\sum_{j=1}^N \frac{G(x_j)}{W^{\prime}(x_j)}\right)\frac{1}{x} +
\left(\sum_{j=1}^N \frac{G(x_j)x_j}{W^{\prime}(x_j)}\right)\frac{1}{x^2}+\dots+ \left(\sum_{j=1}^N \frac{G(x_j)x_j^k}{W^{\prime}(x_j)}\right)\frac{1}{x^{k+1}} + \dots
$$
On the other hand, multiplying the formal expansion
$$
\frac{G(x)}{W(x)}\equiv\frac{c_0}{x}+\frac{c_1}{x^2}+\dots+ \frac{c_k}{x^{k+1}}+ \dots
$$
by $ W(x)\equiv x^N+w_1x^{N-1}+\dots+w_N $, one obtains the following formulas for determining the coefficients $ c_0, c_1, \dots $ recursively:
$$
\begin{array}{ll}
G_0&=c_0, \\
G_1&=c_0w_1+c_1, \\
G_2&=c_0w_2+c_1w_1+c_2, \\
\dots
\end{array}
$$
Comparing the two obtained forms for the expansion of $ G(x)/W(x) $ yields the claimed equalities (\ref{Eu-Lag1}). \qed

Setting $ G(x)\equiv x^k $ in the theorem statement results in

\begin{cor} \label{cor_EL}
The following Euler-Lagrange equalities are valid:
\begin{equation}
\sum_{j=1}^N \frac{x_j^k}{W^{\prime}(x_j)}=\left\{
\begin{array}{ll} 0 & if \ k \in \{1,\dots,N-2\}; \\
1 & if \ k= N-1.
\end{array} \right.
\label{Eu-Lag}
\end{equation}
\end{cor}

\begin{cor} \label{cor_EL1} Let
$$ G(x)\equiv G_0x^{M}+G_1x^{M-1}+\dots+G_{M}, \  F(x) \equiv F_0x^{n}+F_1x^{n-1}+\dots+F_{n},\ G_0\ne 0 , F_0\ne 0, $$
and $ F(x) $ possesses only simple zeros
$ \lambda_1,\dots,\lambda_n $ not coinciding with $ \{x_1,\dots,x_{N}\} $.
Then the following equalities are valid
\begin{equation}
\sum_{j=1}^N \frac{G(x_j)x_j^k}{F(x_j)W^{\prime}(x_j)} =
\left\{
\begin{array}{cl} -  \displaystyle \sum_{\ell=1}^n \frac{G(\lambda_{\ell})\lambda_{\ell}^k}{F^{\prime}(\lambda_{\ell})W(\lambda_{\ell})} & if \ k < N+n-M-1; \\
\displaystyle \frac{G_0}{F_0} -  \displaystyle \sum_{\ell=1}^n \frac{G(\lambda_{\ell})\lambda_{\ell}^{N+n-M-1}}{F^{\prime}(\lambda_{\ell})W(\lambda_{\ell})} & if \ k = N+n-M-1.
\end{array} \right.
\label{Eu-Lag2}
\end{equation}
\end{cor}

\begin{theorem} \label{Th_inter_poly1}
Calculate two sequences of values:
\begin{equation}
\sigma_k = \sum_{j=1}^{N} \frac{x_j^{N+k-1}}{W^{\prime}(x_j)} \quad \mbox{ for } \ k \in \{1,\dots, N \} \label{sigma}
\end{equation}
and
\begin{equation}
\tau_k = \sum_{j=1}^{N} y_j \frac{x_j^{k}}{W^{\prime}(x_j)} \quad \mbox{ for } \ k \in \{1,\dots N-1 \}  \, . \label{tauk}
\end{equation}
The following recursive formulas connect the values  (\ref{sigma}) and (\ref{tauk}) with the coefficients of interpolation polynomial:
\begin{equation}
\tau_0=p_0,\
\tau_k=p_0\sigma_k+p_1\sigma_{k-1}+\dots+p_{k-1}\sigma_1 + p_k  \ \quad \mbox{ for }  k\in \{1,\dots,N-1 \} \enspace .
\label{recurs_interPol}
\end{equation}
\end{theorem}

\textbf{Proof.} Due to equalities (\ref{Eu-Lag}) one has
\begin{eqnarray*}
\tau_k &=& \displaystyle  \sum_{j=1}^{N} \frac{x_j^{k}y_j}{W^{\prime}(x_j)}=
\sum_{j=1}^{N} \frac{(p_0x_j^{N-1}+p_1x_j^{N-2}+\dots+p_kx_j^{N-k-1}+\dots+p_{N-1})x_j^k}
{W^{\prime}(x_j)} \\
&=&p_0\sigma_k+p_1\sigma_{k-1}+\dots+p_{k-1}\sigma_1+p_k  \enspace .
\end{eqnarray*}
\qed

We now suggest an alternative construction for the interpolation polynomial.

\begin{theorem} \label{Th_inter_poly2}
Assume that $ y_j \ne 0  $ for $ j \in \{1,\dots N \} $. Calculate the sequence of values:
\begin{equation}
\widetilde \tau_k = \sum_{j=1}^{N} \frac{1}{y_j} \frac{x_j^{k}}{W^{\prime}(x_j)} \quad \mbox{ for } \ k \in \{1,\dots, 2\,N-2 \}
\enspace .
\label{til_tau}
\end{equation}
Interpolation polynomial can be represented in the   Hankel polynomial form:
\begin{equation}
p(x)= (-1)^{N(N-1)/2} \left(\prod_{j=1}^N y_j \right) \mathcal H_{N-1}(x; \{ \widetilde \tau \})=
(-1)^{N(N-1)/2} \left(\prod_{j=1}^N y_j \right)
\left|
\begin{array}{lllll}
\widetilde \tau_0 & \widetilde \tau_1 & \widetilde \tau_2 & \dots & \widetilde \tau_{N-1} \\
\widetilde \tau_1 & \widetilde \tau_2 & \widetilde \tau_3 & \dots & \widetilde \tau_{N} \\
\vdots & & & & \vdots \\
\widetilde \tau_{N-2} & \widetilde \tau_{N-1} & \widetilde \tau_{N} & \dots & \widetilde \tau_{2N-3} \\
1 & x & x^2 & \dots & x^{N-1}
\end{array}
\right|  \enspace .
\label{interp_Han}
\end{equation}
\end{theorem}

\textbf{Proof.} We first present an underlying idea for the claimed result. Assume first that the interpolation polynomial (\ref{polyF}) exists. Rewrite the equalities (\ref{condit0}) in the form
\begin{equation}
p_{N-1}\frac{1}{y_j}+p_{N-2}\frac{x_j}{y_j}+\dots+p_{1}\frac{x_j^{N-2}}{y_j}+p_0 \frac{x_j^{N-1}}{y_j} = 1
\quad \mbox{ for } \ j \in \{1,\dots, N \}  \enspace .
\label{eq141}
\end{equation}
Multiply each of these equalities by the corresponding multiple $ 1/ W^{\prime}(x_j) $ and sum the obtained results. Due to (\ref{Eu-Lag}), one gets
$$
 p_{N-1} \widetilde \tau_0 +p_{N-2} \widetilde \tau_1+\dots+p_{1} \widetilde \tau_{N-2}+p_0 \widetilde \tau_{N-1} = 0 \enspace .
$$
Similar equalities result from multiplication of (\ref{eq141}) by $ x_j/ W^{\prime}(x_j),x_j^2/ W^{\prime}(x_j),\dots, x_j^{N-2}/ W^{\prime}(x_j) $:
\begin{eqnarray*}
 p_{N-1} \widetilde \tau_1 +p_{N-2} \widetilde \tau_2+\dots+p_{1} \widetilde \tau_{N-1}+p_0 \widetilde \tau_{N} &=& 0 , \\
 p_{N-1} \widetilde \tau_2 +p_{N-2} \widetilde \tau_3+\dots+p_{1} \widetilde \tau_{N}+p_0 \widetilde \tau_{N+1} &=& 0 , \\
 \dots & & \dots , \\
p_{N-1} \widetilde \tau_{N-2} +p_{N-2} \widetilde \tau_{N-1}+\dots+p_{1} \widetilde \tau_{2N-4}+p_0 \widetilde \tau_{2N-3} &=& 0 .
\end{eqnarray*}
Linear combination of (\ref{eq141}) multiplied by $ x_j^{N-1}/ W^{\prime}(x_j)  $ yields something different:
$$
p_{N-1} \widetilde \tau_{N-1} +p_{N-2} \widetilde \tau_{N}+\dots+p_{1} \widetilde \tau_{2N-3}+p_0 \widetilde \tau_{2N-2} = 1 \enspace .
$$
Unifying all the obtained equalities with (\ref{polyF}) one obtains a system of linear equations with respect to
$$ p_{N-1},p_{N-2},\dots,p_{1},p_0  \, . $$
If the interpolation polynomial exists then it should satisfy the following identity:
$$
\left|
\begin{array}{llllll}
\widetilde \tau_0 & \widetilde \tau_1 & \widetilde \tau_2 & \dots & \widetilde \tau_{N-1} & 0\\
\widetilde \tau_1 & \widetilde \tau_2 & \widetilde \tau_3 & \dots & \widetilde \tau_{N} & 0 \\
\vdots & & & & & \vdots \\
\widetilde \tau_{N-2} & \widetilde \tau_{N-1} & \widetilde \tau_{N} & \dots & \widetilde \tau_{2N-3} & 0 \\
\widetilde \tau_{N-1} & \widetilde \tau_{N} & \widetilde \tau_{N+1} & \dots & \widetilde \tau_{2N-2} & 1 \\
1 & x & x^2 & \dots & x^{N-1} & p(x)
\end{array}
\right|\equiv 0 \, .
$$
Expansion of the determinant by its last column gives
\begin{equation}
p(x)\equiv \frac{\mathcal H_{N-1}(x,\{\widetilde \tau\})}{H_N(\{\widetilde \tau\})}
\label{interp_Han1}
\end{equation}
provided that the determinant standing in the denominator does not vanish. To prove this, represent it as a product:
$$
H_N(\{\widetilde \tau\})=
\left|\begin{array}{cccc}
1 & 1 & \dots & 1 \\
x_1 & x_2 & \dots & x_N \\
x_1^2 & x_2^2 & \dots & x_N^2 \\
\vdots & & & \vdots \\
x_1^{N-1} & x_2^{N-1} & \dots & x_N^{N-1}
\end{array}
\right| \cdot
\left| \begin{array}{cccc}
\frac{1}{y_1W^{\prime}(x_1)} & 0 & \dots & 0 \\
0 & \frac{1}{y_2W^{\prime}(x_2)} & \dots & 0 \\
\vdots & & & \vdots \\
0 & 0 & \dots & \frac{1}{y_NW^{\prime}(x_N)}
\end{array}
\right| \cdot
\left|\begin{array}{ccccc}
1 & x_1 & x_1^2 & \dots & x_1^{N-1} \\
1 & x_2 & x_2^2 & \dots & x_2^{N-1} \\
\vdots & & & & \vdots \\
1 & x_{N} & x_N^2 & \dots & x_N^{N-1}
\end{array}
\right|
$$
$$ =\frac{  \displaystyle \prod_{1\le j < k \le N} (x_k-x_j)^2 }{  \displaystyle \prod_{j=1}^N y_j  \prod_{j=1}^N W^{\prime}(x_j)} \, .  $$
It can be easily proved that
$$
\prod_{j=1}^N W^{\prime}(x_j) = (-1)^{N(N-1)/2} \prod_{1\le j < k \le N} (x_k-x_j)^2
$$
and, therefore, the denominator in (\ref{interp_Han1})
\begin{equation}
H_N(\{\widetilde \tau\}) = (-1)^{N(N-1)/2} \Bigg/  \prod_{j=1}^N y_j \, .
\label{H_Ntau_1}
\end{equation}
does not vanish.

One can utilize the last expression for the direct deduction of the validity of the formula (\ref{interp_Han}) as a solution of the polynomial interpolation problem. Indeed, let us substitute $ x=x_1 $ into the numerator of the fraction in the right-hand side of (\ref{interp_Han}) and use an alternative representation of the  Hankel polynomial in the form of the Hankel determinant (\ref{Hkx1}):
\begin{eqnarray*}
\mathcal H_{N-1}(x_1; \{ \widetilde \tau \}) &=&
\left|
\begin{array}{lllll}
\widetilde \tau_0 & \widetilde \tau_1 & \widetilde \tau_2 & \dots & \widetilde \tau_{N-1} \\
\widetilde \tau_1 & \widetilde \tau_2 & \widetilde \tau_3 & \dots & \widetilde \tau_{N} \\
\vdots & & & & \vdots \\
\widetilde \tau_{N-2} & \widetilde \tau_{N-1} & \widetilde \tau_{N} & \dots & \widetilde \tau_{2N-3} \\
1 & x_1 & x_1^2 & \dots & x_1^{N-1}
\end{array}
\right|_{N\times N}  \\
&\stackrel{(\ref{Hkx1})}{\equiv} &
(-1)^{N+1}
\left|
\begin{array}{llll}
 \widetilde \tau_1 - x_1\widetilde \tau_0  & \widetilde \tau_2 - x_1 \widetilde \tau_1 & \dots & \widetilde \tau_{N-1}
-x_1 \widetilde \tau_{N-2} \\
\widetilde \tau_2 - x_1 \widetilde \tau_1 & \widetilde \tau_3 -x_1 \widetilde \tau_2 & \dots & \widetilde \tau_{N}
-x_1 \widetilde \tau_{N-1} \\
\vdots & & & \vdots \\
\widetilde \tau_{N-1} -x_1 \widetilde \tau_{N-2} & \widetilde \tau_{N} - x_1 \widetilde \tau_{N-1} & \dots & \widetilde \tau_{2N-3} -x_1 \widetilde \tau_{2N-4}
\end{array}
\right|_{(N-1)\times (N-1)}
\enspace .
\end{eqnarray*}
Consider an entry of the last determinant
\begin{eqnarray*}
\widetilde \tau_k-x_1  \widetilde \tau_{k-1} &= & \frac{x_1^k}{y_1W^{\prime}(x_{1})}+\frac{x_2^k}{y_2W^{\prime}(x_{2})}+\dots+\frac{x_{N}^k}{y_NW^{\prime}(x_{N})} \\
& & -
\frac{x_1^k}{y_1W^{\prime}(x_{1})}-\frac{x_2^{k-1}x_1}{y_2W^{\prime}(x_{2})}-\dots-\frac{x_N^{k-1}x_1}{y_NW^{\prime}(x_{N})} \\
&=& \frac{x_2^{k-1}(x_2-x_1)}{y_2W^{\prime}(x_{2})}+\dots+ \frac{x_N^{k-1}(x_N-x_1)}{y_NW^{\prime}(x_{N})} \enspace .
\end{eqnarray*}
It can be easily verified that
$$
\frac{x_j-x_1}{W^{\prime}(x_{j})}=\frac{1}{W_1^{\prime}(x_{j})} \quad \mbox{ for } \ j \in \{2,\dots, N \} \quad \mbox{ and } \ W_1(x) \stackrel{(\ref{Wk})}{\equiv}  \prod_{k=2}^N (x-x_k) \, .
$$
Thus,
\begin{equation}
\mathcal H_{N-1}(x_1; \{ \widetilde \tau \})=(-1)^{N-1}
\left|
\begin{array}{llll}
 \widetilde T_0   & \widetilde T_1 & \dots & \widetilde T_{N-2} \\
  \widetilde T_1   & \widetilde T_2 & \dots & \widetilde T_{N-1} \\
  \vdots & & & \vdots \\
  \widetilde T_{N-2}   & \widetilde T_{N-1} & \dots & \widetilde T_{2N-4}
\end{array}
\right| \quad \mbox{ where } \quad  \left\{\widetilde T_k = \sum_{j=2}^N \frac{x_j^k}{y_jW_1^{\prime}(x_{j})} \right\}_{k=0}^{2N-4} \enspace .
\label{HNm1T}
\end{equation}
The last determinant can be evaluated by formula (\ref{H_Ntau_1}):
$$
\mathcal H_{N-1}(x_1; \{ \widetilde \tau \})=(-1)^{N-1}(-1)^{(N-1)(N-2)/2}  \Bigg/  \prod_{j=2}^N y_j \, .
$$
With the aid of this equality and (\ref{H_Ntau_1}), one gets in  (\ref{interp_Han1}): $ p(x_1)=y_1 $.
\qed

At first glance, formula (\ref{interp_Han}) does not have any advantage not only over the algorithm suggested in Theorem \ref{Th_inter_poly1} but even over that one based on  determinant (\ref{Vander_interp}) evaluation. We postpone the justification of this approach till the  next section, and now restrict ourselves with demonstration of the efficient computation for the Hankel polynomial from Theorem  \ref{Th_inter_poly2} with the aid of recursive procedure developed in Section \ref{section-hankel}.

\begin{example} \label{Ex21} Construct the interpolation polynomial for the table:
$$
\begin{array}{c|c|c|c|c|c|c|c}
    x & -2 & -1 & 0 & 1 & 2 & 3 & 4 \\
\hline
    y &  208 & -10 & -8 & -14 & -16 & 478 & 4120
  \end{array}
$$
\end{example}

\textbf{Solution.} Here $ N=7 $. Compute the values (\ref{til_tau})
$$
\widetilde \tau_0=-\frac{7879647}{7168470400}, \ \widetilde \tau_1=-\frac{1359931}{896058800},\  \widetilde \tau_2=-\frac{4508383}{1792117600},\dots,
 \widetilde \tau_{12} =-\frac{499128619}{56003675} \, .
$$
For an expansion of the recursive procedure for the Hankel polynomial computation based on the JJ-identity (\ref{Joach_iden1}), we are in need of initial polynomials, i.e. polynomials of the first and the second order:
$$
\mathcal H_1(x; \{\widetilde \tau \})=-\frac{7879647}{7168470400}x +\frac{1359931}{896058800}\ ,
$$
$$
\mathcal H_2(x; \{\widetilde \tau \})=\underbrace{\frac{21191}{45878210560}}_{\widetilde h_{2,0}}x^2\underbrace{-\frac{396821}{57347763200}}_{\widetilde h_{2,1}}x+\underbrace{\frac{487269}{57347763200}}_{\widetilde h_{2,2}}  \ ,
$$
Now compute $ \mathcal H_3(x; \{\widetilde \tau \}) $:
$$
\mathcal H_3(x; \{\widetilde \tau \}) \equiv -
\left(\frac{\widetilde h_{3,0}}{\widetilde h_{2,0}}\right)^2 \mathcal H_1(x; \{\widetilde \tau \})+ \frac{\widetilde h_{3,0}}{\widetilde h_{2,0}}\left(x-\frac{\widetilde h_{2,1}}{\widetilde h_{2,0}}+\frac{\widetilde h_{3,1}}{\widetilde h_{3,0}} \right)\mathcal H_2(x; \{\widetilde \tau \})
$$
where all the constants are already known except for $\widetilde h_{3,0} $ and $\widetilde h_{3,1} $. To find the latter, utilize the equalities (\ref{hk0hk1})
$$
\widetilde h_{3,0}=H_3=\widetilde \tau_2 \widetilde h_{2,2}+\widetilde \tau_3 \widetilde h_{2,1}+\widetilde \tau_4 \widetilde h_{2,0}=\frac{2141}{57347763200}\, ,
$$
$$
\widetilde h_{3,1}=-(\widetilde \tau_3 \widetilde h_{2,2}+\widetilde \tau_4 \widetilde h_{2,1}+\widetilde \tau_5 \widetilde h_{2,0})=-\frac{1691}{57347763200} \, .
$$
Therefore,
$$
\mathcal H_3(x; \{\widetilde \tau \}) \equiv \frac{2141}{57347763200}x^3-\frac{1691}{57347763200}x^2-\frac{19}{275710400}x-\frac{97}{573477632} \, .
$$
Further computations can be organized in a similar manner up to the polynomial of the order $ 6 $:
$$
\mathcal H_6(x; \{\widetilde \tau \}) =-\frac{1}{7340513689600}(2\,x^6-4\,x^5+2\,x^3-6\,x^2-8)
$$
Since
$$ \prod_{j=1}^7 y_j =7340513689600 \ , $$
formula (\ref{interp_Han}) yields interpolation polynomial in the form
$$ p(x) \equiv 2\,x^6-4\,x^5+2\,x^3-6\,x^2-8 \, . $$
If we expand, just for curiosity, the computation process further  then the next step yields:
$$
\mathcal H_7(x; \{\widetilde \tau \})\equiv -\frac{1}{7340513689600}(x+2)(x+1)x(x-1)(x-2)(x-3)(x-4) \, .
$$
\qed

\begin{theorem} \label{Th_inter_poly3} Under the condition of Theorem \ref{Th_inter_poly2}, one has
\begin{equation}
\mathcal H_{N}(x;\{\widetilde \tau\}) \equiv H_N(\{\widetilde \tau\}) \prod_{j=1}^N (x-x_j) \equiv \frac{(-1)^{N(N-1)/2}}{\prod_{j=1}^N y_j}  \prod_{j=1}^N (x-x_j) \, .
\label{H_Ntau}
\end{equation}
\end{theorem}

\textbf{Proof.} Let us prove that
$$ \mathcal H_{N}(x_j;\{\widetilde \tau\})=0 \quad \mbox{ for } \ j \in \{1,\dots, N \} \, . $$
The linear combination
$$ \sum_{j=1}^N \frac{x_j^k}{W^{\prime}(x_j)} \mathcal H_{N}(x_j;\{\widetilde \tau\}) $$
can be represented in the form of determinant
$$
=
\left|
\begin{array}{llll}
\widetilde \tau_0 & \widetilde \tau_1 &  \dots & \widetilde \tau_{N} \\
\widetilde \tau_1 & \widetilde \tau_2 &  \dots & \widetilde \tau_{N+1}  \\
\vdots & & &  \vdots \\
\widetilde \tau_{N-1} & \widetilde \tau_{N} &  \dots & \widetilde \tau_{2N-1}  \\
\widetilde \tau_{k} & \widetilde \tau_{k+1} &  \dots & \widetilde \tau_{N+k}
\end{array}
\right| \, .
$$
It equals zero for $ k\in \{0,\dots,N-1\} $. Therefore we have $ N $ linear homogeneous equalities
$$ \left\{\sum_{j=1}^N \frac{x_j^{k-1}}{W^{\prime}(x_j)} \mathcal H_{N}(x_j;\{\widetilde \tau\})=0 \right\}_{k=1}^N $$
valid for $ N $ values
\begin{equation}
 \{\mathcal H_{N}(x_j;\{\widetilde \tau\}) \}_{j=1}^N \, .
 \label{valuesH}
\end{equation}
 Since
$$ \det \left[ \frac{x_j^{k-1}}{W^{\prime}(x_j)} \right]_{j,k=1}^N  \ne 0 \, , $$
all the values (\ref{valuesH}) equal to zero. Thus, we know all the zeros of the polynomial $\mathcal H_{N}(x;\{\widetilde \tau\}) $, while its leading coefficient equals to $ H_N(\{\widetilde \tau\}) $. The latter has been already evaluated by (\ref{H_Ntau_1}). This completes the proof. \qed

\section{Polynomial Interpolation: Erroneous Table}\label{SPoly_Error}

\setcounter{equation}{0}
\setcounter{theorem}{0}
\setcounter{example}{0}
\setcounter{cor}{0}

\textbf{Problem 2}. Let the table (\ref{table}) contain up to $ E $ ``erroneous'' values, i.e. there exists a polynomial $ p(x) $ of a degree
$ n < N-1 $  such that
\begin{equation}
p(x_j) = y_j \quad \mbox{ for } \ j \in \{1,\dots N \}\setminus \{e_1,\dots e_E\}
\label{conditD0}
\end{equation}
for some distinct  $ e_1,\dots e_E $ from $ \{1,\dots N \} $.
The exact number of erroneous values and their location are a priori unknown. Find error location and the polynomial $ p(x) $.

Existence and uniqueness of the solution  for the stated problem depend essentially on the relation between the three involved parameters, namely $ N, n $ and $ E $.
One should expect the condition $ N-E > n+1 $ to be a necessary one, i.e. the ``true'' values should be redundant for the polynomial identification.

In order to solve the stated problem, we will first make some experiments aiming at determination the influence of the errors made in the redundant table
(\ref{table}) on the sets of the Hankel polynomials generated by the sequences (\ref{posti}).


\begin{example} \label{Ex22} Construct the sequence of polynomials $ \{ \mathcal H_k(x;\{\widetilde \tau\}) \}_{k=1}^6 $  for the table:
$$
\begin{array}{c|c|c|c|c|c|c|c}
    x & -2 & -1 & 0 & 1 & 2 & 3 & 4 \\
\hline
    y &  30 & 15 & 8 & 9 & 18 & 35 & 60
  \end{array}
$$
\end{example}

\textbf{Solution.} One has:
$$
\mathcal H_1(x; \{\widetilde \tau\}) \equiv-\frac{89}{1814400}x+\frac{211}{226800}, \ \mathcal H_2(x;\{\widetilde \tau\})\equiv-\frac{2}{9568125} (4\,x^2-3\,x+8),
$$
$$
\mathcal H_3(x;\{\widetilde \tau\}) \equiv 0,\ \mathcal H_4(x,\{\widetilde \tau\}) \equiv 0,\ \mathcal H_5(x;\{\widetilde \tau\}) \equiv 0 \, ,
$$
$$
\mathcal H_6(x;\{\widetilde \tau\}) \equiv-\frac{1}{306180000} (4\,x^2-3\,x+8) \, .
$$
It turns out that the given table is generated by the polynomial $ p(x)=4\,x^2-3\,x+8 $ of the second degree and therefore the table is redundant.
\qed

One should pay attention to the fact that, in the previous example, the true expression for the interpolation polynomial has appeared not only at the final step (as has been stated in Theorem \ref{Th_inter_poly2}), but also at an intermediate one of the Hankel polynomial  sequence construction algorithm.

\begin{theorem} \label{Th_inter_poly4} Let the interpolation table (\ref{table}) be generated by the polynomial
$$ p(x)=p_0x^n+\dots+p_n \, ,\ p_0 \ne 0  $$
of the degree $n < N-1 $; let $ y_j \ne 0 $ for $ j\in \{1,\dots, N\} $. One has then:
\begin{equation}
 \mathcal H_n(x; \{\widetilde \tau\}) \equiv \frac{(-1)^{Nn+n(n+1)/2}p_0^{N-n-1}}{\displaystyle \prod_{j=1}^N y_j} p(x),\quad  \mathcal H_{N-1}(x;\{\widetilde \tau\}) \equiv \frac{(-1)^{N(N-1)/2}}{\displaystyle  \prod_{j=1}^N y_j} p(x) \, .
 \label{Redun1}
\end{equation}
If $n < N-2 $ then
\begin{equation}
 \mathcal H_{n+1}(x;\{\widetilde \tau\}) \equiv 0,\dots,  \mathcal H_{N-2}(x;\{\widetilde \tau\}) \equiv 0 \, .
 \label{Redun2}
\end{equation}
\end{theorem}

\textbf{Proof.} We will prove the theorem under an additional assumption that $ p(x) $ possesses only simple zeros. Denote them $ \lambda_1,\dots,\lambda_n $.
Construct a new sequence similar to (\ref{til_tau}), namely
\begin{equation}
 \eta_k= \sum_{\ell=1}^n \frac{\lambda_{\ell}^k}{p^{\prime} (\lambda_{\ell})W(\lambda_{\ell})} \quad \mbox{ for } \ k=0,1,\dots
 \label{eta}
\end{equation}
Due to the relationships (\ref{Eu-Lag2}) one has:
\begin{equation}
\widetilde \tau_k=\sum_{j=1}^{N} \frac{x_j^{k}}{y_jW^{\prime}(x_j)}=\sum_{j=1}^{N} \frac{x_j^{k}}{p(x_j)W^{\prime}(x_j)}=
-\sum_{\ell=1}^n \frac{\lambda_{\ell}^k}{p^{\prime} (\lambda_{\ell})W(\lambda_{\ell})}=-\eta_k \
\mbox{ for} \  k\in \{0,1,\dots,N+n-2\}
\label{tau_eta1}
\end{equation}
while
\begin{equation}
\widetilde \tau_{N+n-1}=\frac{1}{p_0}-\eta_{N+n-1} \, .
\label{tau_eta2}
\end{equation}
With the aid of these relationships, represent first the $ n $th Hankel polynomial as
$$ \mathcal H_n(x; \{\widetilde \tau\})
\equiv (-1)^n
\left|
\begin{array}{llllll}
\eta_0     & \eta_1     & \eta_2 & \ldots & \eta_{n-1} & \eta_{n}  \\
\eta_1     & \eta_2     & \eta_3 & \ldots & \eta_{n} & \eta_{n+1}    \\
\vdots        &    &     & \ddots& &  \vdots    \\
\eta_{n-1} & \eta_{n} & \eta_{n+1} & \ldots  & \eta_{2n-2}  & \eta_{2n-1} \\
1       & x       & x^2 &  \ldots & x^{n-1} & x^{n}
\end{array} \right|
\equiv (-1)^n \mathcal H_n(x; \{ \eta\}) \, .
$$
In exactly the same manner as in the proof of Theorem \ref{Th_inter_poly3}, it can be established that
\begin{equation}
\mathcal H_n(\lambda_j; \{\eta\})=0 \quad \mbox{ for } \ j\in \{1,\dots,n\} \, .
\end{equation}
The leading coefficient of $ \mathcal H_n(x; \{\eta\}) $, i.e.
$$
H_n(\{\eta\})=\left|
\begin{array}{lllll}
\eta_0     & \eta_1     & \eta_2 & \ldots & \eta_{n-1}  \\
\eta_1     & \eta_2     & \eta_3 & \ldots & \eta_{n}     \\
\vdots        &    &     & \ddots&   \vdots    \\
\eta_{n-1} & \eta_{n} & \eta_{n+1} & \ldots  & \eta_{2n-2}
\end{array} \right| \, ,
$$
can be represented as a product
$$
=\left|
\begin{array}{llll}
1     & 1     &  \ldots & 1  \\
\lambda_1     & \lambda_2  &  \ldots & \lambda_{n}     \\
\vdots          &     & \ddots&   \vdots    \\
\lambda_1^{n-1} & \lambda_2^{n-1} & \dots & \lambda_n^{n-1}
\end{array} \right|\cdot
\left|
\begin{array}{cccc}
\frac{1}{p^{\prime}(\lambda_1) W(\lambda_1)}     &   0   &  \dots  &  0  \\
0     & \frac{1}{p^{\prime}(\lambda_2) W(\lambda_2)}  &  \ldots & 0     \\
\vdots          &     & \ddots&   \vdots    \\
0 & 0 & \dots & \frac{1}{p^{\prime}(\lambda_n) W(\lambda_n)}
\end{array} \right| \cdot
\left|
\begin{array}{llll}
1     &  \lambda_1    &  \ldots & \lambda_1^{n-1}  \\
1     & \lambda_2  &  \ldots & \lambda_{2}^{n-1}     \\
\vdots          &     &  &   \vdots    \\
1 & \lambda_n^{n-1} & \dots & \lambda_n^{n-1}
\end{array} \right|
$$
$$
=\frac{\displaystyle \prod_{1\le k < j \le n} (\lambda_j-\lambda_k)^2}{\displaystyle  \prod_{\ell=1}^n p^{\prime}(\lambda_{\ell})  \prod_{\ell=1}^n W(\lambda_{\ell})}   \, .
$$
Since
\begin{equation}
\prod_{1\le k < j \le n} (\lambda_j-\lambda_k)^2= \frac{(-1)^{n(n-1)/2}}{p_0^n}   \prod_{\ell=1}^n p^{\prime}(\lambda_{\ell})
\label{QQ22PP}
\end{equation}
and
\begin{equation}
\prod_{\ell=1}^n W(\lambda_{\ell}) = \prod_{\ell=1}^n \prod_{j=1}^N (\lambda_{\ell}-x_j)  =(-1)^{Nn} \prod_{j=1}^N \prod_{\ell=1}^n  (x_j-\lambda_{\ell}) = (-1)^{Nn} \frac{\prod_{j=1}^N p(x_j)}{p_0^N} \, ,
\label{QQ11PP}
\end{equation}
the validity of the first equality from (\ref{Redun1}) is established.

Next consider the  Hankel polynomial $ \mathcal H_K(x; \{\widetilde \tau\}) $  of the order $ K  \in \{n+1,\dots, \lfloor (N+n-1)/2 \rfloor \} $. With the aid of  (\ref{tau_eta1}), this polynomial can be represented as
$$
\mathcal H_K(x; \{\widetilde \tau\}) \equiv (-1)^K
\left|
\begin{array}{lllll}
\eta_0     & \eta_1     & \eta_2 & \ldots & \eta_{K}  \\
\eta_1     & \eta_2     & \eta_3 & \ldots & \eta_{K+1}    \\
\vdots        &    &     & \ddots& \vdots    \\
\eta_{K-1} & \eta_{K} & \eta_{K+1} & \ldots   & \eta_{2K-1} \\
1       & x       & x^2 &  \ldots & x^{K}
\end{array} \right|
\equiv (-1)^K \mathcal H_K(x, \{\eta\}) \, .
$$
Coefficients of $ \mathcal H_K(x; \{\widetilde \tau\}) $ coincide (up to a sign) with the $ K $th order minors of the Hankel matrix generated by the sequence (\ref{eta}). Due to Theorem \ref{Han_minors}, all these minors equal zero. Therefore,
$$ \mathcal H_{n+1}(x; \{\widetilde \tau\})\equiv 0, \dots, \mathcal H_{\lfloor (n+N-1)/2 \rfloor}(x; \{\widetilde \tau\})\equiv 0 \, . $$

Next consider the case when the Hankel polynomial is of the order $ K \in \{ \lfloor (n+N)/2 \rfloor,\dots, N-2 \} $.
$$
\mathcal H_K(x;\{\widetilde \tau\})=\left|
\begin{array}{llllll}
\widetilde \tau_0     & \widetilde \tau_1   & \ldots   & \ldots & \ldots & \widetilde \tau_{K}  \\
\widetilde \tau_1     & \widetilde \tau_2    & \ldots  & \ldots & \ldots & \widetilde \tau_{K+1}  \\
\vdots & & & & & \vdots \\
\widetilde \tau_{N+n-K-2} & \ldots & & & \ldots & \widetilde \tau_{N+n-2} \\
\widetilde \tau_{N+n-K-1} & & & & \widetilde \tau_{N+n-2} & \widetilde \tau_{N+n-1} \\
\vdots & & & & \vdots \\
\widetilde \tau_{K-1} & \ldots & \widetilde \tau_{N+n-2} & \widetilde \tau_{N+n-1}  & \ldots & \widetilde \tau_{2K-1} \\
1 & x & \ldots & \ldots & & x^K
\end{array} \right| \, .
$$
With the aid of (\ref{tau_eta1})  the entries of the first $ N+n-K-1 $ rows of this determinant can be converted into
$$
\equiv (-1)^{N+n-K-1}
\left|
\begin{array}{llllll}
\eta_0     &  \eta_1   & \ldots   & \ldots & \ldots & \eta_{K}  \\
\eta_1     & \eta_2    & \ldots  & \ldots & \ldots & \eta_{K+1}  \\
\vdots & & & & & \vdots \\
\eta_{N+n-K-2} & \ldots & & & \ldots & \eta_{N+n-2} \\
\widetilde \tau_{N+n-K-1} & & & & \widetilde \tau_{N+n-2} & \widetilde \tau_{N+n-1} \\
\vdots & & & & \vdots \\
\widetilde \tau_{K-1} & \ldots & \widetilde \tau_{N+n-2} & \widetilde \tau_{N+n-1}  & \ldots & \widetilde \tau_{2K-1} \\
1 & x & \ldots & \ldots & x^{K-1} & x^K
\end{array} \right| \, .
$$
The obtained first $ N+n-K-1 $ rows are linearly dependent since all the  $ (N+n-K-1) $th order minors of the matrix composed from these rows equal to zero; the latter statement follows from Theorem \ref{Han_minors}. Therefore $ \mathcal H_K(x;\{\widetilde \tau\}) \equiv 0 $ for the specializations of $ K $ mentioned at the beginning of the present paragraph.

The case $ K=N-1 $ is covered by Theorem \ref{Th_inter_poly2}. \qed

~\\
\indent\textbf{Remark 4.1.} Theorem \ref{Th_inter_poly4} and other theorems of the present section are proved
under additional assumption of the simplicity of zeros of the polynomial $ p(x) $. We do not give
here a proof of the validity of all the statements in the exceptional case of existence of a multiple
zero for $ p(x) $. To deal with this case, we refer to Weil's Principle of the Irrelevance of Algebraic
Inequalities (presented in Appendix). In the proof of a foregoing Theorem \ref{ThResInt} we exemplify
the practice of utilization of this principle for the proof of the statement with similar problem while
treating an exceptional case.

\begin{example} \label{Ex23} Construct the sequence of polynomials $ \{ \mathcal H_k(x;\{\widetilde \tau\}) \}_{k=1}^6 $  for the table
$$
\begin{array}{c|c|c|c|c|c|c|c}
    x & -2 & \mathbf{-1} & 0 & 1 & 2 & 3 & 4 \\
\hline
    y &  30 & \mathbf{12} & 8 & 9 & 18 & 35 & 60
  \end{array}
$$
which differs from that from Example \ref{Ex22}  with a single value at the node $ x_2=-1 $.
\end{example}

\textbf{Solution.}  On excluding this particular value from the treatment,
one gets the table which still keeps the redundancy property for evaluation of the polynomial $ p(x)=4\,x^2-3\,x+8 $ of the second degree.
It is possible to interpret the value $ y_2=12 $ as an \emph{erroneous} one in the interpolation table. We intend to analyze the possible influence of this
error upon the interpolation polynomial construction based on recursive procedure  of the Hankel polynomial computation.
One has:
$$
\mathcal H_1(x;\{\widetilde \tau\}) \equiv-\frac{341}{1814400}x+\frac{359}{453600}, \
\mathcal H_2(x;\{\widetilde \tau\}) \equiv\frac{1}{39191040000}(-19109\,x^2+34323\,x-69448),
$$
$$
\mathcal H_3(x;\{\widetilde \tau\})\equiv \frac{1}{2449440000} (x+1)(4\,x^2-3\,x+8),
$$
$$
\mathcal H_4(x;\{\widetilde \tau\}) \equiv 0,\
$$
$$
\mathcal H_5(x;\{\widetilde \tau\}) \equiv-\frac{1}{39191040000}(x+1)(4\,x^2-3\,x+8) \, ,
$$
$$
\mathcal H_6(x;\{\widetilde \tau\}) \equiv-\frac{1}{979776000}\left( \frac{1}{40}x^6-\frac{1}{5}x^5+\frac{3}{8}x^4+\frac{1}{2}x^3+\frac{21}{10}x^2-\frac{9}{5}x+8\right)
$$
with the interpolation polynomial
$$
\widetilde p(x)\equiv \frac{1}{40}x^6-\frac{1}{5}x^5+\frac{3}{8}x^4+\frac{1}{2}x^3+\frac{21}{10}x^2-\frac{9}{5}x+8
$$
which can be considered as a perturbation of the ``true'' polynomial $ p(x)=4\,x^2-3\,x+8 $:
$$
\equiv p(x) + (y_2-p(x_2)) \frac{W_2(x)}{W^{\prime}(x_2)} \, .
$$
The more surprising happens to be an observation that the Hankel polynomials $ \mathcal H_3(x;\{\widetilde \tau\}) $ and  $ \mathcal H_5(x;\{\widetilde \tau\}) $ coincide up to a constant factor, and they both contain as factors the
expression for ``true'' polynomial $ p(x) $ and the linear polynomial $ x+1 $ whose zero coincide with the value table  where the ``error'' occurs.
\qed

\begin{example} \label{Ex24} Construct the sequence of polynomials $ \{ \mathcal H_k(x;\{\widetilde \tau\}) \}_{k=1}^6 $  for the table
$$
\begin{array}{c|c|c|c|c|c|c|c}
    x & -2 & \mathbf{-1} & 0 & 1 & \mathbf{2} & 3 & 4 \\
\hline
    y &  30 & \mathbf{-7} & 8 & 9 & \mathbf{11} & 35 & 60
  \end{array}
$$
which differs from that from Example \ref{Ex22}  with values at $ x_2=-1 $ and $ x_5=2 $.
\end{example}

\textbf{Solution.}  On excluding both these nodes from the treatment,
one gets the table which still keeps the redundancy property for evaluation of the polynomial $ p(x)=4\,x^2-3\,x+8 $ of the second degree.
One has:
$$
\mathcal H_1(x;\{\widetilde \tau\}) \equiv \frac{1}{19958400}(48569\,x+24016) , \
\mathcal H_2(x;\{\widetilde \tau\}) \equiv\frac{1}{27941760000}(237819\,x^2-515468\,x-654832),
$$
$$
\mathcal H_3(x;\{\widetilde \tau\}) \equiv \frac{1}{111767040000}(-7159\,x^3+27423\,x^2-21498\,x-40400)\, ,
$$
$$
\mathcal H_4(x;\{\widetilde \tau\}) \equiv -\frac{1}{1451520000}(x+1)(x-2)(4\,x^2-3\,x+8) \, ,
$$
$$
\mathcal H_5(x;\{\widetilde \tau\})\equiv \frac{1}{9313920000}\left(-\frac{77}{12}\,x^5+\frac{77}{2}\,x^4-\frac{61}{4}\,x^3-\frac{715}{6}\,x^2+124\,x+\frac{304}{3}\right) ,
$$
$$
\mathcal H_6(x;\{\widetilde \tau\})\equiv \frac{1}{349272000} \bigg(\underbrace{ \frac{3}{80}\,x^6-\frac{59}{80}\,x^5+\frac{51}{16}\,x^4-\frac{9}{16}\,x^3-\frac{409}{40}\,x^2+\frac{93}{10}\,x+8}_{\widetilde p(x)} \bigg) \, .
$$
The interpolation polynomial is as follows:
$$
\widetilde p(x)\equiv p(x) + (y_2-p(x_2)) \frac{W_2(x)}{W^{\prime}(x_2)}+(y_5-p(x_5)) \frac{W_5(x)}{W^{\prime}(x_5)} \, .
$$
 It can be noticed that one of the computed polynomials, namely $ \mathcal H_4(x;\{\widetilde \tau\}) $, keeps the property observed in the previous example: it equals the product of ``true'' interpolation polynomial $ p(x) $  by the \textbf{error locator polynomial}, i.e. the polynomial possessing the set of zeros  coinciding with that one of the table nodes where $ p(x_j)\ne y_j $. \qed

If we continue to further damage the values of $ y $ in the table of the previous example, the observed effect disappears.

\begin{example} \label{Ex2411} Construct the sequence of polynomials $ \{ \mathcal H_k(x;\{\widetilde \tau\}) \}_{k=1}^6 $  for the table
$$
\begin{array}{c|c|c|c|c|c|c|c}
    x & -2 & \mathbf{-1} & 0 & \mathbf{1} & \mathbf{2} & 3 & 4 \\
\hline
    y &  30 & \mathbf{-7} & 8 & \mathbf{-4} & \mathbf{11} & 35 & 60
  \end{array}
$$
which differs from that from Example \ref{Ex22}  with values at $ x_2=-1, x_4=1 $ and $ x_5=2 $.
\end{example}

\textbf{Solution.}  None of the constructed Hankel polynomials is divisible by $ p(x)=4\,x^2-3\,x+8 $. Although the remained undamaged $ 4 $ values are still redundant for computation of this second order polynomial,
 the three spoiled values generate the polynomial $ p_1(x)=9/2\,x^2+3/2\,x-10 $ with the property $ p_1(3)=p(3)=35 $. \qed

\begin{theorem} \label{Th_inter_poly5}  Let polynomial $ p(x)=p_0x^n+\dots+p_n $ be of a degree $ n< N-2 $. Let the table (\ref{table})
satisfy the conditions
\begin{itemize}
  \item[(a)] $ y_j\ne 0 $ for $ j \in \{1,\dots, N\} $,
  \item[(b)] $ y_j=p(x_j) $ for $ j \in \{1,\dots, N\} \setminus \{ e \} $,
  \item[(c)] $ \widehat y_{e}=p(x_{e}) \ne y_{e}  $ and $ \widehat y_{e} \ne 0 $.
\end{itemize}
Then
\begin{equation}
 \mathcal H_{n+1} (x;\{ \widetilde \tau\}) \equiv (-1)^{nN+n(n+1)/2+1} p_0^{N-n-3}\frac{( y_{e} - \widehat y_{e})}{\displaystyle W^{\prime}(x_{e}) \prod_{j=1}^N y_j} (x-x_{e}) p(x) \, .
 \label{TTT1}
\end{equation}
If  $ n< N-3 $ then
\begin{equation}
 \mathcal H_{N-2} (x;\{ \widetilde \tau\}) \equiv \frac{(-1)^{1+(N-n)(N-n+1)/2}}{p_0^{N-n-3}} \mathcal H_{n+1} (x;\{ \widetilde \tau\}) \, .
 \label{TTT11}
\end{equation}
If $ n< N-4 $ then
\begin{equation}
\mathcal H_{n+2} (x;\{ \widetilde \tau\}) \equiv 0, \dots , \mathcal H_{N-3} (x;\{ \widetilde \tau\}) \equiv  0 \, .
 \label{TTT111}
\end{equation}
\end{theorem}

\textbf{Proof.} We assume $ x_{e} = x_1 $. One has:
$$
\widetilde \tau_k=\frac{x_1^k}{y_1W^{\prime}(x_1)}+\frac{x_2^k}{y_2W^{\prime}(x_2)}+\dots+\frac{x_N^k}{y_NW^{\prime}(x_N)}
$$
$$
=\left(\frac{x_1^k}{\widehat y_1W^{\prime}(x_1)}- \frac{\delta x_1^k}{\widehat y_1 W^{\prime}(x_1)}\right) +\frac{x_2^k}{y_2W^{\prime}(x_2)}+\dots+\frac{x_N^k}{y_NW^{\prime}(x_N)}
$$
where
$$
\delta=1 -\widehat y_1/y_1 \, .
$$
Represent the last sum as
$$
\left(\frac{x_1^k}{p(x_1)W^{\prime}(x_1)} +\frac{x_2^k}{p(x_2)W^{\prime}(x_2)}+\dots+\frac{x_N^k}{p(x_N)W^{\prime}(x_N)} \right)- \frac{\delta x_1^k}{\widehat y_1 W^{\prime}(x_1)}
$$
with the polynomial $ p(x) $ introduced in the statement of the theorem. We denote zeros of the latter as $ \lambda_1,\dots, \lambda_n $ and assume that all of them are distinct. For the sum in parentheses, one can apply the result of Corollary \ref{cor_EL1}:
$$
\widetilde \tau_k=-\left( \frac{\lambda_1^k}{p^{\prime}(\lambda_1)W(\lambda_1)} +\frac{\lambda_2^k}{p^{\prime}(\lambda_2)W(\lambda_2)}+\dots+\frac{\lambda_N^k}{p^{\prime}(x_N)W(\lambda_N)} \right) - \frac{\delta x_1^k}{\widehat y_1 W^{\prime}(x_1)} \quad \mbox{ for } \ k \in \{0,1,\dots, N+n-2\} .
$$
and
$$
\widetilde \tau_{N+n-1} = \frac{1}{p_0}-\left( \frac{\lambda_1^k}{p^{\prime}(\lambda_1)W(\lambda_1)} +\frac{\lambda_2^k}{p^{\prime}(\lambda_2)W(\lambda_2)}+\dots+\frac{\lambda_N^k}{p^{\prime}(x_N)W(\lambda_N)} \right) - \frac{\delta x_1^k}{\widehat y_1 W^{\prime}(x_1)} \, .
$$
Let us now transform the denominators of the fractions in the last expression. Introduce the polynomial
$$ F(x) \equiv (x-x_1)p(x) \, , $$
and recall the definition (\ref{Wk}) of the polynomial $ W_1(x) $.
It can be easily proved that
$$
\widehat y_1  W^{\prime}(x_1)=  F^{\prime}(x_1)W_1(x_1) \quad \mbox{ and } \  p^{\prime}(\lambda_j)W(\lambda_j)  = F^{\prime}(\lambda_j)W_1(\lambda_j) \quad \mbox{ for }  \ j\in \{ 1,\dots, n \} \, .
$$
Denote
$$
\widehat \eta_k = \frac{\delta x_1^k}{F^{\prime}(x_1)W_1(x_1)}+\frac{\lambda_1^k}{F^{\prime}(\lambda_1)W_1(\lambda_1)}+
\frac{\lambda_1^k}{F^{\prime}(\lambda_2)W_1(\lambda_2)}+ \dots + \frac{\lambda_n^k}{F^{\prime}(\lambda_n)W_1(\lambda_n)} \quad \mbox{ for } \ k\in \{0,1,\dots \} \, .
$$
We have proved the validity of the following equalities
\begin{equation}
\widetilde \tau_k=- \widehat  \eta_k \quad \mbox{ for } \ k \in \{0,1,\dots, N+n-2\}
\label{taukN}
\end{equation}
and
\begin{equation}
\widetilde \tau_{N+n-1}=1/p_0- \widehat  \eta_{N+n-1} \, .
\label{taukNp1}
\end{equation}
With the equalities (\ref{taukN}), transform the determinant $ \mathcal H_{n+1} (x,\{ \widetilde \tau\}) $:
$$
\mathcal H_{n+1} (x;\{ \widetilde \tau\})=(-1)^{n+1} \mathcal H_{n+1} (x;\{ \widehat  \eta\}) \quad \mbox{ where } \
\mathcal H_{n+1} (x;\{ \widehat  \eta\}) =
\left|
\begin{array}{llllll}
\widehat \eta_0 & \widehat \eta_1 & \widehat \eta_2 & \dots & \widehat \eta_{n} & \widehat \eta_{n+1} \\
\widehat \eta_1 & \widehat \eta_2 & \widehat \eta_3 & \dots & \widehat \eta_{n+1} & \widehat \eta_{n+2} \\
\vdots & & & & \vdots & \vdots \\
\widehat \eta_n & \widehat \eta_{n+1} & \widehat \eta_{n+2} & \dots & \widehat \eta_{2n} & \widehat \eta_{2n+1}  \\
1 & x & x^2 & \dots & x^{n} & x^{n+1}
\end{array}
\right|  \, .
$$
We intend to prove that last determinant vanishes for $ x\in \{x_1,\lambda_1,\dots,\lambda_n\} $. We will do this with the same trick as in the proof of Theorem \ref{Th_inter_poly3}. The linear equalities
\begin{equation}
\frac{\delta x_1^k}{F^{\prime}(x_1)W_1(x_1)} \mathcal H_{n+1} (x_1;\{ \widehat \eta\}) +\sum_{j=1}^n \frac{\lambda_j^k}{F^{\prime}(\lambda_j)W_1(\lambda_j)} \mathcal H_{n+1}(\lambda_j,\{\widehat \eta\}) =0 , \quad k\in \{0,\dots,n\}
\label{lin_r10}
\end{equation}
are valid since the left-hand side  can be represented in the form of determinant
$$
\left|
\begin{array}{llllll}
\widehat \eta_0 & \widehat \eta_1 & \widehat \eta_2 & \dots & \widehat \eta_{n} & \widehat \eta_{n+1} \\
\widehat \eta_1 & \widehat \eta_2 & \widehat \eta_3 & \dots & \widehat \eta_{n+1} & \widehat \eta_{n+2} \\
\vdots & & & & \vdots & \vdots \\
\widehat \eta_n & \widehat \eta_{n+1} & \widehat \eta_{n+2} & \dots & \widehat \eta_{2n} & \widehat \eta_{2n+1}  \\
\widehat \eta_k & \eta_{k+1} & \widehat \eta_{k+2} & \dots & \widehat \eta_{2n} & \widehat \eta_{n+k+1}
\end{array}
\right|
$$
which has equal rows if  $ k\in \{0,\dots,n\} $. The $ n+1 $ linear homogeneous equalities (\ref{lin_r10})
are valid for $ n+1 $ values $ \mathcal H_{n+1}(x_1;\{ \widehat \eta\}), \{\mathcal H_{n+1}(\lambda_j;\{\widehat  \eta\}) \}_{j=1}^n $.
The determinant composed of the coefficients of these values in (\ref{lin_r10}) equals
$$
\frac{\delta}{\widehat y_1 W_1(x_1) \prod_{j=1}^n F^{\prime}(\lambda_j)  \prod_{j=1}^n W_1(\lambda_j)}
\left|
\begin{array}{llll}
1 & 1 & \dots & 1 \\
x_1 & \lambda_1 & \dots & \lambda_n \\
x_1^2 & \lambda_1^2 & \dots & \lambda_n^2 \\
\vdots & & & \vdots \\
x_1^n & \lambda_1^n & \dots & \lambda_n^n
\end{array}
\right|
$$
and is nonzero. Thus
$$ \mathcal H_{n+1}(x_1;\{\widehat  \eta\})=0, \{\mathcal H_{n+1}(\lambda_j;\{\widehat \eta\})=0 \}_{j=1}^n \, . $$
Therefore, we know all the zeros of the polynomial $\mathcal H_{n+1}(x; \{\widehat \eta\}) $. Its leading coefficient
$$
\left|
\begin{array}{lllll}
\widehat \eta_0 & \widehat \eta_1 & \widehat \eta_2 & \dots & \widehat \eta_{n}  \\
\widehat \eta_1 & \widehat \eta_2 & \widehat \eta_3 & \dots & \widehat \eta_{n+1}  \\
\vdots & & & & \vdots  \\
\widehat \eta_n & \widehat \eta_{n+1} & \widehat \eta_{n+2} & \dots & \widehat \eta_{2n}
\end{array}
\right|
$$
can be represented as the product
$$
=\left|
\begin{array}{llll}
1 & 1 & \dots & 1 \\
x_1 & \lambda_1 & \dots & \lambda_n \\
x_1^2 & \lambda_1^2 & \dots & \lambda_n^2 \\
\vdots & & & \vdots \\
x_1^n & \lambda_1^n & \dots & \lambda_n^n
\end{array}
\right| \cdot
\left|
\begin{array}{cccc}
\frac{\delta}{F^{\prime}(x_1) W_1(x_1)}     &   0   &  \dots  &  0  \\
0     & \frac{1}{F^{\prime}(\lambda_1) W_1(\lambda_1)}  &  \ldots & 0     \\
\vdots          &     & \ddots&   \vdots    \\
0 & 0 & \dots & \frac{1}{F^{\prime}(\lambda_n) W_1(\lambda_n)}
\end{array} \right| \cdot
\left|
\begin{array}{lllll}
1     &  x_1    & x_1^2 & \ldots & x_1^{n}  \\
1     & \lambda_1  & \lambda_1^2  & \ldots & \lambda_{1}^{n}     \\
\vdots          &     &  &  &  \vdots    \\
1 & \lambda_n & \lambda_n^2 & \dots & \lambda_n^{n}
\end{array} \right|
$$
$$
=\frac{\delta}{F^{\prime}(x_1) W_1(x_1)} \frac{1}{\prod_{j=1}^n F^{\prime}(\lambda_j)  \prod_{j=1}^n W_1(\lambda_j)} \left(\prod_{j=1}^n (\lambda_j-x_1)\right)^2 \prod_{1\le j < k \le n} (\lambda_j-\lambda_k)^2 \, .
$$
One has
$$
F^{\prime}(x_1) W_1(x_1)=\widehat y_1 W^{\prime}(x_1),\ \left(\prod_{j=1}^n (\lambda_j-x_1)\right)^2=\left(\frac{p(x_1)}{p_0} \right)^2=\frac{\widehat y_1^2}{p_0^2}\, ,
$$
$$
\prod_{j=1}^n F^{\prime}(\lambda_j)  = \prod_{j=1}^n (\lambda_j-x_1)  \prod_{j=1}^n p^{\prime}(\lambda_j) = \left(\frac{(-1)^n \widehat y_1 }{p_0} \right)
\left( (-1)^{n(n-1)/2} p_0^n \prod_{1\le j < k \le n} (\lambda_k-\lambda_j)^2 \right) \, ,
$$
and
$$
\prod_{j=1}^n W_1(\lambda_j)=\frac{(-1)^{n(N-1)} \prod_{j=2}^N y_j }{p_0^{N-1}} \, .
$$
with the validity of the last equality established similar to (\ref{QQ11PP}). Collecting all these expressions, one gets the representation for the
leading coefficient of $ \mathcal H_{n+1} (x;\{ \widetilde \tau\}) $ in the form  corresponding to (\ref{TTT1}).

Assume now that $ n<N-3 $. Next, we are going to prove (\ref{TTT11}).
$$
 \mathcal H_{N-2} (x;\{ \widetilde \tau\}) \equiv \left|
 \begin{array}{llllll}
\widetilde \tau_0     & \widetilde \tau_1   & \ldots   &  & \ldots & \widetilde \tau_{N-2}  \\
\widetilde \tau_1     & \widetilde \tau_2    & \ldots  &  & \ldots & \widetilde \tau_{N-1}  \\
\vdots & & & & & \vdots \\
\widetilde \tau_{n} & \widetilde \tau_{n+1} & \ldots & & \ldots & \widetilde \tau_{N+n-2}  \\
\vdots & & & & & \vdots \\
 \widetilde \tau_{N-3} &   \widetilde \tau_{N-2} & \ldots & & \ldots & \widetilde \tau_{2N-5}  \\
1 & x & \ldots & \ldots & & x^{N-2}
\end{array}
 \right|_{(N-1)\times (N-1)}
$$
\begin{equation}
\stackrel{(\ref{taukN})\atop (\ref{taukNp1})}{\equiv}
\left|
 \begin{array}{cccccccc}
- \widehat \eta_0 & - \widehat \eta_1 & \dots & -\widehat \eta_{n+1} & -\widehat \eta_{n+2} & -\widehat \eta_{n+3} & \dots & -\widehat \eta_{N-2} \\
- \widehat \eta_1 & - \widehat \eta_2 & \dots & -\widehat \eta_{n+2} & -\widehat \eta_{n+3} & -\widehat \eta_{n+4} & \dots & -\widehat \eta_{N-1} \\
\vdots & & & & & & & \vdots \\
- \widehat \eta_n & - \widehat \eta_{n+1} & \dots &  &  &  & \dots & -\widehat \eta_{N+n-2} \\
- \widehat \eta_{n+1} & - \widehat \eta_{n+2} & \dots &  &  &  & \dots & 1/p_0-\widehat \eta_{N+n-1} \\
- \widehat \eta_{n+2} & - \widehat \eta_{n+3} & \dots &  &  &  &  1/p_0-\widehat \eta_{N+n-1} & \ast \\
\vdots & & & & & & & \vdots \\
- \widehat \eta_{N-3} & - \widehat \eta_{N-2} & \dots & - \widehat \eta_{N+n-2} &  1/p_0-\widehat \eta_{N+n-1} & \ast & & \ast \\
1 & x & \ldots &  x^{n+1} & x^{n+2} & \dots & \dots & x^{N-2}
\end{array}
 \right| \, .
 \label{lin_r101}
\end{equation}
Here the entries marked $ \ast $ are unessential.
Represent the last determinant as a sum of two distinguishing in their last rows: the first one contains $ [1,x,\dots,x^{n+1},0,\dots, 0 ] $ while the second ---
$ [0,0,\dots,0,x^{n+2},\dots, x^{N-2}] $:
\begin{equation}
\equiv \left|
\begin{array}{ccccccc}
 & & &    & \ast & & \ast  \\
 & & \mathfrak H^{\top} & & \ast & & \ast  \\
 & &  & & \vdots & & \vdots  \\
  & & &    & \ast & & \ast  \\
1 & x & \dots & x^{n+1} & 0 & \dots & 0
\end{array}
\right|
+\left|
\begin{array}{ccccccc}
 & & &    & \ast & & \ast  \\
 & & \mathfrak H^{\top} & & \ast & & \ast  \\
 & &  & & \vdots & & \vdots  \\
  & & &    & \ast & & \ast  \\
0 & 0 & \dots & 0 & x^{n+2} & \dots &  x^{N-2}
\end{array}
\right|
\label{sum22}
\end{equation}
with
$$
\mathfrak H=
-\left(
 \begin{array}{lllll}
\widehat \eta_0 & \widehat \eta_1 & \dots & \widehat \eta_{N-3} \\
\widehat \eta_1 & \widehat \eta_2 & \dots & \widehat \eta_{N-2} \\
\vdots &  & &  \vdots  \\
\widehat \eta_{n+1} & \widehat \eta_{n+2} & \dots & \widehat \eta_{N+n-2}
\end{array}
\right)_{(n+2)\times (N-2)} \, .
$$
Last matrix has all its minors of the order $ n+2 $ equal to zero due to result of Theorem \ref{Han_minors}. Therefore its rank is lesser than $ n+2 $ and its rows are linearly dependent.
The second determinant in (\ref{sum22}) vanishes. Consequently $  \mathcal H_{N-2} (x;\{ \widetilde \tau\}) $ is a polynomial of the degree at most $ n+1 $. We will prove that its zero set
coincide with $ \{x_1,\lambda_1,\dots,\lambda_n \} $. To do this, we utilize the trick already used in the proof of a similar claim for the polynomial $  \mathcal H_{n+1} (x;\{ \widetilde \tau\}) $.
We make linear combinations of the values  $  \mathcal H_{N-2} (x_1;\{ \widetilde \tau\}), \{\mathcal H_{N-2} (\lambda_j;\{ \widetilde \tau\}) \}_{j=1}^n  $:
\begin{equation}
\frac{\delta x_1^k}{F^{\prime}(x_1)W_1(x_1)} \mathcal H_{N-2} (x_1;\{ \widehat \eta\}) +\sum_{j=1}^n \frac{\lambda_j^k}{F^{\prime}(\lambda_j)W_1(\lambda_j)} \mathcal H_{N-2}(\lambda_j;\{\widehat \eta\})  \ \mbox{ for }  \ k\in \{0,\dots,n\} \, .
\label{lin_r11}
\end{equation}
In the determinantal form this sum can be represented as
$$
\left|
\begin{array}{ccccccc}
 & & &    & \ast & & \ast  \\
 & & \mathfrak H^{\top} & & \ast & & \ast  \\
 & &  & & \vdots & & \vdots  \\
  & & &    & \ast & & \ast  \\
\widehat \eta_{k} & \widehat \eta_{k+1} & \dots & \widehat \eta_{n+k+1} & 0 & \dots & 0
\end{array}
\right| \, .
$$
Matrix
$$
\widetilde{\mathfrak H}=
-\left(
 \begin{array}{llllll}
\widehat \eta_0 & \widehat \eta_1 & \dots & \widehat \eta_{N-3} & -\widehat \eta_{k} \\
\widehat \eta_1 & \widehat \eta_2 & \dots & \widehat \eta_{N-2} & -\widehat \eta_{k+1} \\
\vdots &  & &  & \vdots   \\
\widehat \eta_{n+1} & \widehat \eta_{n+2} & \dots & \widehat \eta_{N+n-2} & -\widehat \eta_{n+k+1}
\end{array}
\right)_{(n+2)\times (N-1)} \, .
$$
has its last column coinciding up to a sign with the $ (k+1) $th. Therefore its rank equals the rank of $\mathfrak H $ and its rows are linearly dependent. Thus all the sums (\ref{lin_r11}) equal zero. From these equalities
it can be deduced that all the values $  \mathcal H_{N-2} (x_1;\{ \widetilde \tau\}), \{\mathcal H_{N-2} (\lambda_j;\{ \widetilde \tau\}) \}_{j=1}^n  $ are zero. Therefore,
$$ \mathcal H_{N-2} (x;\{ \widehat \eta\})\equiv A (x-x_1) p(x) $$
and next task is to evaluate the numerical factor $ A $. We will extract the leading coefficient of $ \mathcal H_{N-2} (x;\{ \widehat \eta\}) $ as the cofactor of the element $ x^{n+1} $ in the last row of the
determinant (\ref{lin_r101}). We first prove that this coefficient equals
\begin{equation}
(-1)^{N+n+1}
\left|
 \begin{array}{llll}
-\widehat \eta_0 & -\widehat \eta_1 & \dots & -\widehat \eta_{n} \\
-\widehat \eta_{1} & -\widehat \eta_{2} & \dots & -\widehat \eta_{n+1} \\
\dots & & & \dots \\
-\widehat \eta_{n} & -\widehat \eta_{n+1} & \dots & -\widehat \eta_{2n}
\end{array}
\right| \cdot
\left|
\begin{array}{lllll}
0 & 0 & \dots & 0 & 1/p_0 \\
0 & 0 & \dots & 1/p_0 & \ast \\
\vdots & & & & \vdots \\
1/p_0 & \ast & \dots &  & \ast
\end{array}
\right|_{(N-n-3)\times (N-n-3)} \, .
\label{lin_r102}
\end{equation}
This will be illustrated by the example $ N=7, n=2 $:
$$
\left|
 \begin{array}{cccccc}
-\widehat \eta_0 & -\widehat \eta_1 & -\widehat \eta_2 & -\widehat \eta_3 & -\widehat \eta_4 & -\widehat \eta_5 \\
-\widehat \eta_1 & -\widehat \eta_2 & -\widehat \eta_3 & -\widehat \eta_4 & -\widehat \eta_5 & -\widehat \eta_6\\
-\widehat \eta_2 & -\widehat \eta_3 & -\widehat \eta_4 & -\widehat \eta_5 & -\widehat \eta_6 & -\widehat \eta_7\\
-\widehat \eta_3 & -\widehat \eta_4 & -\widehat \eta_5 & -\widehat \eta_6 & -\widehat \eta_7 & 1/p_0-\widehat \eta_8 \\
-\widehat \eta_4 & -\widehat \eta_5 & -\widehat \eta_6 & -\widehat \eta_7 & 1/p_0-\widehat \eta_8  & \ast \\
0 & 0 & 0 & 1 & 0  & 0
\end{array}
\right|=(-1)^{7+2+1}
\left|
 \begin{array}{ccccc}
-\widehat \eta_0 & -\widehat \eta_1 & -\widehat \eta_2 &  -\widehat \eta_4 & -\widehat \eta_5 \\
-\widehat \eta_1 & -\widehat \eta_2 & -\widehat \eta_3 &  -\widehat \eta_5 & -\widehat \eta_6\\
-\widehat \eta_2 & -\widehat \eta_3 & -\widehat \eta_4 &  -\widehat \eta_6 & -\widehat \eta_7\\
-\widehat \eta_3 & -\widehat \eta_4 & -\widehat \eta_5 & -\widehat \eta_7 & 1/p_0-\widehat \eta_8 \\
-\widehat \eta_4 & -\widehat \eta_5 & -\widehat \eta_6 & 1/p_0-\widehat \eta_8  & \ast
\end{array}
\right|
$$
$$
=
\left|
 \begin{array}{ccccc}
-\widehat \eta_0 & -\widehat \eta_1 & -\widehat \eta_2 &  -\widehat \eta_4 & -\widehat \eta_5 \\
-\widehat \eta_1 & -\widehat \eta_2 & -\widehat \eta_3 &  -\widehat \eta_5 & -\widehat \eta_6\\
-\widehat \eta_2 & -\widehat \eta_3 & -\widehat \eta_4 &  -\widehat \eta_6 & -\widehat \eta_7\\
0 & 0 & 0 & 0 & 1/p_0 \\
-\widehat \eta_4 & -\widehat \eta_5 & -\widehat \eta_6 & 1/p_0-\widehat \eta_8  & \ast
\end{array}
\right|+
\left|
 \begin{array}{ccccc}
-\widehat \eta_0 & -\widehat \eta_1 & -\widehat \eta_2 &  -\widehat \eta_4 & -\widehat \eta_5 \\
-\widehat \eta_1 & -\widehat \eta_2 & -\widehat \eta_3 &  -\widehat \eta_5 & -\widehat \eta_6\\
-\widehat \eta_2 & -\widehat \eta_3 & -\widehat \eta_4 &  -\widehat \eta_6 & -\widehat \eta_7\\
-\widehat \eta_3 & -\widehat \eta_4 & -\widehat \eta_5 & -\widehat \eta_7 & -\widehat \eta_8 \\
-\widehat \eta_4 & -\widehat \eta_5 & -\widehat \eta_6 & 1/p_0-\widehat \eta_8  & \ast
\end{array}
\right| \, .
$$
The second determinant vanishes since, due to Theorem \ref{Han_minors}, the rank of the matrix
$$
\left(
 \begin{array}{ccccc}
-\widehat \eta_0 & -\widehat \eta_1 & -\widehat \eta_2 &  -\widehat \eta_4 & -\widehat \eta_5 \\
-\widehat \eta_1 & -\widehat \eta_2 & -\widehat \eta_3 &  -\widehat \eta_5 & -\widehat \eta_6\\
-\widehat \eta_2 & -\widehat \eta_3 & -\widehat \eta_4 &  -\widehat \eta_6 & -\widehat \eta_7\\
-\widehat \eta_3 & -\widehat \eta_4 & -\widehat \eta_5 & -\widehat \eta_7 & -\widehat \eta_8
\end{array}
\right)
$$
is at most $ 3 $. Expand the remained determinant by the entries of the fourth row:
$$
=-\frac{1}{p_0}\left|
 \begin{array}{ccccc}
-\widehat \eta_0 & -\widehat \eta_1 & -\widehat \eta_2 &  -\widehat \eta_4 \\
-\widehat \eta_1 & -\widehat \eta_2 & -\widehat \eta_3 &  -\widehat \eta_5 \\
-\widehat \eta_2 & -\widehat \eta_3 & -\widehat \eta_4 &  -\widehat \eta_6 \\
-\widehat \eta_4 & -\widehat \eta_5 & -\widehat \eta_6 & 1/p_0-\widehat \eta_8
\end{array}
\right|=-\frac{1}{p_0}
\left|
 \begin{array}{cccc}
-\widehat \eta_0 & -\widehat \eta_1 & -\widehat \eta_2 &  -\widehat \eta_4 \\
-\widehat \eta_1 & -\widehat \eta_2 & -\widehat \eta_3 &  -\widehat \eta_5 \\
-\widehat \eta_2 & -\widehat \eta_3 & -\widehat \eta_4 &  -\widehat \eta_6 \\
0 & 0 & 0 & 1/p_0
\end{array}
\right|-
\frac{1}{p_0}
\left|
 \begin{array}{cccc}
-\widehat \eta_0 & -\widehat \eta_1 & -\widehat \eta_2 &  -\widehat \eta_4 \\
-\widehat \eta_1 & -\widehat \eta_2 & -\widehat \eta_3 &  -\widehat \eta_5 \\
-\widehat \eta_2 & -\widehat \eta_3 & -\widehat \eta_4 &  -\widehat \eta_6 \\
-\widehat \eta_4 & -\widehat \eta_5 & -\widehat \eta_6 &  -\widehat \eta_8
\end{array}
\right| \, .
$$
The second determinant vanishes due to Theorem \ref{Han_minors}. Therefore, for this example, the
leading coefficient of $ \mathcal H_{7-2} (x;\{ \widehat \eta\}) $ equals
$$
-\frac{1}{p_0^2}
\left|
 \begin{array}{cccc}
-\widehat \eta_0 & -\widehat \eta_1 & -\widehat \eta_2  \\
-\widehat \eta_1 & -\widehat \eta_2 & -\widehat \eta_3  \\
-\widehat \eta_2 & -\widehat \eta_3 & -\widehat \eta_4
\end{array}
\right|=
\left|
 \begin{array}{ccc}
-\widehat \eta_0 & -\widehat \eta_1 & -\widehat \eta_2  \\
-\widehat \eta_1 & -\widehat \eta_2 & -\widehat \eta_3  \\
-\widehat \eta_2 & -\widehat \eta_3 & -\widehat \eta_4
\end{array}
\right| \cdot
\left|
 \begin{array}{cc}
 0 & 1/p_0-\widehat \eta_8 \\
  1/p_0-\widehat \eta_8  & \ast
\end{array}
\right|
$$
what had to be proved.

Next we find the expressions for the determinants in (\ref{lin_r102}).
$$
\left|
\begin{array}{lllll}
0 & 0 & \dots & 0 & 1/p_0 \\
0 & 0 & \dots & 1/p_0 & \ast \\
\vdots & & & & \vdots \\
1/p_0 & \ast & \dots &  & \ast
\end{array}
\right|_{(N-n-3)\times (N-n-3)}=\frac{(-1)^{(N-n-3)(N-n-4)/2}}{p_0^{N-n-3}} \, ,
$$
$$
\left|
 \begin{array}{llll}
-\widehat \eta_0 & -\widehat \eta_1 & \dots & -\widehat \eta_{n} \\
-\widehat \eta_{1} & -\widehat \eta_{2} & \dots & -\widehat \eta_{n+1} \\
\dots & & & \dots \\
-\widehat \eta_{n} & -\widehat \eta_{n+1} & \dots & -\widehat \eta_{2n}
\end{array}
\right|
=(-1)^{n+1}
\left|
 \begin{array}{llll}
\widehat \eta_0 & \widehat \eta_1 & \dots & \widehat \eta_{n} \\
\widehat \eta_{1} & \widehat \eta_{2} & \dots & \widehat \eta_{n+1} \\
\dots & & & \dots \\
\widehat \eta_{n} & \widehat \eta_{n+1} & \dots & \widehat \eta_{2n}
\end{array}
\right|
$$
with the last determinant already evaluated in the previous part of the proof as the leading coefficient of $ \mathcal H_{n+1} (x;\{ \widehat \eta\}) $.
This completes the proof of (\ref{TTT11}).

The equalities (\ref{TTT111}) can be proved in a manner similar to that of their counterparts from Theorem \ref{Th_inter_poly4}. \qed

We now turn to the case of occurrence of several errors in the interpolation table.

\begin{theorem} \label{Th_inter_poly6}  Let $ E \in \{2,3,\dots, \lfloor N/2 \rfloor-1 \} $. Let polynomial $ p(x)=p_0x^n+\dots+p_n $ be of a degree $ n< N-2E $. Let the table (\ref{table})
satisfy the conditions
\begin{itemize}
  \item[(a)] $ y_j\ne 0 $ for $ j \in \{1,\dots, N\} $,
  \item[(b)] $ y_j=p(x_j) $ for $ j \in \{1,\dots, N\} \setminus \{ e_1,\dots,e_E \} $,
  \item[(c)] $ \widehat y_{e_s}=p(x_{e_s}) \ne y_{e_s}  $ and $ \widehat y_{e_s} \ne 0 $ for $ s\in \{1,\dots, E \} $.
\end{itemize}
Then
$$
 \mathcal H_{n+E} (x;\{ \widetilde \tau\})
$$
\begin{equation}
 \equiv (-1)^{nN+n(n+1)/2+E}p_0^{N-n-2E-1} \frac{ \displaystyle \prod_{s=1}^E ( y_{e_s} - \widehat y_{e_s}) \prod_{1\le s  < t \le E } ( x_{e_t} - x_{e_s})^2 }{\displaystyle \prod_{s=1}^E W^{\prime}(x_{e_s}) \prod_{j=1}^N y_j}  p(x) \prod_{s=1}^E (x-x_{e_s}) \, .
 \label{TTTK}
\end{equation}
If  $ n< N-2E-1 $ then
\begin{equation}
 \mathcal H_{N-E-1} (x;\{ \widetilde \tau\}) \equiv \frac{(-1)^{E+(N-n)(N-n+1)/2}}{p_0^{N-n-2E-1}} \mathcal H_{n+E} (x;\{ \widetilde \tau\}) \, .
\label{TTTK11}
\end{equation}
If  $ n< N-2E-2 $ then
\begin{equation}
\mathcal H_{n+E+1} (x;\{ \widetilde \tau\}) \equiv 0, \dots , \mathcal H_{N-E-2} (x;\{ \widetilde \tau\}) \equiv 0 \, .
\label{TTTK111}
\end{equation}
\end{theorem}

\textbf{Proof.} We assume $ \{ e_s=s \}_{s=1}^E $. One has:
$$
\widetilde \tau_k=\sum_{j=1}^N \frac{x_j^k}{y_jW^{\prime}(x_j)}
=\sum_{s=1}^E\left(\frac{x_s^k}{\widehat y_sW^{\prime}(x_s)}- \frac{\delta_s x_s^k}{\widehat y_s W^{\prime}(x_s)}\right) +\sum_{j=E+1}^N \frac{x_j^k}{y_jW^{\prime}(x_j)}
$$
where $ \{\delta_s=1 -\widehat y_s/y_s \}_{s=1}^E $. Represent the last sum as
$$
\sum_{j=1}^N\frac{x_j^k}{p(x_j)W^{\prime}(x_j)}- \sum_{s=1}^E\frac{\delta_s x_s^k}{\widehat y_s W^{\prime}(x_s)}
$$
with the polynomial $ p(x) $ introduced in the statement of the theorem. We denote zeros of the latter as $ \lambda_1,\dots, \lambda_n $ and assume that all of them are distinct.
\begin{equation}
\widetilde \tau_k
\stackrel{(\ref{Eu-Lag2})}{=}
\left\{
\begin{array}{rl} \displaystyle
-   \sum_{s=1}^E\frac{\delta_s x_s^k}{ p(x_s) W^{\prime}(x_s)} - \sum_{\ell=1}^n \frac{\lambda_{\ell}^k}{p^{\prime}(\lambda_{\ell})W(\lambda_{\ell})} & \mbox { if } \  k < N+n-1; \\
\displaystyle \frac{1}{p_0}  -  \sum_{s=1}^E\frac{\delta_s x_s^{N-n-1}}{p(x_s) W^{\prime}(x_s)} - \sum_{\ell=1}^n \frac{\lambda_{\ell}^{N-n-1}}{p^{\prime}(\lambda_{\ell})W(\lambda_{\ell})} & \mbox{ if} \ k = N+n-1.
\end{array} \right.
\label{tauk02}
\end{equation}
Let us now transform the denominators of the fractions. Introduce the polynomials
$$ F(x) \equiv p(x) \prod_{s=1}^E (x-x_s) \, ,\ W_{1,\dots,E}(x)\equiv \frac{W(x)}{\prod_{s=1}^E (x-x_s)}\equiv \prod_{j=E+1}^N (x-x_j) \, . $$
It can be easily proved that
\begin{equation}
\left\{\widehat y_s  W^{\prime}(x_s)=  F^{\prime}(x_s)W_{1,\dots,E}(x_s) \right\}_{s=1}^E \quad \mbox{ and } \quad  \left\{p^{\prime}(\lambda_{\ell})W(\lambda_{\ell})  = F^{\prime}(\lambda_{\ell})W_{1,\dots,E}(\lambda_{\ell}) \right\}_{\ell=1}^n  \, .
\label{ttty}
\end{equation}
With these relations, introduce the numbers
$$
\widehat \eta_k= \sum_{s=1}^E\frac{\delta_s x_s^k}{F^{\prime}(x_s)W_{1,\dots,E}(x_s)}+\sum_{\ell=1}^n \frac{\lambda_{\ell}^k}{F^{\prime}(\lambda_{\ell})W_{1,\dots,E}(\lambda_{\ell})} \quad \mbox{ for } \ k\in \{0,1,\dots \} \, .
$$
and rewrite (\ref{tauk02}) as
\begin{equation}
\widetilde \tau_k
=
\left\{
\begin{array}{cr} - \widehat \eta_k & \mbox { if } \  k < N+n-1; \\
\displaystyle 1/p_0  -  \widehat \eta_{N+n-1} & \mbox{ if} \ k = N+n-1.
\end{array} \right.
\label{tauk03}
\end{equation}
With these equalities, transform the determinant $ \mathcal H_{n+E} (x;\{ \widetilde \tau\}) $:
$$
\mathcal H_{n+E} (x;\{ \widetilde \tau\})\equiv(-1)^{n+E} \mathcal H_{n+E} (x;\{ \widehat  \eta\})
 \equiv (-1)^{n+E}
\left|
\begin{array}{lllll}
\widehat \eta_0 & \widehat \eta_1 & \dots & \widehat \eta_{n+E-1} & \widehat \eta_{n+E} \\
\widehat \eta_1 & \widehat \eta_2  & \dots & \widehat \eta_{n+E} & \widehat \eta_{n+E+1} \\
\vdots &  & & \vdots & \vdots \\
\widehat \eta_{n+E-1} & \widehat \eta_{n+E}  & \dots & \widehat \eta_{2n+2E-2} & \widehat \eta_{2n+2E-1}  \\
1 & x & \dots & x^{n+E-1} & x^{n+E}
\end{array}
\right|  \, .
$$
In a manner similar to that used for the counterpart polynomial from Theorem \ref{Th_inter_poly5}, it can be proved that the zero set of this polynomial coincides with $ \{ x_1,\dots,x_E,\lambda_1,\dots,\lambda_n \} $, and therefore
$ \mathcal H_{n+E} (x;\{ \widetilde \tau\}) $ differs from $ p(x) \prod_{s=1}^E (x-x_s) $ only by a numerical factor. To find it, it is sufficient to compute the determinant
$$
\left|
\begin{array}{lllll}
\widehat \eta_0 & \widehat \eta_1 & \dots & \widehat \eta_{n+E-1} \\
\widehat \eta_1 & \widehat \eta_2  & \dots & \widehat \eta_{n+E}  \\
\vdots &  & & \vdots \\
\widehat \eta_{n+E-1} & \widehat \eta_{n+E}  & \dots & \widehat \eta_{2n+2E-2}
\end{array}
\right| \,  .
$$
On representing it as a product
$$
=\left|
\begin{array}{llllll}
1 &  \dots & 1 & 1 & \dots & 1 \\
x_1 & \dots & x_E & \lambda_1 & \dots & \lambda_n \\
x_1^2 & \dots & x_E^2 & \lambda_1^2 & \dots & \lambda_n^2 \\
\vdots & & \vdots & & & \vdots \\
x_1^{n+E-1} & \dots & x_E^{n+E-1} & \lambda_1^{n+E-1} & \dots & \lambda_n^{n+E-1}
\end{array}
\right|
$$
$$
\times
\left|
\begin{array}{cccccc}
\frac{\delta_1}{F^{\prime}(x_1) W_{1,\dots,E}(x_1)}   & 0   & \dots  &   0   &  \dots  &  0  \\
   & \ddots   & \dots  &      &  \dots  &  0  \\
0 & \dots & \frac{\delta_E}{F^{\prime}(x_E) W_{1,\dots,E}(x_E)}   &    0   &  \dots  &  0  \\
0    & \dots & 0 & \frac{1}{F^{\prime}(\lambda_1) W_{1,\dots,E}(\lambda_1)}  &  \ldots & 0     \\
\vdots    &  &      &     & \ddots&   \vdots    \\
0 & 0 & \dots & & 0 & \frac{1}{F^{\prime}(\lambda_n) W_{1,\dots,E}(\lambda_n)}
\end{array} \right|
$$
$$
\times
\left|
\begin{array}{lllll}
1     &  x_1    & x_1^2 & \ldots & x_1^{n+E-1}  \\
\vdots &  & & & \vdots \\
1     &  x_E    & x_E^2 & \ldots & x_E^{n+E-1}  \\
1     & \lambda_1  & \lambda_1^2  & \ldots & \lambda_{1}^{n+E-1}     \\
\vdots          &     &  &  &  \vdots    \\
1 & \lambda_n & \lambda_n^2 & \dots & \lambda_n^{n+E-1}
\end{array} \right|
$$
we arrive at
$$
\prod_{1 \le j < k\le E} (x_k-x_j)^2  \prod_{1 \le J < K\le n} (\lambda_K-\lambda_J)^2 \prod_{s=1}^E \prod_{\ell=1}^n (\lambda_{\ell}-x_s)^2
$$
$$
\times \frac{\displaystyle \prod_{s=1}^E \delta_s}{\displaystyle \prod_{s=1}^E (F^{\prime}(x_s)  W_{1,\dots,E}(x_s))
\prod_{\ell=1}^n ( F^{\prime}(\lambda_{\ell}) W_{1,\dots,E}(\lambda_{\ell}))} \, .
$$
with the aid of (\ref{QQ22PP}), (\ref{QQ11PP}) and (\ref{ttty}) we transform this result into
$$
=\prod_{1 \le j < k\le E} (x_k-x_j)^2 \cdot \frac{\displaystyle (-1)^{n(n-1)/2}\prod_{\ell=1}^n p^{\prime}(\lambda_{\ell})}{p_0^n}   \left(\frac{\displaystyle \prod_{s=1}^E p(x_s)}{p_0^E} \right)^2
$$
$$
\times \frac{\displaystyle \prod_{s=1}^E \delta_s}{\displaystyle \left(\prod_{s=1}^E \widehat y_s \right) \left(\prod_{s=1}^E W^{\prime}(x_s) \right) \left( \prod_{\ell=1}^n p^{\prime}(\lambda_{\ell}) \right) (-1)^{Nn}  \left(\prod_{j=1}^N p(x_j)\right)\bigg/p_0^N }
$$
$$
=(-1)^{Nn+n(n-1)/2} p_0^{N-n-2E} \frac{\displaystyle \prod_{1 \le j < k\le E} (x_k-x_j)^2 \prod_{s=1}^E \delta_s \prod_{s=1}^E p(x_s)}{\displaystyle \prod_{s=1}^E W^{\prime}(x_s) \prod_{j=1}^N p(x_j)}
$$
$$
=(-1)^{Nn+n(n-1)/2} p_0^{N-n-2E} \frac{\displaystyle  \prod_{1 \le j < k\le E} (x_k-x_j)^2 \prod_{s=1}^E (y_s-\widehat y_s) }{\displaystyle \prod_{s=1}^E W^{\prime}(x_s) \prod_{j=1}^N y_j} \, .
$$
This completes the proof of (\ref{TTTK}).

Let us now prove (\ref{TTTK11}).  To do this, the initial step is quite similar to that for the polynomial $ \mathcal H_{n+E} $, i.e. we conclude that the
degree of the polynomial
$$
\mathcal H_{N-E-1} (x;\{ \widetilde \tau\})\equiv
\left|
 \begin{array}{llll}
\widetilde \tau_0     & \widetilde \tau_1    & \ldots & \widetilde \tau_{N-E-1}  \\
\widetilde \tau_1     & \widetilde \tau_2     & \ldots & \widetilde \tau_{N-E}  \\
\vdots & \vdots  & & \vdots \\
\widetilde \tau_{n+E-1}     & \widetilde \tau_{n+E}    & \ldots & \widetilde \tau_{N+n-2}  \\
\widetilde \tau_{n+E} & \widetilde \tau_{n+E+1} & \ldots & \widetilde \tau_{N+n-1}  \\
\vdots & \vdots  & & \vdots \\
 \widetilde \tau_{N-E-2} &   \widetilde \tau_{N-E-1} & \ldots & \widetilde \tau_{2N-2E-3}  \\
1 & x & \ldots  & x^{N-E-1}
\end{array}
 \right|_{(N-E)\times (N-E)}
$$
$$
\stackrel{(\ref{tauk03})}{\equiv}
\left|
\begin{array}{lllllll}
-\widehat \eta_0 & -\widehat \eta_1 & \dots & -\widehat \eta_{n+E} & -\widehat \eta_{n+E+1} & \dots & - \widehat \eta_{N-E-1}  \\
-\widehat \eta_1 & -\widehat \eta_2 & \dots & -\widehat \eta_{n+E+1} & -\widehat \eta_{n+E+2} & \dots & - \widehat \eta_{N-E}  \\
\vdots &  & & \vdots & & & \vdots \\
-\widehat \eta_{n+E-1} & -\widehat \eta_{n+E} & \dots & & &  &  - \widehat \eta_{N+n-2}  \\
-\widehat \eta_{n+E} & -\widehat \eta_{n+E+1} & \dots &  &  & & 1/p_0- \widehat \eta_{N+n-1}  \\
\vdots &  & & \vdots & &  & \ast \\
-\widehat \eta_{N-E-2} & -\widehat \eta_{N_E-1} & \dots & -\widehat \eta_{N+n-2} &  1/p_0- \widehat \eta_{N+n-1} & \dots \\
1 & x & \dots & x^{n+E} & x^{n+E+1} & \dots & x^{N-E-1}
\end{array}
\right|_{(N-E)\times (N-E)}
$$
does not exceed $ n+E $ and that its zero set coincides with $ \{ x_1,\dots,x_E,\lambda_1,\dots,\lambda_n \} $. Both conclusions are deduced similar to their counterparts for the polynomial $ \mathcal H_{n+E} (x;\{ \widetilde \tau\}) $.  From these it follows that both polynomials
$ \mathcal H_{N-E-1} (x;\{ \widetilde \tau\}) $ and $ \mathcal H_{n+E} (x;\{ \widetilde \tau\}) $ differs only by numerical factor. To find the latter let us compute the leading coefficient of $ \mathcal H_{N-E-1} (x;\{ \widetilde \tau\}) $. Its determinantal representation
$$
(-1)^{N+n+1}\left|
\begin{array}{lllllll}
-\widehat \eta_0 & -\widehat \eta_1 & \dots &  -\widehat \eta_{n+E-1} &  -\widehat \eta_{n+E+1} & \dots & - \widehat \eta_{N-E-1}  \\
-\widehat \eta_1 & -\widehat \eta_2 & \dots &  -\widehat \eta_{n+E} & -\widehat \eta_{n+E+2} & \dots & - \widehat \eta_{N-E}  \\
\vdots &  & & & \vdots & & \vdots \\
-\widehat \eta_{n+E-1} & -\widehat \eta_{n+E} & &  & \dots &   & - \widehat \eta_{N+n-2}  \\
-\widehat \eta_{n+E} & -\widehat \eta_{n+E+1} & &  & \dots &  & 1/p_0- \widehat \eta_{N+n-1}  \\
\vdots &  & & & \vdots &   & \ast \\
-\widehat \eta_{N-E-2} & -\widehat \eta_{N-E-1} & \dots & -\widehat \eta_{N+n-3}  &  1/p_0- \widehat \eta_{N+n-1} & \dots \\
\end{array}
\right|_{(N-E-1)\atop \times (N-E-1)}
$$
can be reduced to
$$
(-1)^{N+n+1}
\left|
\begin{array}{lllll}
-\widehat \eta_0 & -\widehat \eta_1 & \dots & -\widehat \eta_{n+E-1} \\
-\widehat \eta_1 & -\widehat \eta_2  & \dots & -\widehat \eta_{n+E}  \\
\vdots &  & & \vdots \\
-\widehat \eta_{n+E-1} & -\widehat \eta_{n+E}  & \dots & -\widehat \eta_{2n+2E-2}
\end{array}
\right|\cdot
\left|
\begin{array}{lllll}
0 & 0 & \dots & 0 & 1/p_0 \\
0 & 0 & \dots & 1/p_0 & \ast \\
\vdots & & & & \vdots \\
1/p_0 & \ast & \dots &  & \ast
\end{array}
\right|_{(N-n-2E-1)\times (N-n-2E-1)}
$$
The first determinant in this product is the leading coefficient of the polynomial $ \mathcal H_{n+E} (x;\{ \widetilde \tau\}) $. The remained factors constitute that one presented in
(\ref{TTTK11}).

We skip the proof of formulas (\ref{TTTK111}) as it is similar to that of formulas (\ref{TTT111}) in Theorem \ref{Th_inter_poly5}.
\qed

Let us now experiment in construction of the Hankel polynomial sequences $ \{ \mathcal H_k(x;\{\tau\}) \}_{k=1}^{N-1} $
generated by the sequence (\ref{tauk}), i.e.
$$
\left\{\tau_k = \sum_{j=1}^{N} y_j \frac{x_j^{k}}{W^{\prime}(x_j)} \right\}_{k=1}^{N-1}  \, .
$$
As a matter of fact, this construction does not relate to the construction of the interpolant for the table (\ref{table}). Indeed, in accordance with Theorem \ref{Th_inter_poly2}, polynomial $ \mathcal H_{N-1}(x;\{\tau\}) $  coincides, up to a numerical factor, with the interpolant for the table
$$
 \begin{array}{c|c|c|c|c}
    x & x_1 & x_2 & \ldots & x_N \\
\hline
    y & 1/y_1 & 1/y_2 & \ldots & 1/y_N
  \end{array} \, .
$$
Nevertheless, just for curiosity, let us take a look at the expressions for other polynomials $ \mathcal H_k(x;\{\tau\}) $ of the sequence for the case of redundant but erroneous tables.

\begin{example} \label{Ex231} Construct the sequence of polynomials $ \{ \mathcal H_k(x;\{\tau\}) \}_{k=1}^6 $  for the table of Example \ref{Ex23}:
$$
\begin{array}{c|c|c|c|c|c|c|c}
    x & -2 & \mathbf{-1} & 0 & 1 & 2 & 3 & 4 \\
\hline
    y &  30 & \mathbf{12} & 8 & 9 & 18 & 35 & 60
  \end{array}
$$
which is generated by the polynomial $ p(x)=4\,x^2-3\, x+ 8  $  with the exception of a single erroneous value at the node $ x_2=-1 $.
\end{example}

\textbf{Solution.} One gets:
$$
\mathcal H_{1}(x;\{\tau\})\equiv \frac{1}{40}(x+1), \ \mathcal H_{2}(x;\{\tau\})\equiv 0, \ \mathcal H_{3}(x;\{\tau\})\equiv -\frac{2}{5}(x+1), \dots
$$
and one may watch the expression for the error position as a zero of both polynomials $ \mathcal H_{1}(x;\{\tau\}) $ and $ \mathcal H_{3}(x;\{\tau\}) $.
\qed

\begin{theorem} \label{Th_inter_poly7} Let the conditions of Theorem \ref{Th_inter_poly5} be fulfilled. Then
\begin{equation}
 \mathcal H_{1} (x;\{\tau\}) \equiv \frac{( y_{e} - \widehat y_{e})}{W^{\prime}(x_{e})}  (x-x_{e}) \, .
 \label{KKK0}
\end{equation}
If  $ n< N-3 $ then
\begin{equation}
\mathcal H_{N-n-2} (x;\{\tau\}) \equiv (-1)^{1+(N-n)(N-n+1)/2}p_0^{N-n-3} \mathcal H_{1} (x;\{ \tau\}) \, .
 \label{KKK12}
\end{equation}
If $ n< N-4 $ then
\begin{equation}
\mathcal H_{2} (x;\{\tau\}) \equiv 0, \dots , \mathcal H_{N-n-3} (x;\{\tau\}) \equiv 0 \, .
 \label{KKK11}
\end{equation}
\end{theorem}

\textbf{Proof.}  We assume $ x_{e} = x_1 $. One has:
$$
\tau_k=\frac{x_1^ky_1}{W^{\prime}(x_1)}+\frac{x_2^ky_2}{W^{\prime}(x_2)}+\dots+\frac{x_N^ky_N}{W^{\prime}(x_N)}
$$
$$
=\left(\frac{x_1^k\widehat y_1}{W^{\prime}(x_1)}+ \frac{\varepsilon x_1^k}{W^{\prime}(x_1)}\right) +\frac{x_2^k}{y_2W^{\prime}(x_2)}+\dots+\frac{x_N^k}{y_NW^{\prime}(x_N)} \quad \mbox{ where } \  \varepsilon =  y_1 - \widehat y_1
$$
\begin{equation}
=\sum_{j=1}^N \frac{p(x_j)x_j^k}{W^{\prime}(x_j)}+\frac{\varepsilon x_1^k}{W^{\prime}(x_1)}\stackrel{(\ref{Eu-Lag1})}{=}
\left\{
\begin{array}{ll} \varepsilon x_1^k/W^{\prime}(x_1)  & \mbox{ if} \  k < N-n-1; \\
p_0 + \varepsilon x_1^{N-n-1}/W^{\prime}(x_1) & \mbox{ if} \ k = N-n-1.
\end{array} \right.
\label{polezno}
\end{equation}

Thus,
$$
\mathcal H_{1} (x;\{\tau\}) \equiv \left|\begin{array}{cc} \tau_0 & \tau_1 \\ 1 & x \end{array} \right| \equiv
\left|\begin{array}{cc} \varepsilon /W^{\prime}(x_1)  & \varepsilon x_1 /W^{\prime}(x_1) \\ 1 & x \end{array} \right|=\frac{\varepsilon}{W^{\prime}(x_1)}(x-x_1) \, ,
$$
and (\ref{KKK0}) is proved.

The proof of identities (\ref{KKK11}) will be illuminated at the last one.
$$
\mathcal H_{N-n-3} (x;\{\tau\})\equiv
\left|
\begin{array}{llll}
\tau_0     & \tau_1   & \ldots   &  \tau_{N-n-3}  \\
\tau_1     & \tau_2   & \ldots   &  \tau_{N-n-2}  \\
\tau_2     & \tau_3   & \ldots   &  \tau_{N-n-1}  \\
\vdots & & &  \vdots \\
1 & x & \ldots &   x^{N-n-3}
\end{array} \right| \stackrel{(\ref{polezno})}{\equiv}
\left|
\begin{array}{llll}
\frac{\varepsilon}{W^{\prime}(x_1)}     & \frac{\varepsilon x_1}{W^{\prime}(x_1)}   & \ldots   &  \frac{\varepsilon x_1^{N-n-3}}{W^{\prime}(x_1)}  \\
\frac{\varepsilon x_1}{W^{\prime}(x_1)}     & \frac{\varepsilon x_1^2}{W^{\prime}(x_1)}   & \ldots   &  \frac{\varepsilon x_1^{N-n-2}}{W^{\prime}(x_1)}  \\
\tau_2     & \tau_3   & \ldots   &  \tau_{N-n-1}  \\
\vdots & & &  \vdots \\
1 & x & \ldots &   x^{N-n-3}
\end{array} \right| \equiv 0
$$
since the first two rows of the last determinant are proportional.

At last we prove  (\ref{KKK12}). Represent the first two columns and the first row of the determinant
$$
\mathcal H_{N-n-2} (x;\{\tau\}) \equiv
\left|
\begin{array}{lllll}
\tau_0     & \tau_1   & \tau_2 & \ldots   &  \tau_{N-n-2}  \\
\tau_1     & \tau_2   & \tau_3 & \ldots   &  \tau_{N-n-1}  \\
\tau_2     & \tau_3   & \tau_4 &  \ldots   &  \tau_{N-n}  \\
\vdots & & & & \vdots \\
\tau_{N-n-3} & \tau_{N-n-2} & \tau_{N-n-1} & \ldots & \ldots \\
1 & x & x^2 &  \ldots &   x^{N-n-2}
\end{array} \right|
$$
with the aid of (\ref{polezno}):
$$
\equiv\left|
\begin{array}{lllll}
\frac{\varepsilon}{W^{\prime}(x_1)}     & \frac{\varepsilon x_1}{W^{\prime}(x_1)}   & \frac{\varepsilon x_1^2}{W^{\prime}(x_1)} & \ldots   &  \frac{\varepsilon x_1^{N-n-2}}{W^{\prime}(x_1)}  \\
\frac{\varepsilon x_1}{W^{\prime}(x_1)}     & \frac{\varepsilon x_1^2}{W^{\prime}(x_1)}   & \tau_3 & \ldots   &  \tau_{N-n-1}  \\
\frac{\varepsilon x_1^2}{W^{\prime}(x_1)}     & \frac{\varepsilon x_1^3}{W^{\prime}(x_1)}   & \tau_4 &  \ldots   &  \tau_{N-n}  \\
\vdots & & & & \vdots \\
\frac{\varepsilon x_1^{N-n-3}}{W^{\prime}(x_1)} & \frac{\varepsilon x_1^{N-n-2}}{W^{\prime}(x_1)} & \tau_{N-n-1} & \ldots & \ldots \\
1 & x & x^2 &  \ldots &   x^{N-n-2}
\end{array} \right|
$$
It is evident that $ \mathcal H_{N-n-2} (x_1;\{\tau\})=0 $ since substitution $ x=x_1 $ into the last row makes it proportional to the first one.
Represent the determinant as a sum of two:
$$
\equiv
\left|
\begin{array}{lllll}
\frac{\varepsilon}{W^{\prime}(x_1)}     & \frac{\varepsilon x_1}{W^{\prime}(x_1)}   &  \frac{\varepsilon x_1^2}{W^{\prime}(x_1)} & \ldots   &  \frac{\varepsilon x_1^{N-n-2}}{W^{\prime}(x_1)}   \\
\frac{\varepsilon x_1}{W^{\prime}(x_1)}     & \frac{\varepsilon x_1^2}{W^{\prime}(x_1)}   & \tau_3 & \ldots   &  \tau_{N-n-1}  \\
\frac{\varepsilon x_1^2}{W^{\prime}(x_1)}     & \frac{\varepsilon x_1^3}{W^{\prime}(x_1)}   & \tau_4 &  \ldots   &  \tau_{N-n}  \\
\vdots & & & & \vdots \\
\frac{\varepsilon x_1^{N-n-3}}{W^{\prime}(x_1)} & \frac{\varepsilon x_1^{N-n-2}}{W^{\prime}(x_1)} & \tau_{N-n-1} & \ldots & \ldots \\
1 & x & 0 &  \ldots &  0
\end{array} \right|+
\left|
\begin{array}{lllll}
\frac{\varepsilon}{W^{\prime}(x_1)}     & \frac{\varepsilon x_1}{W^{\prime}(x_1)}   &  \frac{\varepsilon x_1^2}{W^{\prime}(x_1)} & \ldots   &  \frac{\varepsilon x_1^{N-n-2}}{W^{\prime}(x_1)}  \\
\frac{\varepsilon x_1}{W^{\prime}(x_1)}     & \frac{\varepsilon x_1^2}{W^{\prime}(x_1)}   & \tau_3 & \ldots   &  \tau_{N-n-1}  \\
\frac{\varepsilon x_1^2}{W^{\prime}(x_1)}     & \frac{\varepsilon x_1^3}{W^{\prime}(x_1)}   & \tau_4 &  \ldots   &  \tau_{N-n}  \\
\vdots & & & & \vdots \\
\frac{\varepsilon x_1^{N-n-3}}{W^{\prime}(x_1)} & \frac{\varepsilon x_1^{N-n-2}}{W^{\prime}(x_1)} & \tau_{N-n-1} & \ldots & \ldots \\
0 & 0 & x^2 &  \ldots &   x^{N-n-2}
\end{array} \right| \, .
$$
The last determinant equals zero since the first two its columns are proportional. Therefore, $ \mathcal H_{N-n-2} (x;\{\tau\}) $ appears to be a linear polynomial with the known zero:
$$ \mathcal H_{N-n-2} (x;\{\tau\}) \equiv \Theta (x-x_1) \, . $$
To find the expression for $ \Theta $ let us compute the cofactor to the element $ 1 $ standing the last row of the remained determinant
$$
\left|
\begin{array}{ccccc}
\frac{\varepsilon}{W^{\prime}(x_1)}     & \frac{\varepsilon x_1}{W^{\prime}(x_1)}   &  \frac{\varepsilon x_1^2}{W^{\prime}(x_1)} & \ldots   &  \frac{\varepsilon x_1^{N-n-2}}{W^{\prime}(x_1)}   \\
\frac{\varepsilon x_1}{W^{\prime}(x_1)}     & \frac{\varepsilon x_1^2}{W^{\prime}(x_1)}   & \frac{\varepsilon x_1^3}{W^{\prime}(x_1)} & \ldots   &  p_0 + \frac{\varepsilon x_1^{N-n-1}}{W^{\prime}(x_1)}   \\
\frac{\varepsilon x_1^2}{W^{\prime}(x_1)}     & \frac{\varepsilon x_1^3}{W^{\prime}(x_1)}   & \frac{\varepsilon x_1^4}{W^{\prime}(x_1)} &  \ldots   &  \ast  \\
\vdots & & & & \ast \\
\frac{\varepsilon x_1^{N-n-3}}{W^{\prime}(x_1)} & \frac{\varepsilon x_1^{N-n-2}}{W^{\prime}(x_1)} & p_0 + \frac{\varepsilon x_1^{N-n-1}}{W^{\prime}(x_1)}   & \ldots & \ast \\
1 & x & 0 &  \ldots &  0
\end{array} \right|_{(N-n-1)\times (N-n-1)}  \, ;
$$
we have just utilized (\ref{polezno}); the entries marked $ \ast $ are unessential. Thus,
$$
-\Theta x_1 =(-1)^{N-n}
\left|
\begin{array}{cccc}
\frac{\varepsilon x_1}{W^{\prime}(x_1)}   &  \frac{\varepsilon x_1^2}{W^{\prime}(x_1)} & \ldots   &  \frac{\varepsilon x_1^{N-n-2}}{W^{\prime}(x_1)}   \\
\frac{\varepsilon x_1^2}{W^{\prime}(x_1)}   & \frac{\varepsilon x_1^3}{W^{\prime}(x_1)} & \ldots   &  p_0 + \frac{\varepsilon x_1^{N-n-1}}{W^{\prime}(x_1)}   \\
\frac{\varepsilon x_1^3}{W^{\prime}(x_1)}   & \frac{\varepsilon x_1^4}{W^{\prime}(x_1)} &  \ldots   &  \ast  \\
\vdots & & & \ast \\
 \frac{\varepsilon x_1^{N-n-2}}{W^{\prime}(x_1)} & p_0 + \frac{\varepsilon x_1^{N-n-1}}{W^{\prime}(x_1)}   & \ldots & \ast
\end{array} \right|_{(N-n-2)\times (N-n-2)} \, .
$$
Make the elementary transformation of the columns of the last determinant:
$$
=(-1)^{N-n}
\left|
\begin{array}{cccc}
\frac{\varepsilon x_1}{W^{\prime}(x_1)}   &  \frac{\varepsilon x_1^2}{W^{\prime}(x_1)} & \ldots   &  \frac{\varepsilon x_1^{N-n-2}}{W^{\prime}(x_1)}   \\
\frac{\varepsilon x_1^2}{W^{\prime}(x_1)}   & \frac{\varepsilon x_1^3}{W^{\prime}(x_1)} & \ldots   &  p_0 + \frac{\varepsilon x_1^{N-n-1}}{W^{\prime}(x_1)}   \\
\frac{\varepsilon x_1^3}{W^{\prime}(x_1)}   & \frac{\varepsilon x_1^4}{W^{\prime}(x_1)} &  \ldots   &  \ast  \\
\vdots & & & \ast \\
 \frac{\varepsilon x_1^{N-n-2}}{W^{\prime}(x_1)} & p_0 + \frac{\varepsilon x_1^{N-n-1}}{W^{\prime}(x_1)}   & \ldots & \ast
\end{array} \right| \cdot
\left|
\begin{array}{ccccc}
1   &  -x_1 & -x_1^2   &  \dots & -x_1^{N-n-3}   \\
0  & 1 & 0 & \dots   &  0   \\
0 & 0 & 1 &  \ldots   &  0  \\
\vdots & & & \ddots & \vdots \\
 0 & 0 & 0 & \dots & 1
\end{array} \right|
$$
$$
=
(-1)^{N-n}
\left|
\begin{array}{ccccc}
\frac{\varepsilon x_1}{W^{\prime}(x_1)}   &  0 & \ldots  & 0  &  0   \\
\frac{\varepsilon x_1^2}{W^{\prime}(x_1)}   & 0 & \ldots  & 0 &  p_0   \\
\frac{\varepsilon x_1^3}{W^{\prime}(x_1)}   & 0 &  \ldots  & p_0 &  \ast  \\
\vdots & & &  \ast & \ast \\
 \frac{\varepsilon x_1^{N-n-2}}{W^{\prime}(x_1)} & p_0  & \ldots & \ast & \ast
\end{array} \right|=
(-1)^{N-n}
\frac{\varepsilon x_1}{W^{\prime}(x_1)}
\left|
\begin{array}{cccc}
  0 & \ldots  & 0 &  p_0   \\
 0 &  \ldots  & p_0 &  \ast  \\
\vdots & &   \ast & \ast \\
p_0  & \ldots & \ast & \ast
\end{array} \right|_{(N-n-3)\times (N-n-3)} \, .
$$
Finally
$$
\Theta=(-1)^{N-n-1}\frac{\varepsilon }{W^{\prime}(x_1)} p_0^{N-n-3} (-1)^{(N-n-3)(N-n-4)/2} \, ,
$$
and this completes the proof of (\ref{KKK12}). \qed

Consider now the case of occurrence of  several errors.

\begin{example} \label{Ex241} Construct the sequence of polynomials $ \{ \mathcal H_k(x;\{\tau\}) \}_{k=1}^6 $  for the table of Example \ref{Ex24}
$$
\begin{array}{c|c|c|c|c|c|c|c}
    x & -2 & \mathbf{-1} & 0 & 1 & \mathbf{2} & 3 & 4 \\
\hline
    y &  30 & \mathbf{-7} & 8 & 9 & \mathbf{11} & 35 & 60
  \end{array}
$$
which is generated by the polynomial $ p(x)=4\,x^2-3\, x+ 8  $  with the exception of two erroneous value at $ x_2=-1 $ and $ x_5=2 $.
\end{example}

\textbf{Solution.} Here
$$
\mathcal H_{1}(x;\{\tau\})\equiv \frac{1}{80}(3\,x+38), \  \mathcal H_{2}(x;\{\tau\})\equiv -\frac{77}{320}(x+1)(x-2), \
\mathcal H_{3}(x;\{\tau\})\equiv -\frac{77}{80}x^3+\frac{1617}{320}x^2-\frac{177}{64}x-\frac{505}{32}, \dots
$$
and this time the erroneous nodes are detected as the zeros of polynomial $ \mathcal H_{2}(x;\{\tau\}) $. \qed

\begin{theorem} \label{Th_inter_poly8} Let the conditions of Theorem \ref{Th_inter_poly6} be fulfilled. Then
\begin{equation}
 \mathcal H_{E} (x;\{\tau\}) \equiv \frac{ \displaystyle \prod_{s=1}^E ( y_{e_s} - \widehat y_{e_s}) \prod_{1\le s  < t \le E } ( x_{e_t} - x_{e_s})^2 }{\displaystyle \prod_{s=1}^E W^{\prime}(x_{e_s})}  \prod_{s=1}^E (x-x_{e_s}) \, .
 \label{KKK1}
\end{equation}
If  $ n< N-2E-1 $ then
$$ \mathcal H_{N-n-E-1} (x;\{\tau\}) \equiv (-1)^{E+(N-n)(N-n+1)/2}p_0^{N-n-2E-1} \mathcal H_{E} (x;\{ \tau\}) \, . $$
If  $ n< N-2E-2 $ then
$$
\mathcal H_{E+1} (x;\{\tau\}) \equiv 0, \dots , \mathcal H_{N-n-E-2} (x;\{\tau\}) \equiv 0 \, .
$$
\end{theorem}

\textbf{Proof.} We will prove only (\ref{KKK1}) since the proofs of the rest claims of the theorem are similar to their counterparts from theorems \ref{Th_inter_poly5}-\ref{Th_inter_poly7}.
Assume $ \{ e_s=s \}_{s=1}^E $. Denote
$$ \theta_k=\sum_{s=1}^E \frac{\varepsilon_sx_s^k}{W^{\prime}(x_s)} \quad \mbox{ where } \varepsilon_j=y_j-\widehat y_j \mbox{ for } \ j\in \{1,\dots E \}, k\in \{0,1,2,\dots\} \, . $$
One has:
$$
\tau_k=\sum_{s=1}^E \frac{ \varepsilon_s x_s^k }{W^{\prime}(x_s)}+ \sum_{j=1}^N\frac{p(x_j)x_j^k}{W^{\prime}(x_j)}  \stackrel{(\ref{Eu-Lag1})}{=}
 \left\{ \begin{array}{ll}
 \theta_k & \mbox{ if } \ k \in \{0,\dots, N-n-2\}, \\
 p_0+ \theta_{N-n-1} & \mbox{ if } \  k=N-n-1 \, .
 \end{array} \right.
$$
Rewrite the expression for $ \mathcal H_{E} (x;\{\tau\}) $:
$$
\mathcal H_{E} (x;\{\tau\})\equiv \mathcal H_{E} (x;\{\theta\})
\equiv \left|\begin{array}{lllll}
\theta_0 & \theta_1 & \dots & \theta_{E-1} & \theta_{E} \\
\theta_1 & \theta_2 & \dots & \theta_{E} & \theta_{E+1} \\
\vdots & & & & \vdots \\
\theta_{E-1} & \theta_E & \dots & \theta_{2E-2}  & \theta_{2E-1} \\
1 & x & \dots & x^{E-1} & x^E
\end{array}
\right| \, .
$$
The set of zeros of $ \mathcal H_{E} (x;\{\theta\}) $ coincide with $ \{x_1,\dots,x_E\} $. This follows from the equalities
$$
\sum_{s=1}^E \frac{\varepsilon_s x_s^k}{W^{\prime}(x_s)} \mathcal H_{E} (x_s;\{\theta\})=
\left|\begin{array}{lllll}
\theta_0 & \theta_1 & \dots & \theta_{E-1} & \theta_{E} \\
\theta_1 & \theta_2 & \dots & \theta_{E} & \theta_{E+1} \\
\vdots & & & & \vdots \\
\theta_{E-1} & \theta_E & \dots & \theta_{2E-2} & \theta_{2E-1} \\
\theta_k & \theta_{k+1} & \dots & \theta^{k+E-1} & \theta^{k+E}
\end{array}
\right|=0 \quad \mbox{ for } \ k\in \{0,\dots,E-1\} \, .
$$
The leading coefficient of $ \mathcal H_{E} (x;\{\theta\}) $ is evaluated as follows
$$
\left|\begin{array}{llll}
\theta_0 & \theta_1 & \dots & \theta_{E-1}  \\
\theta_1 & \theta_2 & \dots & \theta_{E}  \\
\vdots & & & \vdots \\
\theta_{E-1} & \theta_E & \dots & \theta_{2E-2}
\end{array}
\right|
$$
$$
=
\left|\begin{array}{cccc}
1 & 1 & \dots & 1  \\
x_1 & x_2 & \dots & x_{E}  \\
\vdots & & & \vdots \\
x_1^{E-1} & x_2^{E-1} & \dots & x_E^{E-1}
\end{array}
\right| \cdot
\left|\begin{array}{cccc}
\varepsilon_1/W^{\prime}(x_1)  & 0 & \dots & 0  \\
 & \varepsilon_2/W^{\prime}(x_2) & \dots & 0  \\
\vdots & & & \vdots \\
0 & 0 & \dots & \varepsilon_E/W^{\prime}(x_E)
\end{array}
\right|\cdot
\left|\begin{array}{cccc}
1 & x_1 & \dots & x_1^{E-1}  \\
1 & x_2 & \dots & x_2^{E-1}  \\
\vdots & & & \vdots \\
1 & x_E & \dots & x_E^{E-1}
\end{array}
\right|
$$
$$
=\frac{\displaystyle \prod_{1\le s  < t \le E } ( x_{t} - x_{s})^2 \prod_{s=1}^E \varepsilon_s}{\prod_{s=1}^E W^{\prime}(x_s)} \, .
$$
\qed

To conclude the present section, we will address the problem of recovering of the ``true'' interpolation polynomial from the table probably containing erroneous values. In other words, how close are the
conditions of Theorems \ref{Th_inter_poly5} and \ref{Th_inter_poly6} to sufficient ones with regard to the problem of existence of the given number of erroneous values?

\textbf{Inverse Problem 2}. Given the system of the Hankel polynomials $ \{ \mathcal H_k(x; \{ \widetilde \tau \}) \}_{k=1}^{N-1} $ constructed for some table (\ref{table}), is it possible to conclude the existence of a positive integer $ E < N $ and a polynomial $ p(x) $ of the given degree $ n < N-1 $ such that  at least $ N-E $ of the $ N $ equalities $ \{ p(x_j)=y_j\}_{j=1}^N $ are valid?

Our successes in resolving this problem are restricted to a single error case.

\begin{theorem} Assume $ \{ y_j \ne 0 \}_{j=1}^N $. Let the polynomials
$ \mathcal H_{N-2}(x; \{ \widetilde \tau \}) $ and $ \mathcal H_{N-1}(x; \{ \widetilde \tau \}) $
constructed for the table (\ref{table}) satisfy the conditions
\begin{itemize}
  \item[(a)] $ \deg \mathcal H_{N-1}(x; \{ \widetilde \tau \}) = N-1 $;
  \item[(b)] $  \mathcal H_{N-2}(x; \{ \widetilde \tau \}) $ be  factorizable as
$$ \mathcal H_{N-2}(x; \{ \widetilde \tau \}) \equiv (x-x_{e}) \breve p(x) $$
for some $ e \in \{1,\dots, N \} $ and for polynomial $ \breve p(x) $ such that  $ \deg  \breve p(x) = N-3 $ and $ \left\{  \breve p(x_j) \ne 0 \right\}_{j=1}^N $.
\end{itemize}
There exists a number $ A \ne 0 $ such that the polynomial
$ A\breve p(x) $ satisfies the conditions
$$ A \breve p(x_j)= y_j \quad \mbox{ for } j \in \{1,\dots, N\} \setminus \{e\} \, . $$
\end{theorem}

\textbf{Proof.} We assume $ x_e=x_1 $. Consider three polynomials $ \mathcal H_{N-2}(x; \{ \widetilde \tau \}) $, $ \mathcal H_{N-1}(x; \{ \widetilde \tau \}) $ and $ \mathcal H_{N}(x; \{ \widetilde \tau \}) $ constructed for the table (\ref{table}). The structure of two last ones has been established in
Theorems \ref{Th_inter_poly2} and \ref{Th_inter_poly3}.

This triple is connected by the JJ-identity (\ref{Joach_iden}) which we rewrite in the form
$$
\mathcal H_{N-2}(x; \{ \widetilde \tau \}) \equiv (Bx+C) \widetilde p(x)+ D
\mathcal H_{N}(x; \{ \widetilde \tau \})  \,
$$
Here $\widetilde p(x) $ stands for the interpolation polynomial for the table (\ref{table}), i.e.
$ \{ \widetilde p(x_j)=y_j \}_{j=1}^N $, and   $ B \ne 0 $ due to assumption \emph{(a)} of the theorem.
On substituting $ x=x_j $  in this identity one gets
\begin{equation}
(x_j-x_{1}) \breve p(x_j) = (Bx_j+C)y_j \quad \mbox{ for }  j\in \{2,\dots, N \}
\label{LLLL11}
\end{equation}
and
$$
0=(Bx_1+C)y_1 \, .
$$
Since $ y_1 \ne 0 $, the latter equality results in $ C/B=-x_1 $. Substitution this into (\ref{LLLL11}) yields
$$  \breve p(x_j) = B y_j \quad \mbox{ for }  j\in \{2,\dots, N \} \, . $$
\qed


\section{Rational Interpolation} \label{rat-interp}

\setcounter{equation}{0}
\setcounter{theorem}{0}
\setcounter{example}{0}
\setcounter{cor}{0}

\textbf{Problem 3}. Find a rational function of the form
\begin{equation}
r(x)=\frac{p(x)}{q(x)}
\label{ratF}
\end{equation}
satisfying the table (\ref{table}), i.e.
\begin{equation}
r(x_j) = y_j \quad \mbox{ for } \ j \in \{1,\dots N \} \, .
\label{condit1}
\end{equation}
Here
$$
p(x)= p_{0}x^n + p_{1}x^{n-1} + \ldots + p_n,\
q(x) = q_{0}x^m + q_{1}x^{m-1} + \ldots + q_m, \ p_0\ne 0, q_0\ne 0 ,
$$
and
\begin{equation}
N=n+m+1
\label{numpoints}
\end{equation}
Hereinafter we do not distinguish the solutions to the problem with numerator and denominator multiplied by a common \underline{numerical} factor.

The first solution to the problem was proposed by Cauchy \cite{Cauchy}.

\begin{theorem}[Cauchy] \label{CauchyT}
Denote
$$ W_{\overline{j_1j_2\dots j_{\ell}}} (x)=\prod_{s=1}^{\ell} (x-x_{j_s}) \quad \mbox{ and } \quad  W_{j_1j_2\dots j_{\ell}} (x)= \frac{W(x)}{W_{\overline{j_1j_2\dots j_{\ell}}}(x)} \, .
$$
Solution to Problem 3 is given by the formulas
$$
p(x)=\sum_{(j_1,j_2,\dots,j_{m+1})} \frac{\prod_{s=1}^{m+1} y_{j_s}}{\prod_{s=1}^{m+1}  W_{j_1j_2\dots j_{m+1}} (x_{j_s}) } W_{j_1j_2\dots j_{m+1}} (x)
$$
and
$$
q(x)=(-1)^{mn}\sum_{(j_1,j_2,\dots,j_{m})} \frac{\prod_{\ell=1}^{m} y_{j_{\ell}}}{\displaystyle \prod_{j\in\{1,\dots,N\}\setminus \{j_1,\dots,j_m\}}  W_{\overline{j_1j_2\dots j_{m}}} (x_{j}) } W_{\overline{j_1j_2\dots j_{m}}} (x) \, .
$$
Both sums are taken with respect to all combinations $ m+1 $ and respectively $ m $ integers from the set $ \{1,\dots, N\} $.
\end{theorem}

Being valid generically, Cauchy's solution fails for some particular choices of interpolation table.
Whereas the polynomial interpolation problem always has a solution, the rational interpolation one is not always resolvable.
This defect was first discovered by Kronecker \cite{Kronecker81_2}, and later discussed by Netto \cite{Netto}.
To exemplify this, we first generate from the condition (\ref{condit1}) the system of equations
\begin{equation}
 p(x_j)=y_jq(x_j) \quad \mbox{ for } \ j \in \{1,\dots N \} \,
 \label{2411}
\end{equation}
or, equivalently,
\begin{equation}
p_n+p_{n-1}x_j +\dots+ p_1 x_j^{n-1}+p_0 x_j^n=q_m y_j +q_{m-1}x_jy_j+\dots+ q_1  x_j^{m-1}y_j+ q_0 x_j^{m}y_j \quad \mbox{ for } j \in \{1,\dots,N\}
\label{241}
\end{equation}
which is linear with respect to the $ N+1 $ coefficients of $ p(x) $ and $ q(x) $.
The principal solvability of this system can be established with the aid of Linear Algebra methods, like, for instance via Gaussian elimination procedure.

\begin{example} \label{Ex-2} Given the table
$$
\begin{array}{c|c|c|c|c|c}
    x &   -1 & 0 & 1 & 2 & 3  \\
\hline
    y &  1  & 1 & 1/3 & 3 & 1/13
  \end{array}
$$
find the  rational functions $ r(x)=p(x)/q(x) $ with $ \deg p(x)=1,\deg q(x) =3 $ satisfying  it.
\end{example}

\textbf{Solution.}  Resolving the system (\ref{241}),  one gets the expressions:
$$p(x)\equiv x-2, \quad q(x)\equiv x^3-x^2-x-2 \, . $$
However $ p(2)=0 $ and $ q(2)=0 $, and therefore the condition $ r(2)=3 $ is not satisfied.
It is not satisfied even if cancel the numerator and denominator by the common linear factor. \qed

Explanation for this phenomenon consists in nonequivalence of the passage from (\ref{condit1}) to (\ref{2411}) since for some node $ x_j $ one might
obtain a solution for the linear system (\ref{241}) satisfying both conditions $ p(x_j)=0 $ and $ q(x_j)=0 $.

On the other hand, solution to Problem 3 might be not unique.

\begin{example} \label{Ex-1} For the table
$$
\begin{array}{c|c|c|c|c|c}
    x &   -1 & 0 & 1 & 2 & 3  \\
\hline
    y &  1  & 1 & 1/3 & 1/7 & 1/13
  \end{array}
$$
generated by the rational function $ 1/(x^2+x+1) $ there exists infinitely many solutions
for Problem 3  with $ \deg p(x)=1,\deg q(x) =3 $ in the form $ (x-\lambda)/((x-\lambda)(x^2+x+1)) $ where $ \lambda \not\in \{-1,0,1,2,3\} $.
\end{example}

We now pass to development of an alternative approach to the problem, due to Jacobi. We assume that the functions $ W(x) $ and $ W_k(x) $ are kept being defined by (\ref{W})  and (\ref{Wk}) correspondingly.

\begin{theorem}\label{ThRatInt} Let $ y_j \ne 0 $ for $ j \in \{1,\dots, N\} $. Compute the values
\begin{equation}
\tau_k=\sum_{j=1}^N y_j \frac{x_j^k}{W^{\prime}(x_j)} \quad \mbox{ for } k \in \{0,\dots,2m\}
\label{AF1}
\end{equation}
and
\begin{equation}
\widetilde \tau_k=\sum_{j=1}^N \frac{1}{y_j} \frac{x_j^k}{W^{\prime}(x_j)} \quad \mbox{ for } k \in \{0,\dots,2n-2 \} \, ,
\label{AF2}
\end{equation}
and generate the corresponding Hankel polynomials $ \mathcal H_m (x;\{\tau\}) $ and $ \mathcal H_n (x;\{\widetilde \tau\}) $.
If
\begin{equation}
 H_{n}(\{\widetilde \tau\}) \ne 0
 \label{RIcond1}
\end{equation}
and
\begin{equation}
\mathcal H_m (x_j;\{\tau\})\ne 0 \quad \mbox{ for } \  j \in \{1,\dots, N\}
 \label{RIcond2}
\end{equation}
then there exists a unique solution  to Problem 3 where $ \deg p(x)=n, \deg q(x) \le m=N-n-1 $. It can be expressed as:
\begin{equation}
p(x) = H_{m+1}(\{\tau\}) \mathcal H_n (x;\{\widetilde \tau\}) = \left|
\begin{array}{llll}
\tau_0 & \tau_1 & \dots & \tau_m \\
\tau_1 & \tau_2 & \dots & \tau_{m+1} \\
\dots & & & \dots \\
\tau_{m-1} & \tau_{m} & \dots & \tau_{2m-1} \\
\tau_{m} & \tau_{m+1} & \dots & \tau_{2m}
\end{array}
\right| \cdot
\left|
\begin{array}{cccc}
\widetilde \tau_0 & \widetilde \tau_1 & \dots & \widetilde \tau_n \\
\widetilde \tau_1 & \widetilde \tau_2 & \dots & \widetilde \tau_{n+1} \\
\dots & & & \dots \\
\widetilde \tau_{n-1} & \widetilde \tau_n & \dots & \widetilde \tau_{2n-1} \\
1 & x & \dots & x^n
\end{array}
\right| \, ,
\label{RatNum}
\end{equation}
\begin{equation}
q(x) = H_{n}(\{\widetilde \tau\}) \mathcal H_m (x;\{\tau\})  = \left|
\begin{array}{llll}
\widetilde \tau_0 & \widetilde \tau_1 & \dots & \widetilde \tau_{n-1} \\
\widetilde \tau_1 & \widetilde \tau_2 & \dots & \widetilde \tau_{n} \\
\dots & & & \dots \\
\widetilde \tau_{n-1} & \widetilde \tau_n & \dots & \widetilde \tau_{2n-2}
\end{array}
\right| \cdot
\left|
\begin{array}{cccc}
\tau_0 & \tau_1 & \dots & \tau_m \\
\tau_1 & \tau_2 & \dots & \tau_{m+1} \\
\dots & & & \dots \\
\tau_{m-1} & \tau_{m} & \dots & \tau_{2m-1} \\
1 & x & \dots & x^m
\end{array}
\right|
 \, .
\label{RatDen}
\end{equation}
\end{theorem}

\textbf{Proof.} We first prove the uniqueness. If a solution to the Problem 3 exists then the equalities (\ref{241}) are valid. Multiply
$ j $th equality by $ x_j^k/W^{\prime}(x_j) $ for $ k \in \{0,\dots, m-1 \} $ and sum the obtained equalities by $ j $. Due to the  Euler-Lagrange equalities (\ref{Eu-Lag}), one arrives at a system of equations
\begin{eqnarray*}
q_m \tau_0 + q_{m-1} \tau_1 + \dots + q_1 \tau_{m-1} + q_0 \tau_m &=&0, \\
q_m \tau_1 + q_{m-1} \tau_2 + \dots + q_1 \tau_{m} + q_0 \tau_{m+1} &=&0, \\
\dots & & \dots , \\
q_m \tau_{m-1} + q_{m-1} \tau_m + \dots + q_1 \tau_{2m-2} + q_0 \tau_{2m-1} &=&0.
\end{eqnarray*}
Therefore, the denominator of the fraction should satisfy the relation
$$
Aq(x) \equiv \left|
\begin{array}{llll}
\tau_0 & \tau_1 & \dots & \tau_m \\
\tau_1 & \tau_2 & \dots & \tau_{m+1} \\
\dots & & & \dots \\
\tau_{m-1} & \tau_{m} & \dots & \tau_{2m-1} \\
1 & x & \dots & x^m
\end{array}
\right|
$$
for some constant factor $ A $.

In a similar way, multiplying the equalities
$$
p_n\frac{1}{y_j}+p_{n-1}\frac{x_j}{y_j} +\dots+ p_1 \frac{x_j^{n-1}}{y_j} +p_0  \frac{x_j^{n}}{y_j}=q_m  +q_{m-1}x_j+\dots+ q_1  x_j^{m-1}+ q_0 x_j^{m} \quad, \quad j \in \{1,\dots,N\}
$$
by $ x_j^{\ell}/W^{\prime}(x_j) $ for $ \ell \in \{0,\dots, n-1 \} $ and summarizing by $ j $, one gets the equality for the numerator:
$$
Bp(x) \equiv \left|
\begin{array}{llll}
\widetilde \tau_0 &  \widetilde \tau_1 & \dots & \widetilde \tau_n \\
\widetilde \tau_1 & \widetilde \tau_2 & \dots & \widetilde \tau_{n+1} \\
\dots & & & \dots \\
\widetilde \tau_{n-1} & \widetilde \tau_{n} & \dots & \widetilde \tau_{2n-1} \\
1 & x & \dots & x^n
\end{array}
\right|\equiv H_n(\{\widetilde \tau\}) x^n + \dots
$$
for some constant factor $ B $.  Due to assumption (\ref{RIcond1}), $ B \ne 0 $ and $ \deg p(x) = n $.

To evaluate the factors $ A $ and $ B $ substitute the obtained expressions into (\ref{2411}):
$$
A \mathcal  H_n(x_j;\{\widetilde \tau\}) = B y_j \mathcal  H_m(x_j; \{\tau\})   \quad \mbox{ for } \ j \in \{1,\dots N \} \, .
$$
Due to assumption (\ref{RIcond2}), $ A \ne 0 $ and $ \{ \mathcal  H_n(x_j;\{\widetilde \tau\}) \ne 0 \}_{j=1}^N $ .
Multiply each of these equalities by $ x_j^m/W^{\prime}(x_j) $ and sum the obtained results.
Due to the linear property of the determinant, one has:
$$
\sum_{j=1}^N \frac{ \mathcal  H_n(x_j;\{\widetilde \tau\}) x_j^m}{W^{\prime}(x_j)}
$$
$$
=
\left|\begin{array}{ccccc}
\widetilde \tau_0 &  \widetilde \tau_1 & \dots & \widetilde \tau_{n-1} & \widetilde \tau_n \\
\widetilde \tau_1 & \widetilde \tau_2 & \dots & \widetilde \tau_n & \widetilde \tau_{n+1} \\
\vdots & & & \vdots \\
\widetilde \tau_{n-1} & \widetilde \tau_{n} & \dots & \widetilde \tau_{2n-2} & \widetilde \tau_{2n-1} \\
\displaystyle \sum_{j=1}^N \frac{ x_j^m}{W^{\prime}(x_j)} &  \displaystyle \sum_{j=1}^N\frac{x_j^{m+1}}{W^{\prime}(x_j)} & \dots & \displaystyle \sum_{j=1}^N \frac{ x_j^{m+n-1}}{W^{\prime}(x_j)} & \displaystyle \sum_{j=1}^N \frac{ x_j^{m+n}}{W^{\prime}(x_j)}
\end{array}
\right|
\stackrel{(\ref{Eu-Lag})}{=}
\left|\begin{array}{lllll}
\widetilde \tau_0 &  \widetilde \tau_1 & \dots & \widetilde \tau_{n-1} & \widetilde \tau_n \\
\widetilde \tau_1 & \widetilde \tau_2 & \dots & \widetilde \tau_n & \widetilde \tau_{n+1} \\
\vdots & & & & \vdots \\
\widetilde \tau_{n-1} & \widetilde \tau_{n} & \dots & \widetilde \tau_{2n-2} & \widetilde \tau_{2n-1} \\
0 & 0 & \dots & 0 & 1
\end{array}
\right| \, .
$$
Similarly:
$$
\sum_{j=1}^N \frac{\mathcal  H_m(x_j;\{\widetilde \tau \}) y_jx_j^m}{W^{\prime}(x_j)}=
\left|\begin{array}{cccc}
\tau_0 &  \tau_1 & \dots & \tau_m \\
\tau_1 & \tau_2 & \dots &  \tau_{m+1} \\
\vdots & & & \vdots \\
\tau_{m-1} & \tau_{m} & \dots & \tau_{2m-1} \\
\displaystyle \sum_{j=1}^N \frac{ y_jx_j^m}{W^{\prime}(x_j)} & \displaystyle  \sum_{j=1}^N\frac{y_jx_j^{m+1}}{W^{\prime}(x_j)} & \dots & \displaystyle \sum_{j=1}^N \frac{ y_jx_j^{2m}}{W^{\prime}(x_j)}
\end{array}
\right| =
\left|
\begin{array}{llll}
\tau_0 & \tau_1 & \dots & \tau_m \\
\tau_1 & \tau_2 & \dots & \tau_{m+1} \\
\vdots & & & \vdots \\
\tau_{m-1} & \tau_{m} & \dots & \tau_{2m-1} \\
\tau_{m} & \tau_{m+1} & \dots & \tau_{2m}
\end{array}
\right| \, .
$$
Thus,
$$
AH_n(\{\widetilde \tau \})=BH_{m+1}(\{ \tau \}) \, .
$$
Since $ A \ne 0 $ and  $ H_n(\{\widetilde \tau \}) $, one has $ H_{m+1}(\{ \tau \}) \ne 0 $, and the last equality completes the proof of the uniqueness claim of the theorem.

To prove that the polynomials (\ref{RatNum}) and (\ref{RatDen}) indeed satisfy the equalities (\ref{2411}), we first deduce the following relationship
\begin{equation}
H_{m+1}(\{\tau\})= (-1)^{N(N-1)/2}H_{n}(\{ \widetilde \tau\}) \prod_{j=1}^N y_j \enspace .
\label{A&B}
\end{equation}
We outline here only an idea of the proof for the particular case $ n=2, m=3,N=6 $, i.e. we prove the validity of
\begin{equation}
\left| \begin{array}{cccc}
\displaystyle  \sum_{j=1}^6\frac{y_j}{W^{\prime}(x_j)} & \displaystyle \sum_{j=1}^6\frac{x_jy_j}{W^{\prime}(x_j)} & \displaystyle \sum_{j=1}^6\frac{x_j^2y_j}{W^{\prime}(x_j)} & \displaystyle \sum_{j=1}^6\frac{x_j^3y_j}{W^{\prime}(x_j)} \\
\displaystyle \sum_{j=1}^6\frac{x_jy_j}{W^{\prime}(x_j)} & \displaystyle \sum_{j=1}^6\frac{x_j^2y_j}{W^{\prime}(x_j)} & \displaystyle \sum_{j=1}^6\frac{x_j^3y_j}{W^{\prime}(x_j)} & \displaystyle \sum_{j=1}^6\frac{x_j^4y_j}{W^{\prime}(x_j)} \\
\displaystyle \sum_{j=1}^6\frac{x_j^2y_j}{W^{\prime}(x_j)} & \displaystyle \sum_{j=1}^6\frac{x_j^3y_j}{W^{\prime}(x_j)} & \displaystyle \sum_{j=1}^6\frac{x_j^4y_j}{W^{\prime}(x_j)} & \displaystyle \sum_{j=1}^6\frac{x_j^5y_j}{W^{\prime}(x_j)} \\
\displaystyle \sum_{j=1}^6\frac{x_j^3y_j}{W^{\prime}(x_j)} & \displaystyle \sum_{j=1}^6\frac{x_j^4y_j}{W^{\prime}(x_j)} & \displaystyle \sum_{j=1}^6\frac{x_j^5y_j}{W^{\prime}(x_j)}
 & \displaystyle \sum_{j=1}^6\frac{x_j^6y_j}{W^{\prime}(x_j)}
 \end{array}
\right|= -\prod_{j=1}^6 y_j
\left| \begin{array}{cc}
\displaystyle \sum_{j=1}^6\frac{1}{y_jW^{\prime}(x_j)} & \displaystyle \sum_{j=1}^6\frac{x_j}{y_jW^{\prime}(x_j)}  \\
\displaystyle \sum_{j=1}^6\frac{x_j}{y_jW^{\prime}(x_j)} & \displaystyle \sum_{j=1}^6\frac{x_j^2}{y_jW^{\prime}(x_j)}
 \end{array}
\right|
\label{A&B1}
\end{equation}
Consider the matrices standing under the determinant signs in the equality (\ref{A&B1}). The left one can be represented as a product:
$$
\left(\begin{array}{cccc}
1 & 1 & \dots & 1 \\
x_1 & x_2 & \dots & x_6 \\
x_1^2 & x_2^2 & \dots & x_6^2 \\
x_1^3 & x_2^3 & \dots & x_6^3
\end{array}
\right)
\left(\begin{array}{cccc}
y_1/W^{\prime}(x_1) & x_1y_1/W^{\prime}(x_1) & x_1^2y_1/W^{\prime}(x_1) & x_1^3y_1/W^{\prime}(x_1) \\
y_2/W^{\prime}(x_2) & x_2y_2/W^{\prime}(x_2) & x_2^2y_2/W^{\prime}(x_2) & x_2^3y_2/W^{\prime}(x_2) \\
\vdots & & & \vdots \\
y_6/W^{\prime}(x_6) & x_6y_6/W^{\prime}(x_6) & x_6^2y_6/W^{\prime}(x_6) & x_6^3y_6/W^{\prime}(x_6)
\end{array}
\right)
$$
while the right one as
$$
\left(\begin{array}{cccc}
1 & 1 & \dots & 1 \\
x_1 & x_2 & \dots & x_6
\end{array}
\right)
\left(\begin{array}{cc}
1/(y_1W^{\prime}(x_1)) & x_1/(y_1W^{\prime}(x_1)) \\
1/(y_2W^{\prime}(x_2)) & x_2/(y_2W^{\prime}(x_2)) \\
\vdots &  \vdots \\
1/(y_6W^{\prime}(x_6)) & x_6/(y_6W^{\prime}(x_6))
\end{array}
\right) \enspace .
$$
Express the determinants of these products with the aid of Cauchy-–Binet formula:
$$
\sum_{1\le j_1 < j_2 < j_3 < j_4 \le 6}
\left|\begin{array}{cccc}
1 & 1 & 1 & 1 \\
x_{j_1} & x_{j_2} & x_{j_3} & x_{j_4} \\
x_{j_1}^2 & x_{j_2}^2 & x_{j_3}^2 & x_{j_4}^2 \\
x_{j_1}^3 & x_{j_2}^3 & x_{j_3}^3 & x_{j_4}^3 \\
\end{array}
\right| \cdot
\left|\begin{array}{cccc}
y_{j_1}/W^{\prime}(x_{j_1}) & x_{j_1}y_{j_1}/W^{\prime}(x_{j_1}) & x_{j_1}^2y_{j_1}/W^{\prime}(x_{j_1}) & x_{j_1}^3y_{j_1}/W^{\prime}(x_{j_1}) \\
y_{j_2}/W^{\prime}(x_{j_2}) & x_{j_2}y_{j_2}/W^{\prime}(x_{j_2}) & x_{j_2}^2y_{j_2}/W^{\prime}(x_{j_2}) & x_{j_2}^3y_{j_2}/W^{\prime}(x_{j_2}) \\
y_{j_3}/W^{\prime}(x_{j_3}) & x_{j_3}y_{j_3}/W^{\prime}(x_{j_3}) & x_{j_3}^2y_{j_3}/W^{\prime}(x_{j_3}) & x_{j_3}^3y_{j_3}/W^{\prime}(x_{j_3}) \\
y_{j_4}/W^{\prime}(x_{j_4}) & x_{j_4}y_{j_4}/W^{\prime}(x_{j_4}) & x_{j_4}^2y_{j_4}/W^{\prime}(x_{j_4}) & x_{j_4}^3y_{j_4}/W^{\prime}(x_{j_4})
\end{array}
\right|
$$
$$
=\sum_{1\le j_1 < j_2 < j_3 < j_4 \le 6} y_{j_1} y_{j_2} y_{j_3} y_{j_4} \frac{\displaystyle \prod_{1\le k < \ell \le 4} (x_{j_{\ell}}-x_{j_{k}})^2}{W^{\prime}(x_{j_1})W^{\prime}(x_{j_2})W^{\prime}(x_{j_3})W^{\prime}(x_{j_4})}
$$
and
$$
\sum_{1\le k_1 < k_2 \le 6}
\left|\begin{array}{cc}
1 & 1  \\
x_{k_1} & x_{k_2}
\end{array}
\right| \cdot
\left|\begin{array}{cc}
1/(y_{k_1}W^{\prime}(x_{k_1})) & x_{k_1}/(y_{k_1}W^{\prime}(x_{k_1})) \\
1/(y_{k_2}W^{\prime}(x_{k_2})) & x_{k_2}/(y_{k_2}W^{\prime}(x_{k_2}))
\end{array}
\right| = \sum_{1\le k_1 < k_2 \le 6} \frac{1}{y_{k_1} y_{k_2}}  \frac{(x_{k_2}-x_{k_1})^2}{W^{\prime}(x_{k_1})W^{\prime}(x_{k_2})} \enspace .
$$
Both sums contain $ \left( \begin{array}{c} 6 \\ 4 \end{array}  \right) = \left( \begin{array}{c} 6 \\ 2 \end{array}  \right) = 10 $ summands. It turns out that the corresponding summands in these sums are equal up to a sign:
$$
y_{j_1} y_{j_2} y_{j_3} y_{j_4} \frac{\displaystyle \prod_{1\le k < \ell \le 4} (x_{j_{\ell}}-x_{j_{k}})^2}{W^{\prime}(x_{j_1})W^{\prime}(x_{j_2})W^{\prime}(x_{j_3})W^{\prime}(x_{j_4})}
=-\frac{\displaystyle \prod_{j=1}^6 y_j  }{y_{k_1} y_{k_2}} \cdot \frac{(x_{k_2}-x_{k_1})^2}{W^{\prime}(x_{k_1})W^{\prime}(x_{k_2})}
$$
for $ \{k_1,k_2\} = \{1,2,3,4,5,6\} \setminus \{j_1,j_2,j_3,j_4\} $. This proves (\ref{A&B1}).

With the aid of this formula, let us demonstrate now that $p(x_1)=y_1 q(x_1) $ with $ p(x) $ and $ q(x) $ given by (\ref{RatNum}) and (\ref{RatDen}) correspondingly. In view of (\ref{A&B}), it is sufficient to prove that
\begin{equation}
\mathcal  H_m(x_1;\{\tau\}) = (-1)^{N(N-1)/2}\mathcal  H_n(x_1;\{\widetilde \tau\}) \prod_{j=2}^N y_j \enspace .
\label{p&q}
\end{equation}
To evaluate the determinants in both sides of this equality, we will utilize the trick used in the proof of Theorem \ref{Th_inter_poly2}. Similarly to the formula (\ref{HNm1T}),  it can be proved that
$$
\mathcal H_n(x_1;\{\widetilde \tau\}) =(-1)^n
\left|
\begin{array}{llll}
 \widetilde T_0   & \widetilde T_1 & \dots & \widetilde T_{n-1} \\
  \widetilde T_1   & \widetilde T_2 & \dots & \widetilde T_{n} \\
  \vdots & & & \vdots \\
  \widetilde T_{n-1}   & \widetilde T_{n} & \dots & \widetilde T_{2n-2}
\end{array}
\right|= (-1)^n H_n (\{\widetilde T\}) \ \mbox{ where } \  \left\{\widetilde T_k = \sum_{j=2}^N \frac{x_j^k}{y_jW_1^{\prime}(x_{j})} \right\}_{k=0}^{2n-2} \enspace .
$$
Similar arguments work for the left-hand side:
$$
\mathcal H_m(x_1;\{\tau\}) =(-1)^m
\left|
\begin{array}{llll}
 T_0   & T_1 & \dots & T_{m-1} \\
 T_1   & T_2 & \dots & T_{m} \\
  \vdots & & & \vdots \\
T_{m-1}   & T_{m} & \dots & T_{2m-2}
\end{array}
\right| = (-1)^m H_m (\{T\})  \ \mbox{ where } \  \left\{T_k = \sum_{j=2}^N \frac{x_j^k}{y_jW_1^{\prime}(x_{j})} \right\}_{k=0}^{2m-2} \enspace .
$$
One may now utilize the equality (\ref{A&B}):
$$ H_m (\{T\}) / H_n (\{\widetilde T\}) = (-1)^{(N-1)(N-2)/2} \prod_{j=2}^N y_j \, . $$
Wherefrom follows (\ref{p&q}). \qed

\begin{cor} The following relationship is valid:
\begin{equation}
H_{n}(\{\widetilde \tau\}) H_{m}(\{\tau\})=H_{n+1}(\{\widetilde \tau\}) H_{m+1}(\{\tau\}) \enspace .
\label{rec_eq}
\end{equation}
\end{cor}

\textbf{Remark 5.1.} Formulation of Theorem \ref{ThRatInt} is due to the present authors.  Jacobi did not bother himself in \cite{Jacobi46} with the questions of existence or uniqueness of solution to the interpolation problem. He just only suggested that the denominator of the (potential candidate) rational interpolant can be represented in the form of the Hankel polynomial $ \mathcal H_m (x;\{\tau\})  $. On its computation, the
rational interpolation problem is reduced to the polynomial interpolation one for the numerator. Jacobi did not care either on computational aspects of the problem like those discussed in the solution of the following example
--- with the major one exploiting the result of his own preceding work!

\begin{example} \label{Ex3} Given the table
$$
\begin{array}{c|c|c|c|c|c|c|c}
    x &  -2 & -1 & 0 & 1 & 2 & 3 & 4 \\
\hline
    y & 26/51 & 2  & -1/2 & 1/6 & -4/7 & 16/31 & 7/36
  \end{array}
$$
find all the  rational functions $ r(x)=p(x)/q(x) $ with $ \deg p(x)+\deg q(x) \le 6 $ satisfying  it.
\end{example}

\textbf{Solution.} Since we do not know a priori the degrees of the numerator and the denominator of $ r(x) $,  we have to compute the values (\ref{AF1}) and (\ref{AF2}) for the maximal possible indices, i.e.
$$ \tau_k = \sum_{j=1}^7  \frac{y_jx_j^k}{W^{\prime}(x_j)} \quad \mbox{ and } \quad  \widetilde \tau_k = \sum_{j=1}^7  \frac{x_j^k}{y_jW^{\prime}(x_j)} \quad  \mbox{ for }  k\in \{0,\dots,12\} \, . $$
$$ \tau_0=-\frac{-897683}{19123776},\ \tau_1=-\frac{119579}{4780944},\ \tau_2=-\frac{240175}{2390472}, \ \tau_3= -\frac{448717}{2390472}, \dots, \tau_{12}=\frac{5257205447}{2390472};
$$
$$
\widetilde \tau_0=-\frac{2973}{11648},\ \widetilde \tau_1=-\frac{3037}{11648}, \widetilde \tau_2=-\frac{3923}{11648},  \widetilde \tau_3=-\frac{5297}{11648},
 \dots, \widetilde \tau_{12}=\frac{1294306589}{11648} \, .
$$
Now compute the Hankel polynomials of the first and the second order:
$$
\mathcal H_1(x;\{\tau\})=-\frac{897683}{19123776}x+\frac{119579}{4780944}, \ \mathcal H_2(x;\{\tau\})=\underbrace{\frac{208609}{50996736}}_{h_{2,0}}x^2\underbrace{-\frac{321193}{50996736}}_{h_{2,1}}x
\underbrace{-\frac{7649}{1416576}}_{h_{2,2}} \, .
$$
Computation of  $ \mathcal H_3(x;\{\tau\}) $ can be organized with  the aid of the JJ-identity (\ref{Joach_iden1}):
$$
\mathcal H_3(x;\{\tau\}) \equiv -
\left(\frac{h_{3,0}}{h_{2,0}}\right)^2 \mathcal H_1(x;\{\tau\})+ \frac{h_{3,0}}{h_{2,0}}\left(x-\frac{h_{2,1}}{h_{2,0}}+\frac{h_{3,1}}{h_{3,0}} \right)\mathcal H_2(x;\{ \tau\})
$$
where all the constants are already known except for $ h_{3,0}=H_3(\{\tau\}) $ and $ h_{3,1} $. To find the latter,
utilize the equalities (\ref{hk0hk1})
$$
h_{3,0}=H_3(\{\tau \})=\tau_4 h_{2,0}+\tau_3 h_{2,1}+\tau_2 h_{2,2}=-\frac{4037}{16998912} \, ,
$$
$$
h_{3,1}=-(\tau_5 h_{2,0}+\tau_4 h_{2,1}+\tau_3 h_{2,2})= \frac{36263}{50996736}\, .
$$
Therefore,
$$
\mathcal H_3(x;\{\tau\}) \equiv-\frac{4037}{16998912}x^3+\frac{36263}{50996736}x^2-\frac{767}{12749184}x-\frac{41}{75888} \, .
$$
Continuing the recursive utilization of the JJ-identity (\ref{Joach_iden1}) we get further:
$$
\mathcal H_4(x;\{\tau\}) \equiv -\frac{1915}{50996736}x^4+\frac{1915}{25498368}x^3+\frac{9575}{152990208}x+\frac{1915}{38247552} \, ,
$$
$$
\mathcal H_5(x;\{\tau\})
\equiv -\frac{1915}{21855744}x^5+\frac{36385}{76495104}x^4+\frac{6229}{8999424}x^3-\frac{40711}{10927872}x^2-
\frac{359}{6374592}x+\frac{3037}{1195236}\, ,
$$
$$
\mathcal H_6(x;\{\tau\}) \equiv
\underbrace{\frac{991}{796824}}_{h_{6,0}}x^6\underbrace{-\frac{8887}{1195236}}_{h_{6,1}}x^5+\frac{3475}{2390472}x^4+\frac{51575}{1195236}x^3-\frac{8450}{298809}x^2
-\frac{4892}{99603}x+\underbrace{\frac{416}{42687}}_{h_{6,6}} \, .
$$
We need one extra computation, namely
$$
H_7(\{\tau\})=\tau_{12}h_{6,0}+ \tau_{11}h_{6,1} + \dots + \tau_{6} h_{6,6}=-\frac{208}{42687} \, .
$$
Thus, all the potential denominators of the interpolation fractions are computed. In parallel, similar recursive
procedure can be organized for the numerator computation:
$$
\mathcal H_1(x;\{\widetilde\tau\}) =-\frac{2973}{11648}x+\frac{3037}{11648},\
\mathcal H_2(x;\{\widetilde\tau\})=\underbrace{\frac{1915}{106496}}_{\widetilde h_{2,0}=H_2(\{\widetilde \tau\})}\,x^2\underbrace{-\frac{21065}{745472}}_{\widetilde h_{2,1}}\, x+\underbrace{\frac{1915}{372736}}_{\widetilde h_{2,2}}\, ,
$$
$$
\widetilde h_{3,0}=H_3(\{\widetilde \tau \})= \widetilde \tau_4 \widetilde h_{2,0}+ \widetilde \tau_3 \widetilde h_{2,1}+ \widetilde \tau_2 \widetilde h_{2,2}=\frac{5745}{745472} \, ,
$$
$$
\widetilde h_{3,1}=-(\widetilde  \tau_5 \widetilde  h_{2,0}+\widetilde \tau_4 \widetilde  h_{2,1}+\widetilde \tau_3 \widetilde  h_{2,2})=-\frac{28725}{745472} \, ,
$$
$$
\mathcal H_3(x;\{\widetilde\tau\}) \equiv -
\left(\frac{\widetilde h_{3,0}}{\widetilde h_{2,0}}\right)^2 \mathcal H_1(x;\{\widetilde\tau\})+ \frac{\widetilde h_{3,0}}{\widetilde h_{2,0}}\left(x-\frac{\widetilde h_{2,1}}{\widetilde h_{2,0}}+\frac{\widetilde h_{3,1}}{\widetilde h_{3,0}} \right)\mathcal H_2(x;\{\widetilde \tau\})
$$
$$
\equiv \frac{5745}{745472}x^3-\frac{28725}{745472}x^2+\frac{33771}{372736}x-\frac{369}{6656} \, ,
$$
$$
\mathcal H_4(x;\{\widetilde\tau\}) \equiv
\frac{36333}{745472}x^4-\frac{72771}{372736}x^3-\frac{11139}{28672}x^2+\frac{1005843}{745472}x-\frac{206523}{372736} \, ,
$$
$$
\mathcal H_5(x;\{\widetilde\tau\}) \equiv
-\frac{625827}{745472}x^5+\frac{1708605}{372736}x^4-\frac{362367}{745472}x^3-\frac{2007361}{93184}x^2
+\frac{3068941}{186368}x+\frac{119579}{46592} \, ,
$$
$$
\mathcal H_6(x;\{\widetilde\tau\}) \equiv
\frac{897683}{93184}x^6-\frac{5805465}{93184}x^5+\frac{373613}{7168}x^4+\frac{24907053}{93184}x^3
-\frac{9008491}{23296}x^2-\frac{392865}{23296}x+\frac{42687}{416} \, .
$$
Now we are able to compose the set of interpolation rational functions:
$$
r_{0,6}(x)=\frac{H_7(\{\tau\})}{\mathcal H_6(x;\{\tau\})} \equiv
-\frac{11648}{2973\,x^6-17774\,x^5+3475\,x^4+103150\,x^3-67600\,x^2-117408\,x+23296}
$$
$$
r_{1,5}(x)=\frac{h_{6,0}\mathcal H_1(x;\{\widetilde \tau\})}{\widetilde h_{1,0} \mathcal H_5(x;\{\tau\})}\equiv -\frac{64(2973\,x-3037)}{13405\,x^5-72770\,x^4-105893\,x^3+569954\,x^2+8616\,x-388736} \, ,
$$
$$
r_{2,4}(x)=\frac{h_{5,0}\mathcal H_2(x;\{\widetilde \tau\})}{\widetilde h_{2,0} \mathcal H_4(x;\{\tau\})} \equiv \frac{7\,x^2-11\,x+2}{3\,x^4-6\,x^3-5\,x-4} \ ;
$$
$$ \cdots ; $$
$$
r_{6,0}(x)=\frac{h_{1,0}\mathcal H_6(x;\{\widetilde \tau\})}{\widetilde h_{6,0}}
$$
$$
\equiv
-\frac{897683}{19123776}x^6+\frac{1935155}{6374592}x^5-\frac{4856969}{19123776}x^4
-\frac{8302351}{6374592}x^3+\frac{9008491}{4780944}x^2+\frac{130955}{1593648}x-\frac{1}{2} \, .
$$
\qed

To conclude the present section, we mention an alternative representation for the rational interpolation given in \cite{Jacobi46}; the following result is presented in the original Jacobi's formulation\footnote{Therefore, the reader should keep in mind the comments on Jacobi's standards of rigor mentioned in Remark 5.1.}.

\begin{theorem}[Jacobi] \label{ThJacobi2} Compute the functions
\begin{equation}
\xi_k(x)=\sum_{j=1}^N \frac{x_j^ky_j}{W^{\prime}(x_j)} \cdot \frac{1}{x_j-x}\quad \mbox{ for } k \in \{0,\dots,2m \}
\label{AF4}
\end{equation}
and
\begin{equation}
\zeta_k(x)=\sum_{j=1}^N \frac{x_j^ky_j}{W^{\prime}(x_j)}(x_j-x) =\tau_{k+1}-x \tau_{k} \quad \mbox{ for } k \in \{0,\dots,2m-2\} \, ,
\label{AF3}
\end{equation}
and for $ \{\tau_k\}_{k=0}^{2m-2} $ defined by (\ref{AF1}).
Solution to Problem 3  is given by the polynomials:
\begin{equation}
p(x)\equiv  -W(x) H_{m+1}(\{\xi_k(x)\}) \equiv
-W(x)\left|\begin{array}{lllll}
\xi_0(x) & \xi_1(x) & \dots & \xi_{m-1}(x) & \xi_m(x) \\
\xi_1(x) & \xi_2(x) & \dots & \xi_{m}(x) & \xi_{m+1}(x) \\
\vdots & & \ddots &  & \vdots \\
\xi_{m-1}(x) & \xi_{m}(x) & \dots & \xi_{2m-2}(x) & \xi_{2m-1}(x) \\
\xi_m(x) & \xi_{m+1}(x) & \dots & \xi_{2m-1}(x) & \xi_{2m}(x)
\end{array}
\right|
\label{JacobiAlterN}
\end{equation}
and
\begin{equation}
q(x)\equiv H_{m}(\{\zeta_k(x)\}) \equiv \left|\begin{array}{llll}
\zeta_0(x) & \zeta_1(x) & \dots & \zeta_{m-1}(x)  \\
\zeta_1(x) & \zeta_2(x) & \dots & \zeta_{m}(x)  \\
\vdots & & \ddots &  \vdots \\
\zeta_{m-1}(x) & \zeta_{m}(x) & \dots &  \zeta_{2m-2}(x)
\end{array} \right|
\stackrel{(\ref{Hkx1})}{\equiv} (-1)^m \mathcal H_{m} (x;\{\tau\}) \, . \label{JacobiAlterD}
\end{equation}
\end{theorem}

\textbf{Remark 5.2.} The expression for the denominator (\ref{JacobiAlterD}) coincides up to a numerical factor with its counterpart (\ref{RatDen}) from Theorem \ref{ThRatInt}. As for the
expression (\ref{JacobiAlterN}) for the numerator, it looks like more complicated, from the computational point of view, in comparison with (\ref{RatNum}) --- at least for the case if the fraction is proper one. Jacobi proved that solution given in Theorem \ref{ThJacobi2} is equivalent to Cauchy's solution from Theorem \ref{CauchyT}, and extended his approach to the general interpolation problem --- when at any node the values of some derivatives of the function are specified along with its value.

We next intend to discover what assumption from those posed in Theorem \ref{ThRatInt}
is responsible for the uniqueness of the solution to the rational interpolation problem, i.e. the one that prevents the cases like that in Example \ref{Ex-1}.

\section{Resultant Interpolation} \label{ResInterp}

\setcounter{equation}{0}
\setcounter{theorem}{0}
\setcounter{example}{0}

Let us clarify the meaning of the Hankel determinants $ H_{m+1}(\{\tau\})  $ and $ H_{n}(\{\widetilde \tau\}) $  appeared in Theorem  \ref{ThRatInt}. They are related to the main object of Elimination Theory known as \textbf{the resultant} of the polynomials. We first recall the definition \cite{Bocher,Uspensky,VZG}.
For the polynomials
$$
p(x)=p_0x^n+p_1x^{n-1}+\dots+p_n \quad \mbox{ and } \quad q(x)=q_0x^m+q_1x^{m-1}+\dots+q_m
$$
with $ p_0 \ne 0, q_0\ne 0, n\ge 1, m \ge 1 $ their resultant is formally defined as
\begin{equation}
\mathcal R (p(x),q(x))= p_0^m \prod_{j=1}^n q(\lambda_j)
\label{Res1}
\end{equation}
where $ \{\lambda_j\}_{j=1}^n $ denote the zeros of $ p(x) $ (counted with their multiplicities). Equivalently, the resultant can be defined as
\begin{equation}
\mathcal R (p(x),q(x))= (-1)^{mn}q_0^n \prod_{\ell=1}^m p(\mu_{\ell})
\label{Res2}
\end{equation}
where $ \{\mu_{\ell}\}_{\ell=1}^m $ denote the zeros of $ q(x) $ (counted with their multiplicities). As for the constructive methods of computing the resultant as a polynomial function of the coefficients of $p(x) $ and $ q(x) $, this can be done with the aid of several determinantal representation (like Sylvester's, B\'ezout's or Kronecker's).

\begin{example} For the polynomials
$$ p(x)=p_0x^3+p_1x^2+p_2x+p_3 \quad \mbox{ and } \quad  q(x)= q_0x^5+ \dots +  q_5 $$
with $ p_0 \ne 0, q_0 \ne 0 $, their resultant in Sylvester's form is the $ (3+5) $-order determinant\footnote{Not indicated entries of the determinant are assigned to zero.}:
$$
\mathcal R (p(x),q(x))=-
\left|\begin{array}[l]{cccccccc}
p_0&p_1&p_2&p_3&&&&\\
& p_0&p_1&p_2&p_3&&&\\
&& p_0&p_1&p_2&p_3&&\\
&&& p_0&p_1&p_2&p_3&\\
&&&& p_0&p_1&p_2&p_3\\
&& q_0 & q_1 & q_2 & q_3 & q_4 & q_5 \\
& q_0 & q_1 & q_2 & q_3 & q_4 & q_5 &  \\
q_0 & q_1 & q_2 & q_3 & q_4 & q_5 & &  \\
\end{array}\right|
\begin{array}{l}
\left.\begin{array}{r}
\\ \\ \\ \\ \\
\end{array}\right\}5
\\
\left.\begin{array}{r}
\\ \\ \\
\end{array}\right\} 3
\end{array}
$$

\end{example}

\begin{theorem}  Polynomials $ p(x) $ and $ q(x) $ possess a common zero if and only if their resultant vanishes: $ \mathcal R (p(x),q(x))= 0 $.
\end{theorem}

An important particular case related to the resultant of a polynomial and its derivative, gives rise to a special notion: the expression
$$ \mathcal D(p(x)) = \frac{(-1)^{n(n-1)/2}}{p_0} \mathcal R (p(x),p^{\prime}(x))  $$
is known as the \textbf{discriminant} of the polynomial $ p(x) $. It is a polynomial function in the coefficients $ p_0,\dots, p_n $.

\begin{cor} \label{CorDiscr} $ \mathcal D(p(x)) \ne 0 $ if and only if all the zeros of polynomial $ p(x) $ are distinct.
\end{cor}

Compute now the values
\begin{equation}
p(x_j), q(x_j)  \quad \mbox{ for } \quad j\in \{1,\dots, N=m+n+1 \}
\label{pqmnogo}
\end{equation}
and assume that none of them is zero. Let $ W(x) $ be defined by (\ref{W}). Compute the values
\begin{equation}
\tau_k = \sum_{j=1}^{N} \frac{p(x_j)}{q(x_j)} \frac{x_j^{k}}{W^{\prime}(x_j)} \quad \mbox{ for } \ k \in \{0,\dots, 2m  \}
 \label{tau}
\end{equation}
and
\begin{equation}
\widetilde \tau_k = \sum_{j=1}^{N} \frac{q(x_j)}{p(x_j)} \frac{x_j^{k}}{W^{\prime}(x_j)} \quad \mbox{ for } \ k \in \{0,\dots, 2n  \} \, . \label{wtau}
\end{equation}
Compose the Hankel matrices:
$$
H({\tau}) = \left[\tau_{i+j-2} \right]_{i,j=1}^{m+1}, \quad H({\widetilde \tau}) = \left[\widetilde \tau_{i+j-2} \right]_{i,j=1}^{n+1} \, .
$$
Denote their $ k $th leading principal minors as $ H_k(\{\tau\}) $ and $ H_k(\{\widetilde \tau\}) $ correspondingly.

\begin{theorem} \label{ThResInt} The following equalities are valid
\begin{equation}
H_{m}(\{\tau\})=\frac{ (-1)^{m(m+1)/2} q_0}{\prod_{j=1}^N q(x_j)} \mathcal R(p(x),q(x))\, ,\quad H_{n}(\{\widetilde \tau\})= \frac{(-1)^{mn+n(n+1)/2} p_0}{\prod_{j=1}^N p(x_j)} \mathcal R(p(x),q(x)) \, ;
\label{HmHn}
\end{equation}
\begin{equation}
H_{m+1}(\{\tau\})=\frac{(-1)^{m(m+1)/2}p_0}{\prod_{j=1}^N q(x_j)} \mathcal R(p(x),q(x)) ,\quad H_{n+1}(\{\widetilde \tau\})=\frac{(-1)^{mn+n(n+1)/2} q_0}{\prod_{j=1}^N p(x_j)} \mathcal R(p(x),q(x))  \, .
\label{Hm+1Hn+1}
\end{equation}
\end{theorem}

\textbf{Proof} will be illuminated for a particular case $ n=3,m=5 $. Consider first the case where $ p(x) $ possesses only simple zeros; denote them by $ \lambda_1,\lambda_2,\lambda_3 $. Construct a new sequence:
\begin{equation}
\eta_{k} = \sum_{\ell=1}^3 \frac{q(\lambda_{\ell})\lambda_{\ell}^k}{p^{\prime}(\lambda_{\ell})W(\lambda_{\ell})} \quad
\mbox{ for } \quad k \in \{0,1,\dots \} \, .
\label{tau_lam00}
\end{equation}
Similarly to the proof of the relationships (\ref{tau_eta1}) one can deduce that
$$
\widetilde \tau_k = - \eta_{k} \quad \mbox{ for } \ k \in \{0,1,2,3,4,5\} .
$$
With the aid this formula, rewrite  the expression for the determinant $ H_{3}(\{\widetilde \tau\}) $:
$$
\left|
\begin{array}{ccc}
\widetilde \tau_{0} & \widetilde \tau_{1} & \widetilde \tau_{2} \\
\widetilde \tau_{1} & \widetilde \tau_{2} & \widetilde \tau_{3} \\
\widetilde \tau_{2} & \widetilde \tau_{3} & \widetilde \tau_{4}
\end{array}
\right|
=-
\left|
\begin{array}{ccc}
\eta_{0} & \eta_{1} & \eta_{2} \\
\eta_{1} & \eta_{2} & \eta_{3} \\
\eta_{2} & \eta_{3} & \eta_{4}
\end{array}
\right|=
$$
$$
=-
\left|
\begin{array}{ccc}
1 & 1 & 1 \\
\lambda_1 & \lambda_2 & \lambda_3 \\
\lambda_1^2 & \lambda_2^2 & \lambda_3^2
\end{array}
\right|\cdot
\left|
\begin{array}{ccc}
 \frac{q(\lambda_{1})}{p^{\prime}(\lambda_1)W(\lambda_{1})} &  0 & 0 \\
0 & \frac{q(\lambda_{2})}{p^{\prime}(\lambda_2)W(\lambda_{2})} & 0 \\
0 & 0 & \frac{q(\lambda_{3})}{p^{\prime}(\lambda_3)W(\lambda_{3})}
\end{array}
\right|
\cdot
\left|
\begin{array}{ccc}
1 & \lambda_1 & \lambda_1^2 \\
1 & \lambda_2 & \lambda_2^2 \\
1 & \lambda_3 & \lambda_3^2
\end{array}
\right|
$$
$$
=-\prod_{1\le j< k\le 3} (\lambda_j-\lambda_k)^2 \frac{ \displaystyle \prod_{\ell=1}^3 q(\lambda_{\ell})}{ \displaystyle \prod_{\ell=1}^3 W(\lambda_{\ell}) \prod_{\ell=1}^3 p^{\prime}(\lambda_{\ell}) } \, .
$$
Then use the definition of the resultant in the form (\ref{Res1}):
$$
=-\prod_{1\le j< k\le 3} (\lambda_j-\lambda_k)^2 \frac{ \displaystyle \prod_{\ell=1}^3 q(\lambda_{\ell})}{ \displaystyle \frac{ \mathcal R(p(x),W(x))}{p_0^9} p_0^3 (-1)^3 \prod_{1\le j< k\le 3} (\lambda_j-\lambda_k)^2  }
$$
Use now the alternative definition of the resultant (\ref{Res2}):
$$
=\frac{ \displaystyle p_0^6 \prod_{\ell=1}^3 q(\lambda_{\ell})}{ \displaystyle \mathcal R(p(x),W(x))   }= \frac{ p_0 \mathcal R(p(x),q(x))}{(-1)^{9 \times 3} \displaystyle \prod_{j=1}^9 p(x_j)} \, .
$$
Thus, the second equality from (\ref{HmHn}) is true.

We just proved this equality under the additional assumption that polynomial $ p(x) $ has all its zeros distinct. To extend this result to the general case, the traditional trick consists in application of the
Weil's Principle of the irrelevance of algebraic inequalities (presented in Appendix). By Corollary \ref{CorDiscr}, the condition of distinction (simplicity) of zeros of polynomial with symbolic (indeterminate) coefficients can be expressed as an algebraic inequality with respect to these coefficients. In accordance with  the referred principle, the algebraic identity which is valid under an extra assumption in the form of algebraic inequality, should be valid for all the values of indeterminates.

To prove the second equality from (\ref{Hm+1Hn+1}), multiply the determinant
$$
H_{4}(\{\widetilde \tau\})=
\left|
\begin{array}{cccc}
\widetilde \tau_{0} & \widetilde \tau_{1} & \widetilde \tau_{2} & \widetilde \tau_{3} \\
\widetilde \tau_{1} & \widetilde \tau_{2} & \widetilde \tau_{3} & \widetilde \tau_{4} \\
\widetilde \tau_{2} & \widetilde \tau_{3} & \widetilde \tau_{4} & \widetilde \tau_{5} \\
\widetilde \tau_{3} & \widetilde \tau_{4} & \widetilde \tau_{5} & \widetilde \tau_{6}
\end{array}
\right|
$$
from the right-hand side by the determinant
$$
\left|
\begin{array}{cccc}
1 & 0 & 0 & p_3/p_0 \\
0 & 1 & 0 & p_2/p_0 \\
0 & 0 & 1 & p_1/p_0 \\
0 & 0 & 0 & 1
\end{array}
\right| \ ,
$$
evidently equal to $ 1 $. This results in the determinant which differs from the initial one only in the last column:
$$
\frac{1}{p_0}
\left|
\begin{array}{cccc}
\widetilde \tau_{0} & \widetilde \tau_{1} & \widetilde \tau_{2} & p_0\widetilde \tau_{3}+p_1 \widetilde \tau_{2} + p_2 \widetilde \tau_{1} + p_3  \widetilde \tau_{0} \\
\widetilde \tau_{1} & \widetilde \tau_{2} & \widetilde \tau_{3} & p_0\widetilde \tau_{4}+p_1 \widetilde \tau_{3} + p_2 \widetilde \tau_{2} + p_3 \widetilde \tau_{1} \\
\widetilde \tau_{2} & \widetilde \tau_{3} & \widetilde \tau_{4} & p_0\widetilde \tau_{5}+p_1 \widetilde \tau_{4} + p_2 \widetilde \tau_{3} + p_3 \widetilde \tau_{2} \\ \widetilde \tau_{3} &  \widetilde \tau_{4} & \widetilde \tau_{5} & p_0\widetilde \tau_{6}+p_1 \widetilde \tau_{5} + p_2 \widetilde \tau_{4} + p_3 \widetilde \tau_{3}
\end{array}
\right| \, .
$$
Consequently:
$$
p_0\widetilde \tau_{3}+p_1 \widetilde \tau_{2} + p_2 \widetilde \tau_{1} + p_3 \tau_{0} = \sum_{j=1}^9 \frac{q(x_j)}{p(x_j)} \frac{p_0x_j^3+p_1x_j^2+p_2x_j+p_3}{W^{\prime}(x_j)} =
\sum_{j=1}^9 \frac{q(x_j)}{p(x_j)} \frac{p(x_j)}{W^{\prime}(x_j)}
=
\sum_{j=1}^9 \frac{q(x_j)}{W^{\prime}(x_j)}\stackrel{(\ref{Eu-Lag1})}{=} 0 \, .
$$
Similarly the validity of equalities
$$
p_0\widetilde \tau_{4}+p_1 \widetilde \tau_{3} + p_2 \widetilde \tau_{2} + p_3  \widetilde \tau_{1}=
\sum_{j=1}^9 \frac{q(x_j)x_j}{W^{\prime}(x_j)}=0
$$
and
$$
p_0\widetilde \tau_{5}+p_1 \widetilde \tau_{4} + p_2 \widetilde \tau_{3} + p_3   \widetilde \tau_{2}
=
\sum_{j=1}^9 \frac{q(x_j)x_j^2}{W^{\prime}(x_j)}=0 \ ,
$$
can be established whereas
$$
p_0\widetilde \tau_{6}+p_1 \widetilde \tau_{5} + p_2 \widetilde \tau_{4} + p_3  \widetilde \tau_{3}
=
\sum_{j=1}^9 \frac{q(x_j)x_j^3}{W^{\prime}(x_j)}=q_0 \ .
$$
Therefore,
$$
H_{4}(\{\widetilde \tau\})=
\frac{1}{p_0}
\left|
\begin{array}{cccc}
\widetilde \tau_{0} & \widetilde \tau_{1} & \widetilde \tau_{2} & 0 \\
\widetilde \tau_{1} & \widetilde \tau_{2} & \widetilde \tau_{3} & 0\\
\widetilde \tau_{2} & \widetilde \tau_{3} & \widetilde \tau_{4} & 0 \\
\widetilde \tau_{3} &  \widetilde \tau_{4} & \widetilde \tau_{5} & q_0
\end{array}
\right|=
\frac{q_0}{p_0} H_{3}(\{\widetilde \tau\}) \, .
$$
wherefrom follows the second equality from (\ref{Hm+1Hn+1}).
\qed

~\\
\indent\textbf{Remark 6.1.} The number of interpolation values (\ref{pqmnogo}) exceeds twice the number of coefficients of both polynomials $ p(x) $ and $ q(x) $, i.e. the set of interpolation values is redundant for the resultant evaluation. However, one can notice from the statement of the previous theorem that only the set $ \{p(x_j)/q(x_j)\}_{j=1}^{n+m+1} $ of ratios (or their reciprocals) is involved in the resultant computation\footnote{As for some extra multiples in the right-hand sides of formulas (\ref{HmHn}) and (\ref{Hm+1Hn+1}), one may ignore the case of their vanishing as a \emph{practically improbable} if the polynomials are randomly chosen over \underline{infinite} fields.}. This set is not redundant.

\begin{example}
Given the interpolation table for the values of $ p(x)/q(x) $ at the nodes
$$   \{ -7, -4, -3, 2, 4, 6, 9, 11, 12 \} $$
find the values for the parameter $ \alpha $ under which the polynomials
$$ p(x)= 4\,x^3-3\,x^2+ (-8+20\sqrt{3})x-7 \quad \mbox { and } \ q(x) = x^5+\alpha\,x^4+21\,x^3-6\,x^2+4\,x-1 $$
possess a common zero.
\end{example}

\textbf{Solution.} From Theorem \ref{ThResInt}, let us take the resultant representation in the form of the determinant of the lowest possible order, namely $ H_3(\{ \widetilde \tau \}) $.

Compute the values (\ref{wtau}):
$$
\widetilde \tau_0=\frac{1}{{\scriptstyle 10919077355342296572660820982567357300658307770}}
$$
$$
\times \bigg(
{\scriptstyle (48383784371515509269913418094982273540012
-27491338967063529790699709261310825633099\sqrt{3})}\alpha
$$
$$
-{\scriptstyle 338097744094156301245854735877467728763783 \sqrt{3} +590927606274219600412492089059592472570254} \bigg), \dots
$$
The resultant $ \mathcal R(p(x),q(x)) $ up to a numerical factor  equals
$$
H_3(\{\widetilde \tau\})=
\left|
\begin{array}{ccc}
\widetilde \tau_0 & \widetilde \tau_1 & \widetilde \tau_2 \\
\widetilde \tau_1 & \widetilde \tau_2 & \widetilde \tau_3 \\
\widetilde \tau_2 & \widetilde \tau_3 & \widetilde \tau_4
\end{array} \right|=
\frac{5305630018587757052\sqrt{3}-9179929579442616601}{737750572976743797348793688224543583031540}
$$
$$
\times (\alpha+9)(9604\alpha^2+(6995120\sqrt{3}-12080761)\alpha+135361627-77980120\sqrt{3}) \, .
$$
It vanishes for $ 3 $ values of the parameter, with one of them $ \alpha=-9 $. For any of these values, the common zero for $ p(x) $ and $ q(x) $ can be evaluated by the following formula:
$$
\lambda= \frac{3}{4}-\left|\begin{array}{cc} \widetilde \tau_0 & \widetilde \tau_2 \\ \widetilde \tau_1 & \widetilde \tau_3  \end{array} \right| \bigg/
\left|\begin{array}{cc} \widetilde \tau_0 & \widetilde \tau_1 \\ \widetilde \tau_1 & \widetilde \tau_2  \end{array} \right| \, .
$$
Thus, for $ \alpha=-9 $ one gets: $ \lambda=2-\sqrt{3} $. \qed

~\\
\indent\textbf{Remark 6.2.} We do not give here any justification for the last step in the solution of the previous example. The general formula for  common zero evaluation\footnote{In case of its uniqueness.} is composed from two coefficients of the polynomial $ p(x) $ and two coefficients of the Hankel polynomial $ \mathcal H_{n-1}(x; \{\widetilde \tau\}) $:
$$
\lambda= -\frac{p_1}{p_0} + \frac{\widetilde h_{n-1,1}}{H_{n-1}(\{\widetilde \tau\})}
\stackrel{(\ref{RatNum})}{=}
-\frac{\widetilde h_{n1}}{H_n(\{\widetilde \tau\})} + \frac{\widetilde h_{n-1,1}}{H_{n-1}(\{\widetilde \tau\})} \,  .
$$
The second equality is to be applied only for the case of symbolic parameter dependent polynomials (like those treated in the example); for this case, we assign to the fraction  $ \widetilde h_{n1}/H_n(\{\widetilde \tau\}) $
the value of its limit when the parameter tends to a value annihilating $ H_n(\{\widetilde \tau\}) $.
The explanation needs a preliminary definition of an extra notion from Elimination Theory, namely the \textbf{first subresultant} \cite{Bocher,VZG} of the resultant $ \mathcal R (p(x),q(x)) $. Although the minor $ H_{n-1}(\{\widetilde \tau\}) $ \underline{does not coincide} formally with the subresultant but it possesses similar properties.

\section{Conclusions}

We have developed an approach for the solution of the polynomial and rational interpolation problem originated in the paper by Jacobi. It consists in representing the interpolant by virtue of appropriate Hankel polynomials. The appearance of the latter in relation to the interpolation problem should not be taken as an unexpected. Indeed, they have been naturally appeared in the problem of interpolation by sum of exponentials \cite{Henrici}, where the interpolation function for the table (\ref{table}) has to be found in the form $ f(x) = \sum_{j=1}^{m} a_j e^{k_j x} $. Application of the results by the XIX century scholars to old stated problem in modernized versions

Our investigation can, by no means, be considered as complete. Among the several problems remained for further investigation, the most fascinating one is that addressed in Section \ref{SPoly_Error}. 
Given the erroneous data set how to distinguish systematic errors from the non-systematic ones?

\section{Appendix: Weil's Principle} \label{Appen}
\setcounter{equation}{0}
\setcounter{theorem}{0}
\setcounter{example}{0}

In 1946 Hermann Weil stated the following theorem known as the Principle of the Irrelevance of Algebraic Inequalities \cite{Weil1946}:

\begin{theorem}[Weil H.] Let $ \mathfrak R $ be an infinite integral domain  with $ n $ independent indeterminates $ x_1,\dots,x_n $. Let $ p \not\equiv 0 $ and $ q $ be polynomials in $
\mathfrak R [x_1,\dots,x_n ] $ such that if $ p(\mathfrak r_1,\dots, \mathfrak r_n)\ne 0 $, for some $ \mathfrak r_i $ in $ \mathfrak R $, then $ q(\mathfrak r_1,\dots, \mathfrak r_n)= 0 $.
Then $ q \equiv 0 $.
\end{theorem}



\newpage
\begin{thebibliography}{99}

\bibitem{Becker&Lab} Beckermann B., Labahn G.  \emph{Fraction-free computation of matrix rational interpolants and matrix GCD's.} 2000. SIAM J. Matrix Anal. Appl. V. 22 (1), pp. 114--144


\bibitem{Ber&Wel} Berlekamp E., Welch L.  Error Correction of Algebraic Block Codes. US Patent Number 4\,633\,470, 1986.



\bibitem{blahut} Blahut R.  Fast Algorithms for Digital Signal Processing. Addison-Wesley, 1985.

\bibitem{Bocher} B\^ocher M.  Introduction to Higher Algebra. NY. Macmillan, 1907

\bibitem{Cauchy} Cauchy A.-L. Cours d'Analyse de l'\'Ecole Royale Polytechnique: Part I: Analyse Alg\'ebrique. Paris, France: L'Imprimerie Royale, 1821, pt. 1. Annotated English Translation:
Bradley R.E., Sandifer C.E. Cauchy's Course d'analyse. NY. Springer, 2009

\bibitem{DAndrea15} D'Andrea C., Krick T., Szanto A. \emph{Subresultants, Sylvester sums and the rational interpolation problem.} J.Symbolic Comput. 2015. V. 68, pp. 72--83

\bibitem{Gantmacher} Gantmacher F.R. The Theory of Matrices. NY. Chelsea, 1959

\bibitem{Henrici} Henrici P.  Applied and Computational Complex Analysis. V. 1. NY. Wiley, 1974


\bibitem{Jacobi36} Jacobi C.G.J.  \emph{De eliminatione variabilis e duabus aequationibus algebraicis.} J.reine angew. Math. 1836. V. 15, pp. 101--124

\bibitem{Jacobi46} Jacobi C.G.J.  \emph{\"Uber die Darstellung einer Reihe gegebner Werthe durch eine gebrochne rationale Function.} J.reine angew. Math. 1846. V. 30, pp. 127--156

\bibitem{Joach} Joachimsthal F. \emph{Bemerkungen \"uber den Sturm'schen Satz.} J.reine angew. Math. 1854. V. 48, pp. 386--416


\bibitem{Kronecker81_1} Kronecker L. \emph{\"Uber einreihige Determinanten.} Nachrichten den K\"oniglichen Gesellschaft der Wissenschaften zu G\"ottingen. 1881, V. 9, pp.271--279

\bibitem{Kronecker81_2} Kronecker L. \emph{Zur Theorie der Elimination einer Variabeln aus zwei algebraischen Gleichungen.}
Monatsberichte der K\"oniglichen Preussische Akademie des Wissenschaften zu Berlin. 1881, Juni, pp. 535--600

\bibitem{Mejiering} Meijering E. \emph{A chronology of interpolation: from ancient astronomy to modern signal and image processing.}
Proc. IEEE. 2002, V.90, No. 3,  pp. 319--342

\bibitem{Netto} Netto E. \emph{Zur Cauchy'schen Interpolationsaufgabe.} Math. Ann. 1893, V.42 (3), pp. 453--456

\bibitem{Uspensky} Uspensky J.V. Theory of Equations. New York. McGraw-Hill. 1948.

\bibitem{VZG} von zur Gathen J., Gerhard J. Modern Computer Algebra. Cambridge. Cambridge University Press,  2003.

\bibitem{Weil1946} Weil H. The Classical Groups, their Invariants and Representations. Princeton, N.J., Princeton University Press; London, G. Cumberlege, Oxford University Press. 1946


\end{thebibliography}
\end{document}